\documentclass[apjl]{emulateapj} 
\usepackage{amsmath}
\usepackage{xcolor}
\def\alwaysmath#1{\ifmmode{#1}\else{$#1$}\fi} 
\newcommand{\teff}{$T_{\rm{eff}}$}
\newcommand{\logg}{$\log{g} $}
\newcommand{\microt}{$\xi_{\rm{t}}$}

\newcommand{\feh}{[M/H]}
\newcommand{\fehw}{[Fe/H]}
\newcommand{\cfe}{[C/M]}
\newcommand{\nfe}{[N/M]}
\newcommand{\afe}{[$\alpha$/M]}
\newcommand{\ferre}{{\sc ferre}}
\newcommand{\ctwo}{$\chi^2$}
\newcommand{\kms}{km~s$^{-1} $}

\lefthead{Garc\'{\i}a P\'erez et al.}

\begin{document}

\title{ASPCAP: the APOGEE Stellar Parameter and Chemical Abundances Pipeline}

\author{Ana E. Garc\'{\i}a P\'erez\altaffilmark{1,2,3}, Carlos Allende Prieto\altaffilmark{2,3}, Jon A. Holtzman\altaffilmark{4}, Matthew Shetrone\altaffilmark{5}, Szabolcs M\'esz\'aros\altaffilmark{6}, Dmitry Bizyaev\altaffilmark{7,8}, Ricardo Carrera\altaffilmark{2,3}, Katia Cunha\altaffilmark{9,10}, D. A. Garc\'{\i}a-Hern\'andez\altaffilmark{2,3}, Jennifer A. Johnson\altaffilmark{11}, Steven R. Majewski\altaffilmark{1}, David L. Nidever\altaffilmark{12}, Ricardo P. Schiavon\altaffilmark{13}, Neville Shane\altaffilmark{1}, Verne V. Smith\altaffilmark{14}, Jennifer Sobeck\altaffilmark{1}, Nicholas Troup\altaffilmark{1}, Olga Zamora\altaffilmark{2,3}, 
Jo Bovy\altaffilmark{15}, Daniel J. Eisenstein\altaffilmark{16}, Diane Feuillet\altaffilmark{4}, Peter M. Frinchaboy\altaffilmark{17}, Michael R. Hayden\altaffilmark{4}, Fred R. Hearty\altaffilmark{18}, Duy C. Nguyen\altaffilmark{19}, Robert W. O'Connell\altaffilmark{1}, Marc H. Pinsonneault\altaffilmark{11}, David H. Weinberg\altaffilmark{11}, John C. Wilson\altaffilmark{1}, Gail Zasowski\altaffilmark{20}}

\altaffiltext{1}{Department of Astronomy, University of Virginia, Charlottesville, VA 22904-4325, USA}
\altaffiltext{2}{Instituto de Astrof\'{\i}sica de Canarias, 38205 La Laguna, Tenerife, Spain (agp@iac.es)}
\altaffiltext{3}{Departamento de Astrof\'{\i}sica, Universidad de La Laguna, 38206 La Laguna, Tenerife, Spain}
\altaffiltext{4}{New Mexico State University, Las Cruces, NM 88003, USA}
\altaffiltext{5}{University of Texas at Austin, McDonald Observatory, Fort Davis, TX 79734, USA}
\altaffiltext{6}{ELTE Gothard Astrophysical Observatory, H-9704 Szombathely, Szent Imre Herceg st. 112, Hungary} 
\altaffiltext{7}{Apache Point Observatory, P.O. Box 59, Sunspot, NM 88349-0059, USA)}
\altaffiltext{8}{Sternberg Astronomical Institute, Moscow State University, Moscow, Russia}
\altaffiltext{9}{Observat\'orio Nacional, S\~ao Crist\'ov\~ao, Rio de Janeiro, Brazil}
\altaffiltext{10}{Steward Observatory, University of Arizona, Tucson 85719} 
\altaffiltext{11}{Department of Astronomy, The Ohio State University, Columbus, OH 43210, USA}
\altaffiltext{12}{Department of Astronomy, University of Michigan, Ann Arbor, MI 48109, USA} 
\altaffiltext{13}{Astrophysics Research Institute, Liverpool John Moores University, Egerton Wharf, Birkenhead, Wirral CH41 1LD, UK}
\altaffiltext{14}{National Optical Astronomy Observatories, Tucson, AZ 85719, USA}
\altaffiltext{15}{Institute for Advanced Study, Einstein Drive, Princeton, NJ 08540, USA}
\altaffiltext{16}{Harvard-Smithsonian Center for Astrophysics, Cambridge, MA 02138, USA}
\altaffiltext{17}{Department of Physics and Astronomy, Texas Christian University, Fort Worth, TX 76129, USA}
\altaffiltext{18}{Department of Astronomy and Astrophysics, The Pennsylvania State University, University Park, PA 16802, USA}
\altaffiltext{19}{Dunlap Institute for Astronomy and Astrophysics, University of Toronto, Toronto, ON, M5S 3H4, Canada}
\altaffiltext{20}{Johns Hopkins University, Department of Physics and Astronomy, Baltimore, MD, 21218, USA}

\begin{abstract}

The Apache Point Observatory Galactic Evolution Experiment (APOGEE) 
has built the largest moderately high-resolution ($R\approx 22,500$) 
spectroscopic map of the stars across the Milky Way, and including dust-obscured areas. 
The APOGEE Stellar Parameter and Chemical Abundances Pipeline (ASPCAP) is the software 
developed for the automated analysis of these spectra. ASPCAP determines atmospheric 
parameters and chemical abundances from observed spectra 
by comparing observed spectra to libraries of theoretical spectra, using {\ctwo} minimization in a multidimensional 
parameter space. The package consists of a {\sc fortran90} code that does 
the actual minimization and a wrapper IDL code for book-keeping and data handling. 
This paper explains in detail the ASPCAP components and functionality, and 
presents results from a number of tests designed to check its performance.
ASPCAP provides stellar 
effective temperatures, surface gravities, and metallicities 
precise to 2\%, 0.1~dex, and 0.05~dex, respectively, for most APOGEE stars, which are predominantly giants. 
It also provides abundances for up to 15 chemical elements with various
levels of precision, typically under 0.1~dex. The final data release (DR12) of the 
Sloan Digital Sky Survey III contains an APOGEE database of more than 150,000 stars.
ASPCAP development continues in the SDSS-IV APOGEE-2 survey.

\end{abstract} 

\section{Introduction}

The Apache Point Observatory Galactic Evolution Experiment (APOGEE\footnote{http://www.sdss.org/surveys/apogee/.}, Majewski et al. 2015) is one of the 
three projects of the Sloan Digital Sky Survey III \citep[SDSS-III;][]{Eisenstein11}. 
Between 2011-2014, the survey obtained high-resolution, 
near infrared (IR) spectra of over $150,000$ stars using the APOGEE multi-object spectrograph \citep{Wilson12} 
attached to the Sloan 2.5-m telescope \citep{Gunn06}. Observations will 
 continue through 2020 in the framework of the APOGEE-2 survey, part of SDSS-IV.
The three main Galactic stellar components (bulge, disk, and halo) are mapped using the kinematical and 
chemical information derived from an automated spectral analysis. The unparalleled APOGEE stellar sample and associated data products represent a powerful means to understand the origins 
and evolution of the Milky Way.

APOGEE targeted red giant stars selected from the 2MASS Point Source Catalog \citep{Skrutskie06}, 
employing de-reddened photometry and a simple color cut ($7 \le H \le 13.8$ and $[J-K]_0 \ge 0.5$; for more details, see \citealt{Gail13}). 
The majority of targets have effective temperatures in the range 3500\ $<T_{\rm eff}<5500$ K, 
although warmer (telluric) stars were also targeted to correct for absorption lines 
produced in the atmosphere of the Earth (H$_{2}$O, CO$_2$, and CH$_4$). 
About $20\%$ of the stars in APOGEE are dwarfs. 
 
APOGEE performs a detailed characterization of the inner Galaxy via the near-infrared observation of a large 
numbers of stars and the accompanying derivation of their kinematical and chemical properties at high precision. 
The $H$-band (1.51-1.70~${\mu}$m) APOGEE observations are acquired at high resolution ($R=22,500$) and high signal-to-noise ($S/N \geq $ 100 per half-resolution element or $\sim$ per pixel). They are also 
rich in chemical information. At least 15 individual element abundances can be measured, and the $S/N$ ratio is high enough to
allow typical abundance precisions better than 0.1 dex.
Such multi-dimensional study requires an automated, detailed, and accurate spectral analysis pipeline.
This level of automated analysis would be challenging under any circumstances, and it is
particularly challenging for the $H$-band wavelength regime, where many features
are blended (e.g., by molecular line contaminants) and which has not been studied
as extensively as optical ranges.

Several optical surveys have already created automated spectral analysis software used for the extraction of atmospheric parameters and chemical abundances, including: the Sloan Extension for Galactic
Understanding and Exploration \citep[SEGUE;][]{Yanny09}, the Radial Velocity Experiment \citep[RAVE; e.g,][]{Steinmetz06}, the Large Sky Area Multi-Object Fiber Spectroscopic Telescope (LAMOST) Experiment for Galactic Understanding and Exploration,\cite[LEGUE;][]{Zhao12}, the Abundances and Radial Velocity Galactic Origins Survey \citep[ARGOS;][]{Freeman13}, and the 
Gaia-ESO Survey \citep[GES;][]{Gilmore12}. Notably, SEGUE \citep{Lee08} and LEGUE \citep{Xiang15} generate data products through fully-automated, completely self-contained analysis ``pipelines" (named the SEGUE Stellar Parameter Pipeline [SSPP] and the LAMOST Stellar Parameter Pipeline [LASP]).

APOGEE has developed its own pipeline for parameter determinations: the APOGEE Stellar Parameter and Chemical Abundances Pipeline (ASPCAP). This pipeline operates on combined visit or individual-visit spectra processed by the APOGEE data reduction pipeline \citep{Nidever15}. ASPCAP is innovative in the use of the $H$-band to extract abundances accurately for a large number of elements (up to 15) in an immense stellar sample ($>10^{5}$ targets). 
ASPCAP performs spectral analysis over a wide wavelength range ($\sim200$~nm) and consequently, manipulates a large volume of data (approximately $10^4$ wavelength points). Further complicating the analysis is the presence in typical APOGEE targets of numerous molecular features (from CO, CN and OH lines) that can affect the determination of the spectroscopic parameters via the contribution of these features to the molecular equilibrium and continuous opacity.

The first year of APOGEE observations and ASPCAP results were released in the Sloan Digital Sky Survey 
 $10^\mathrm{th}$ data release\footnote{http://www.sdss3.org/dr10/.} (DR10; \citealt{DR10}), while the full (three years) APOGEE 
 database of more than 150,000 stars is now publicly available in DR12\footnote{http://www.sdss.org/dr12/.} 
(\citealt{DR12}). The APOGEE reduction and analysis software is also released through the SDSS repository.\footnote{http://www.sdss.org/dr12/software/products/.}

This paper provides a thorough description of the ASPCAP software as well as relays the results of numerous performance and reliability tests. Section 2 presents the overall structure of the ASPCAP software. In Section 3, the model spectra employed in the ASPCAP analysis of APOGEE data are discussed. Sections 4 and 5 contain detailed descriptions of the {\ctwo} minimization code {\ferre} and the IDL wrapper, respectively. Section 6 is devoted to the testing of the ASPCAP algorithms and the software. Finally, Section 7 reviews the performance of the ASPCAP pipeline with actual APOGEE Survey data.

\section{Overall ASPCAP Structure}

\begin{figure*}
\figurenum{1}
\begin{center}
\includegraphics[page=1,angle=0,trim=0cm 2cm 0cm 0cm, clip,scale=0.5]{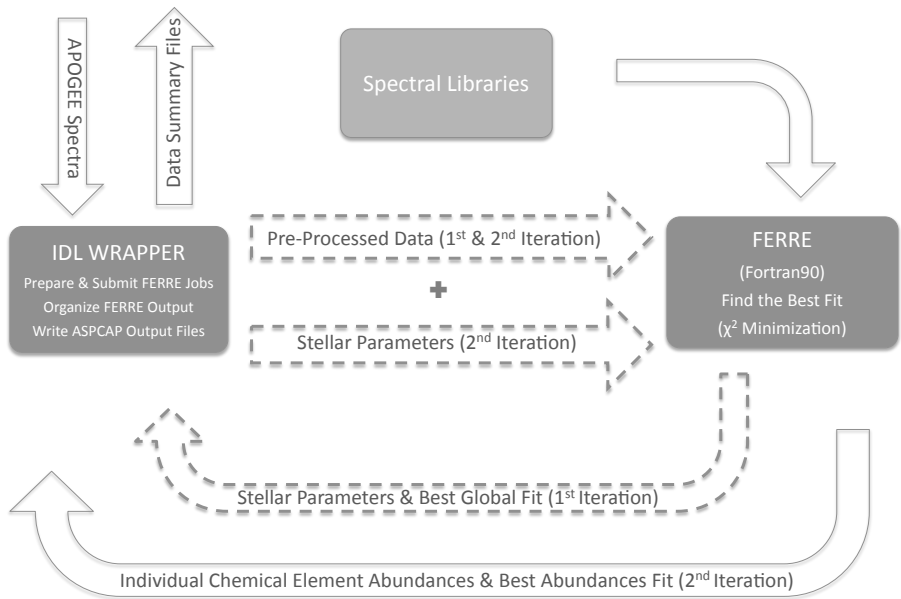}
\end{center}
\caption{Overview of ASPCAP workflow. The IDL wrapper pre-processes the APOGEE spectra for FERRE,
which identifies the best six- or seven-parameter fit (depending on whether microturbulence is
free or fixed), using the model spectral libraries. On the second iteration, FERRE is run
fitting one elemental abundance at a time, with windows used to select the portions of the
spectrum that are sensitive to that element. The IDL wrapper writes output files based on
the FERRE results.}
\label{flowchart}
\end{figure*}

The ASPCAP software has two main functional components:  a {\sc fortran90} code ({\ferre}, Section~\ref{ferre}), and an IDL wrapper (Section~\ref{wrapper}). The general schematic of ASPCAP is displayed in Figure~\ref{flowchart}. The IDL wrapper is multifunctional in that it reads the input APOGEE spectra and prepares them for analysis as well as performs the overall ``bookkeeping", which entails multiple calls to the {\sc fortran90} optimization code. The workhorse of ASPCAP is the {\sc fortran90} code, which compares the APOGEE observed spectra to a library of synthetic spectra and, subsequently, identifies 
the set of atmospheric parameters and abundances that yields the best fit spectrum. Specifically, during the observed spectrum fitting process, the {\sc fortran90} code performs an interpolation in the synthetic spectral grid and generates a best-fit (interpolated) synthetic spectrum, which then allows for the final parameter extraction.  

\begin{figure*}
\figurenum{2}
\begin{center}
\includegraphics[page=1,angle=-270,trim=2.5cm 2.5cm 3.5cm 14.2cm, clip,scale=0.99]{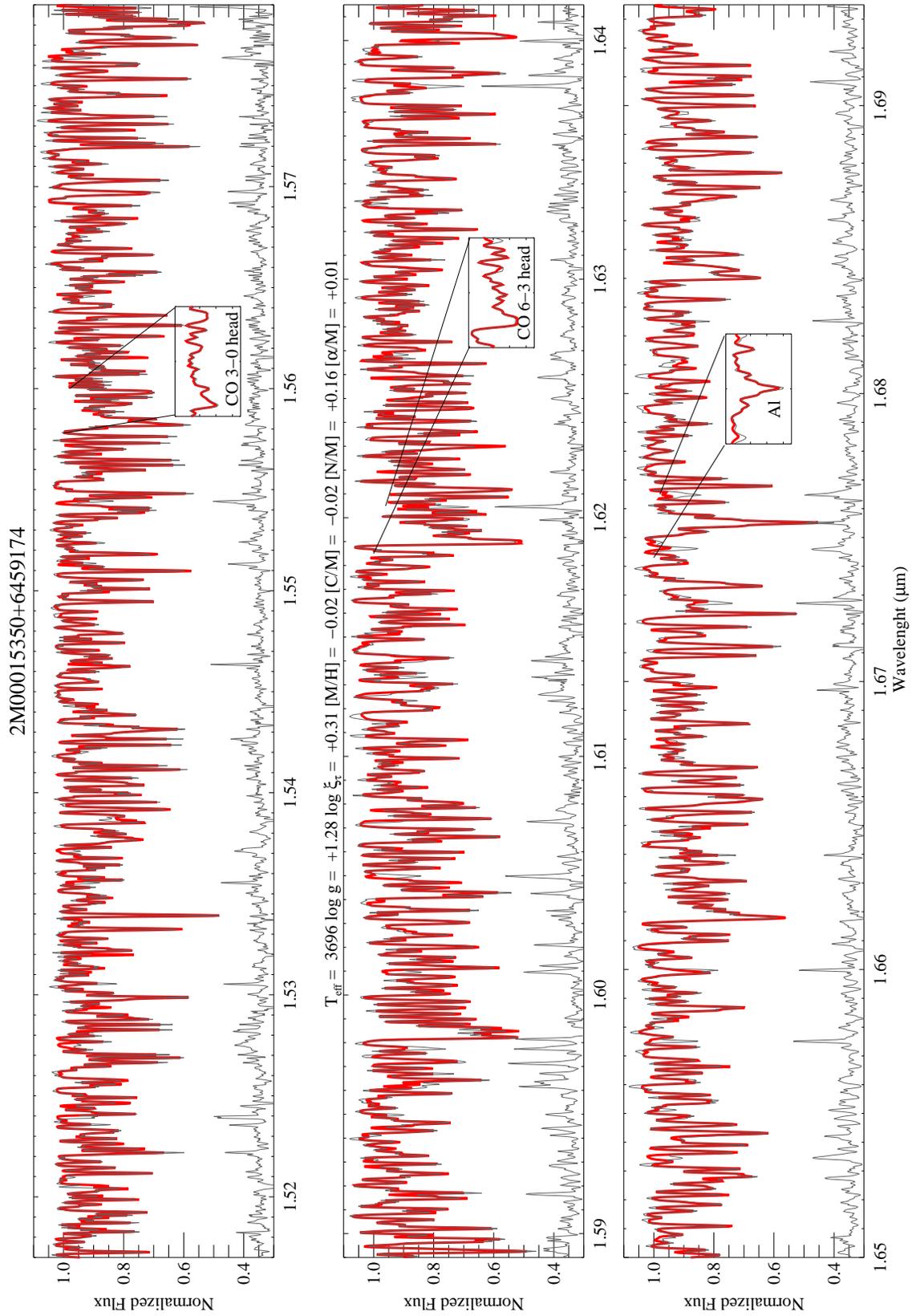}
\includegraphics[page=2,angle=-270,trim=2.5cm 2.5cm 3.5cm 14.0cm, clip,scale=0.99]{figure2.pdf}
\includegraphics[page=3,angle=-270,trim=2.5cm 2.0cm 3.5cm 14.5cm, clip,scale=0.99]{figure2.pdf}
\end{center}
\caption{The observed normalized spectrum (black) of the cool, solar metallicity star 2M00015350-6459174 (\feh$\ =-0.02$), its ASPCAP best spectral fit (red), 
and the residuals (relative differences), shifted by 0.35, are shown at the bottom of each panel. The spectrum from each of APOGEE's three detectors is shown in a 
separate panel.}
\label{synthesis}
\end{figure*}

As shown in Figure~\ref{flowchart}, the iterative determination of ASPCAP proceeds in a two-step fashion. The first step is to assign the input APOGEE observed spectrum a set of fundamental atmospheric parameters (effective temperature, \teff; surface gravity, $\log{g}$; microturbulent velocity, \microt; scaled-solar general metallicity, \feh) from a first-pass fit of the entire APOGEE spectrum. In conjunction with these parameters, the abundances of C, N and the $\alpha$-elements (O, Mg, Si, S, Ca, and Ti) are also allowed to vary around the scaled-solar values due to their significant spectral contribution in the $H$-band. Figure~\ref{synthesis} gives an illustration of an (example) APOGEE spectrum for a cool, solar metallicity giant and its best ASPCAP global fit. The second step of ASPCAP is to extract the individual element abundances, one at a time, from the fitting of spectral windows. These windows have been optimized for each element.

\section{Model Spectra}
\label{models}

The model synthetic spectra are generated by solving the radiative transfer equation for a grid of model atmospheres over the APOGEE portion of the 
$H$-band wavelength regime. In this section we provide details on the model atmospheres, the atomic and molecular line lists, and the spectral synthesis calculations. 
Some information on the structure of the ASPCAP databases is also provided.\footnote{Spectral libraries can be downloaded from http://data.sdss3.org/sas/{\it datarelease}/apogee/spectro/redux/speclib/ with {\it datarelease} being dr10 or dr12.}

\subsection{Model Atmospheres}

ASPCAP uses a set of model atmospheres specifically generated for APOGEE; the APOGEE ATLAS9 models {\citep{Meszaros12}}, which are 
based on the ATLAS9 model atmosphere code from \citet{Castelli04}. These models are one-dimensional, assume local themodynamic equilibrium (LTE), and use no 
convective overshooting. In DR10, ASPCAP results relied upon on ATLAS9 models with scaled-solar compositions and the set of solar reference photospheric abundances from 
\citet{Grevesse98}. For DR12, the results were based upon customized abundances (a set of varied C, N, and $\alpha$ contents at a given metallicity) and a more recent set of solar reference abundances 
(\citealt{Asplund05}, the photospheric abundance column of their Table~1). The grid steps for the atmospheric parameters and (C, N, $\alpha$) abundances as well as the associated ranges are given in Table~\ref{libraries} (see Zamora et al. 2015 for library names nomenclature). Note that the step sizes are small enough to minimize interpolation uncertainties and also allow for efficient ASPCAP computation and performance.

\subsection{Line List}

The input atomic and molecular data are essential for accurate determination of the atmospheric parameters and abundances. The base 
line list originated from the R. Kurucz website (\url{http://kurucz.harvard.edu/}), which provides wavelengths, excitation potentials, oscillator strengths, and 
hyperfine structure information. Best effort was then made to update line list values with laboratory measurements of wavelength and (most critically) oscillator strengths. When possible, van der 
Waals damping constants based on the study of Barklem, Anstee, \& O'Mara (1998) were used. Molecular data for CO, OH, CN, C$_2$, H$_2$ and SiH were included.
Finally, astrophysical inversion from matching the spectra of the Sun and Arcturus ($\alpha$ Boo) was employed to fine tune line list values (i.e., gf's and C6 constants). For a complete description of the line list assembly, consult Shetrone et al. (2015).  The final line list adopted for DR10 was m201105101120 and for DR12, m201312161124.

\subsection{Spectral Synthesis}

The synthetic spectral library was generated with the line transfer code ASS$\epsilon$T \citep{Koesterke09, Koesterke08}. The solar photospheric reference abundances from \cite{Asplund05} were adopted and 
a terrestrial isotopic composition for C, N, and O was used. Initially, synthetic spectra were computed at very high spectral resolution (1--2 km s$^{-1}$) and then, later re-smoothed to account for instrumental broadening. Three spectral regions, which correspond to the spectral coverage of the three APOGEE detectors, are synthesized: $\lambda=$1.51681--1.57923~${\mu}$m, 1.58814--1.64166~${\mu}$m, and 1.64995--1.69367~${\mu}$m (vacuum wavelengths). The resampling of the synthetic spectra to match that of the APOGEE observed spectra was done with a constant step size in $\log{\lambda}$---see 
Table~\ref{wavsol}. In that table, $N_\mathrm{pixels}$ and $p_i$ are the number of pixels and the pixel {\it ith} number for each detector synthesized spectrum, respectively. Additional details regarding the available synthetic spectral libraries can be found in Zamora et al. (2015).

The spectra in DR10 were convolved with a Gaussian kernel to bring the resolving power $\lambda/$FWHM$\equiv~R=22,500$. For DR12, the convolution employed a more
realistic, empirical kernel (\citealt{Holtzman15}, Nidever et al. 2015).  For the APOGEE instrumental set-up, spectral resolution variations as high as 10--15\% can occur for different fibers as well as across the APOGEE wavelength regime.  In DR12, accounting for some of these LSF variations \citep{Nidever15} was done by averaging the LSF's of five fibers that were located at equidistant steps along the pseudoslit and then fitting them with a Gauss-Hermite function that varied with wavelength. 
The impact of LSF treatment is discussed in \S\ref{lsfsub}.
DR10 ignored macroturbulence, which is usually 
significantly smaller than the instrumental broadening. However, DR12 used a constant value of 6~{\kms}(FWHM) for the macroturbulence, modeled with a Gaussian kernel.  Note that neither DR10 nor DR12 considered rotational broadening, which could compromise the quality of the derived ASPCAP values for fast rotating stars.

\setlength{\tabcolsep}{2pt}
\begin{table*}
 \tablenum{1}
 \label{libraries}
 \begin{center}
 \caption{Synthetic Spectra Libraries}
 \begin{tabular}{lrrrrrrrrrrrrrrrrrrc}
\hline
\hline
 Name& \multicolumn{3}{c}{\teff} &\multicolumn{3}{c}{\logg} & \multicolumn{3}{c}{\feh}&\multicolumn{3}{c}{\cfe} &\multicolumn{3}{c}{\nfe} &\multicolumn{3}{c}{\afe}  &  Data Release \\
  & & & & & & & & & & \multicolumn{4}{c}{Low High Step} &  & & & & & \\
\hline
p\_aps23k0821\_w123  & 3500 & 5000& 250 & 0 & 5 & 0.5 & $-2.5$ & +0.5 & 0.5 & $-1$ & +1 & 0.25 &$-1$ & +1 & 0.5&$-1$ & +1 & 0.25 & DR10 \\ 
 p\_aps23k0921\_w123 & 4750 & 6000 & 250 & 0 & 5 & 0.5 & $-2.5$ & +0.5 & 0.5  &$-1$ & +1 & 0.25 &$-1$ & +1 & 0.5&$-1$ & +1 & 0.25 & DR10 \\
 n\_aps23k2121\_w123 & 6000 & 10000 & 1000 & 2 & 5 & 1.0 & $-2.5$ & +0.5 & 0.5  &+0 & +0 & 0.00 & +0 & +0 & 0.0 & +0 & +0 & 0.00 & DR10 \\
 n\_aps23k3121\_w123 & 8000 & 15000 & 1000 & 3 & 5 & 1.0 & $-1.0$ & +0.0 & 1.0 & +0 & +0 & 0.00 & +0 & +0 & 0.0 & +0 & +0 & 0.00 & DR10 \\
p6\_apsasGK\_131216\_lsfcombo5v6 & 3500 & 6000 & 250 & 0 & 5 & 0.5 & $-2.5$ & +0.5 & 0.5 & $-1$ & $+1$ & 0.25 & $-1$ & $+1$ & 0.5  &  $-1$ & $+1$ & 0.25  & DR12 \\
p6\_apsasF\_131216\_lsfcombo5v6 &  5500 & 8000 & 250 & 1 & 5 & 0.5 & $-2.5$ & +0.5 & 0.5 & $-1$ & $+1$ & 0.25 & $-1$ & $+1$ & 0.5  & $-1$ & $+1$ & 0.25  & DR12 \\

\hline

\end{tabular}
\end{center}
\end{table*}

\begin{table}
\tablenum{2}
\label{wavsol}
\begin{center}
\caption{Library wavelength scale ($\log\lambda=\mathrm{a}_{0}+\mathrm{a}_{1}*p_{i}$)}
\begin{tabular} {ccc}
\hline
\hline
Npixels & a$_{0}$ & a$_1$ \\
\hline

2920 &4.180932 & 6.000000E-06 \\
2400 & 4.200888 & 6.000000E-06 \\
1894  & 4.217472 & 6.000000E-06  \\

\hline
\end{tabular}
\end{center}
\end{table}

\subsection{Principal Component Analysis Compression}

\label{pcasub}

The cool-star libraries used in ASPCAP have typical 
sizes of tens of gigabytes. There are 5--11 nodes per dimension, 6 or 7 dimensions 
(3 dimensions for hot stars) per library, and of order 10$^4$ wavelengths.
Accessing the libraries on a hard drive, as a direct-access file,
is much slower than holding them in RAM, but even when they do fit in the
computer's RAM, accessing the data becomes slower as the arrays grow in size.

Reducing the size of libraries has many advantages, and we achieve
that by applying principal component analysis compression (PCA; Pearson 1901) 
to identify correlations between the fluxes at different wavelengths 
and compress the model spectra.

The full arrays are too large to perform PCA on them.
We split the arrays into several dozen contiguous wavelength intervals 
(30 pieces with $\sim300$ wavelengths each) and run PCA on 
those. We retain the first 30 components for each, creating arrays of PCA coefficients 
that are the concatenation of the coefficients for each wavelength interval 
(900 coefficients in total). This procedure is very effective, reducing
the size of the libraries by nearly a factor of ten. Compression is 
done on the library nodes independently of ASPCAP runs. The analysis of simulations 
with PCA libraries works well at the metallicities typical of APOGEE targets but can cause 
some problems at low metallicities---see Section~\ref{PCAsubtest}.

\subsection{Database Preparation}
\label{database}

ASPCAP searches in grids of pre-computed, normalized, convolved, and PCA-compressed synthetic spectra 
that cover the stellar spectral classes from early-M to F in DR12 (3500--8000~K), and to 
B (3500--15000~K) in DR10---see Table~\ref{libraries}. There are separate grids per spectral class. 
ASPCAP analysis is not optimized for early-type stars, thus they are not analyzed in DR12. 
The parameter space searched includes the atmospheric stellar parameters \teff, \logg, and \feh, and, 
in most cases, the C, N, and $\alpha$-element abundances. 
Molecular features are used to derive the C, N, and O abundances, but
these features disappear in the spectra of hot stars.
In the DR10 analysis of stars hotter than 6000~K, the number of searched parameters
was reduced by requiring solar scaled abundances, \afe$=$\cfe$=$\nfe$=0$. 

In principle, atmospheric microturbulence is an additional parameter to be considered
in the fitting of the APOGEE spectra, however we found it more effective to 
reduce the dimensionality of the problem by adopting a linear relationship between 
microturbulence and surface gravity:

\begin{equation}
\begin{split}
\xi_{\rm t} &= 2.240-0.300\log g \quad{\rm (DR10)} \\
\xi_{\rm t} &= 2.478-0.325\log g \quad{\rm (DR12)},
\end{split}
\end{equation}
and a \microt$\ =2$~{\kms} in the DR10 analysis of stars hotter than 6000~K. 
The {\microt}-{\logg} relations were derived using a subsample of the APOGEE data analyzed with a library in 
which the microturbulence is a free parameter. 

\section{FERRE}
\label{ferre}

{\ferre} is the optimization code that finds the parameters of the model spectrum that best matches an observed spectrum. 
The code is written in {\sc fortran90} and can take advantage of 
multi-core processors, performing optimizations for several spectra in parallel using OpenMP. 
{\ferre} has been applied to a number of data sets before APOGEE 
(e.g., SDSS SEGUE and BOSS, as illustrated in \citealt{Allende06,Allende14}).\footnote{{\ferre} is publicly available from http://hebe.as.utexas.edu/ferre/.} 
{\ferre} can be used with different configurations that allow a choice of different search algorithms 
or interpolation schemes, which are chosen in a control file. 

\subsection{Algorithm}

Observed spectra are matched against a grid of synthetic spectra 
to search for the best fit. The search algorithms compare observations and 
model spectra using the {\ctwo} as a merit function
\begin{equation}
\chi^2 = \sum_{\lambda} \frac{(O_{\lambda} -F_{\lambda})^2}{\sigma_{\lambda}^2},
\end{equation}
\noindent where $O_{\lambda}$ are the observed fluxes, $F_{\lambda}$ the model fluxes, and
$1/\sigma_{\lambda}^2$ the weights. 

The weights are calculated directly from the 
error bars for the fluxes computed during data reduction, increasing 
artificially the uncertainties in regions severely affected by sky emission lines. When deriving abundances, the $\chi^2$ merit function also
takes into account the sensitivity of different spectral features 
to changes in the abundance of the elements of interest, the lack
of sensitivity to other elements, and the level of agreement between model spectra and actual APOGEE
observations as a function of wavelength---see Section~\ref{abndet}.

Searches are initialized at specific locations. Currently for the global parameter fit, 
we are working with 12 searches, which are symmetrically distributed over the parameter
space, at the centers of the 6D cells resulting from dividing the ranges in [C/M],
[N/M], and [$\alpha$/M] in one bin, the ranges in [M/H] and \logg\ in two bins,
and the range  \teff\ in three bins. The optimization is carried out using the 
\cite{NelderMead65} algorithm, which evaluates and compares the {\ctwo} at the test points of the simplex 
(a triangle in multi-dimensions). As search continues, the simplex moves to a series
of rules, which can shift the search off the nodes of the grid in an attempt to reach regions where 
the chi-squared is lower. The algorithm typically 
requires a few hundred evaluations of the {\ctwo} for a 6D search in our cool-star databases. 
The search stops when the convergence criterion is satisfied:
a standard deviation below $10^{-4}$ for the values of {\ctwo} evaluated at 
the test points of the simplex. There is no special treatment in {\ferre} for dealing with a flat 
{\ctwo} surface, e.g., for the cases of non-detection of a spectral feature in abundance
determinations.  The minimization that yields the lowest $\chi^2$ among the 12 searches
is accepted as the best fit.

Model fluxes need to be interpolated to evaluate fluxes at points off the grid nodes; 
the interpolation in fluxes is more accurate than interpolations of atmospheric structures \citep{Meszaros13b}. 
Early APOGEE ASPCAP analyses and interpolation tests (see Section~\ref{intsub}) give more accurate 
results for higher polynomial orders. ASPCAP used cubic B\`ezier 
interpolation for both DR10 and DR12. Solving for abundances from the global fit with linear or quadratic interpolation leads 
to solutions that systematically cluster the results around the spectral library nodes. This effect is 
more significant for spectra with $S/N < 50$, 
and while this effect becomes very small at $S/N > 70$, smaller steps when implemented 
in the grid (0.25~dex instead of 0.5~dex for \cfe\ and \afe) help to minimize the problem.

{\ferre} has an option to use masks to block spectral windows, or more 
generally to use weights that depend on wavelength in the $\chi^2$ evaluation. 
ASPCAP uses this capability to ignore bad and/or contaminated pixels 
in the fitting process and in the determination of individual elemental 
abundances.

\subsection{Errors}
\label{errors}

The internal random errors in the retrieved parameters can be estimated 
from the inverse of the curvature matrix. This was the method used in DR10 
and DR12, and typically used in {\ferre}. DR10 assumed these internal errors with a factor of 15 
enhancement. The matrix elements of the curvature matrix ($\beta_{ij}$) are calculated from the partial derivatives 
of the synthetic spectra ($F_{\lambda}$): 
\begin{equation}
\beta_{ij} = \sum_{\lambda} \frac{1}{\sigma_{\lambda}^2} 
\frac{\partial F_{\lambda}}{\partial P_i} \frac{\partial F_{\lambda}}{\partial P_j},
\end{equation}
\noindent where $P_i$ are the different parameters/abundances
considered in the optimization \citep{Press07} and $\sigma_{\lambda}$ the flux error.
The inverse of the curvature matrix gives the parameter errors and their 
covariance under the assumption that the likelihood of the data is
described by $\chi^2$ (${\cal L} \propto e^{-\chi^2/2})$ and that the
model gives a correct description of the data (up to observational errors)
for some choice of its parameters.
{\ferre} can also estimate errors by searching for the parameter solution 
multiple times after adding random noise to the observed spectra, according to the 
uncertainty in the observations.
As discussed in \S\ref{errsub}, tests on synthetic spectra show that these
two methods give comparable error estimates and that they reasonably capture
the uncertainty associated with observational noise.
However, empirical estimates of abundance uncertainties based on star clusters
indicate that the true abundance errors are larger than these
internal ASPCAP estimates \citep{Holtzman15}, probably because (unsurprisingly)
the model atmospheres and synthetic spectra remain an imperfect representation
of the true spectra (see \S\ref{errsub} for further discussion).

\subsection{Derivation of Elemental Abundances}
\label{abndet}

\begin{figure*} 
\figurenum{3}
\begin{center}
\includegraphics[angle=0,trim=1cm 0cm 0cm 0cm, clip,scale=0.6]{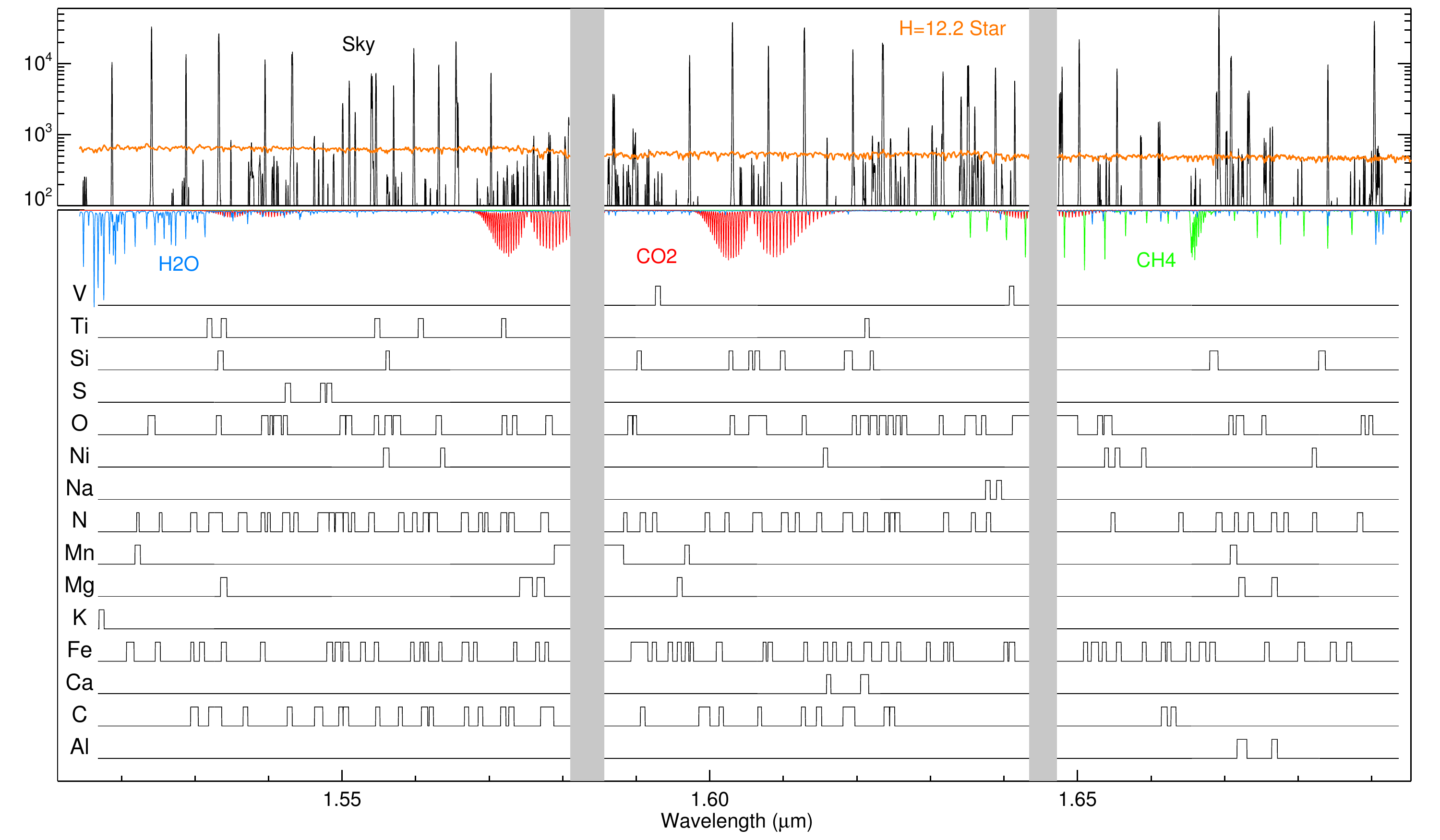}
\end{center}

\caption{The ASPCAP spectral windows for the 15 APOGEE chemical elements along 
with examples of APOGEE sky (black), telluric (blue, red and green), and stellar spectra (orange). 
To aid visibility, 
the spectral windows 
are broadened by $\pm 30$~{\kms}, and all weights are set to the same value.
Full information is available in electronic Table~\ref{weights}.
}
\label{figwind}
\end{figure*}

After the first {\ferre} pass to derive atmospheric parameters from the entire 
APOGEE spectrum, we perform
a series of new runs in which all parameters but that of the dimension 
used for the abundance of the element of interest remain fixed. Only specific spectral windows are 
fitted---see Figure~\ref{figwind}. In these fittings, the same databases of synthetic spectra are used, 
but the abundances of individual $\alpha$-elements are derived 
	by varying the {\afe} dimension of the grid, 
	the abundance of carbon and nitrogen by varying the {\cfe} and {\nfe}
	dimensions, respectively, and the abundances of all other elements by varying 
	the {\feh} dimension. The weights for the {\ctwo} calculations are also 
changed so that we only consider spectral features that are 
sensitive to the element of interest.

Deriving the relevant pixel weights for each element is equivalent to identifying the
transitions to be used for each element. This is accomplished by first using an 
algorithm that evaluates the derivatives of the model fluxes with respect to
each elemental abundance for a star with {\teff}$\ = 4000$~K, {\logg}$\ =1.0$, and three
different metallicities ([M/H]$\ = +0.0, -1.0,$ and $-2.0$). Wavelengths are 
assumed to be sensitive to abundance changes of a given element
for each metallicity, if the modulus of the derivative is larger than three times
the standard deviation of all points in the spectrum. Weights are normalized to 
the value of the most sensitive point. 
Therefore the weight at $\lambda$ for element $i$ at a
metallicity {\feh} is proportional to the change of the flux with the abundance at that wavelength:
\begin{equation}
w^{\prime}_{\lambda,i,\mathrm{[M/H]}}\propto\frac{\partial F_{\lambda,\mathrm{[M/H]}}}{\partial A_i}.
\end{equation}

If a wavelength is sensitive to the abundance of another element except for Fe, 
the weight of this other element is subtracted. This procedure can yield negative values, 
which are fixed to zero. 

\begin{equation}
w_{\lambda,i, \mathrm{[M/H]}}=w^{\prime}_{\lambda,i,\rm{[M/H]}}-\sum\limits_{j\neq i,Fe}w^{\prime}_{\lambda,j,\mathrm{[M/H]}}.
\end{equation}

All the weights obtained for each metallicity are combined in the form

\begin{equation}
\begin{split}
w_{\lambda,i}=0.3\times w_{\lambda,i,\mathrm{[M/H]}=+0.0} \\
 +0.3\times w_{\lambda,i,\mathrm{[M/H]}=-1.0}  \\
 +0.4\times w_{\lambda,i,\mathrm{[M/H]}=-2.0}.
\end{split}
\end{equation}

 This favors
those wavelengths that at low metallicity are sensitive to abundance changes,
since significant variations of the modulus of the derivative are more difficult to detect in that metallicity
regime. Weights are adjusted with a multiplicative factor $\alpha_{\lambda}$ that takes into account how
well the model spectrum for Arcturus reproduces an actual (\citealt{Hinkle95}) observation of this star. This multiplicative factor goes from one, when the ratio between
model spectrum and atlas is lower than three times the sigma of the distribution, to zero
for the most deviant points. A second multiplicative factor $\beta_{\lambda}$, takes into account
how well APOGEE spectra are reproduced by the model fluxes, using the median
residuals at each wavelength after fitting the entire APOGEE sample. Therefore
the final weight for each element is in the form:
\begin{equation}
W_{\lambda,i}=\alpha_{\lambda}\times\beta_{\lambda}\times w_{\lambda,i}. 
\end{equation}
The whole procedure
makes it possible to use only parts of a line profile, e.g., the red and/or
blue wings, when the
core is removed owing to any criteria described above (mostly blends) . Finally, a few regions were removed
after a visual inspection of the fits for each element in the set of reference stars defined by Smith et al. (2013).

Table~\ref{weights}, available electronically, gives the weights as a function of wavelength, for the 15 APOGEE
chemical elements. 
In the short portion of Table~\ref{weights} shown in the text, only K has non-zero weight. The number of features used in the abundance determinations 
varies from element to element: there are dozens for C (mainly CO and CN), N (CN), O (OH), and Fe 
but only a handful for Na, Mg, Al, Si, S, K, Ca, Ti, V, Mn, and Ni. Most of the features are neutral versions of elements. 
Figure~\ref{figwind} shows the location of the spectral windows for each of the elements.

\begin{table*}
\tablenum{3}
\caption{ASPCAP Spectral Windows for Chemical Abundances Determinations} 
\centering
\begin{tabular}{cccccccccccccccc}
\hline
\hline
Wavelength &\multicolumn{15}{c}{Abundance Weights}\\
& Fe & C & N & 0 & Na & Mg & Al & Si & S & K & Ca & Ti & V & Mn & Ni\\
{($\mu$m)}&\\
\hline
1.516812988281 & 0.0000  & 0.0000  & 0.0000  & 0.0000  & 0.0000  & 0.0000  & 0.0000  & 0.0000  & 0.0000  & 0.0143  & 0.0000  & 0.0000  & 0.0000  & 0.0000  & 0.0000 \\
1.516834667969 & 0.0000  & 0.0000  & 0.0000  & 0.0000  & 0.0000  & 0.0000  & 0.0000  & 0.0000  & 0.0000  & 0.0008  & 0.0000  & 0.0000  & 0.0000  & 0.0000  & 0.0000 \\
1.516854687500 & 0.0000  & 0.0000  & 0.0000  & 0.0000  & 0.0000  & 0.0000  & 0.0000  & 0.0000  & 0.0000  & 0.0000  & 0.0000  & 0.0000  & 0.0000  & 0.0000  & 0.0000 \\
1.516876269531 & 0.0000  & 0.0000  & 0.0000  & 0.0000  & 0.0000  & 0.0000  & 0.0000  & 0.0000  & 0.0000  & 0.0000  & 0.0000  & 0.0000  & 0.0000  & 0.0000  & 0.0000 \\
1.516896289062 & 0.0000  & 0.0000  & 0.0000  & 0.0000  & 0.0000  & 0.0000  & 0.0000  & 0.0000  & 0.0000  & 0.0000  & 0.0000  & 0.0000  & 0.0000  & 0.0000  & 0.0000 \\
\hline
\end{tabular}
\label{weights}
\tablecomments{Table~3 is published in its entirety in the electronic edition of ApJ. A portion is shown here for guidance regarding its form and content.}
\end{table*}

\section{IDL wrapper}
\label{wrapper}

While {\ferre} performs the search for the optimal set of parameters
for each observed spectrum, there are many other tasks that need
to be done before and after the optimization. A suite of IDL programs called 
the IDL Wrapper\footnote{The software is available in http://www.sdss3.org/svn/repo/apogee/aspcap/idlwrap/.} 
performs those other tasks.

The wrapper works in blocks of observations defined by fields. APOGEE fields are typically 
defined by their Galactic coordinates ($l,b$) or their location ID ($locID$, 
a unique four digit number assigned to each APOGEE field). The reading of the data is done 
separately for each individual field, and the pre-processing and analysis runs are done 
independently for each individual stellar spectral class.

\subsection{Data Preparation}
\label{dataprep}

Observations are compared to synthetic spectra in the stellar rest frame. 
The data reduction pipeline corrects the observed wavelengths for Doppler 
shifts associated with the stellar radial velocities estimated by the pipeline itself, 
and using a sinc interpolation places the observed spectra in the wavelength 
scale of the synthesis \citep[for more details, see][]{Nidever15}.

The comparison of observations to synthetic spectra uses continuum-normalized fluxes 
to minimize differences associated with reddening and the instrumental 
response function. The normalization 
 of the observed and synthetic spectra should be the same to minimize systematic differences. 
 In the case of individual abundances derivations, ASPCAP employs the normalized observed spectra used for the global fit. 

We have opted for a simple normalization procedure, based on a
 pseudo rather than a real continuum, to facilitate consistency with observations. 
 The normalization consists of 
a repeated least-squares polynomial fit ($O_{\lambda,\mathrm{fit}}=\Sigma a_{i}*\lambda^{i}$) 
after successive sigma clippings. Each APOGEE detector spectrum is normalized independently, as done 
for the library spectra. The library continuum information is stored in the 
 library header and used by ASPCAP. The parameters of the fit are: 
the polynomial order, the number of iterations, and the rejection levels, which are given 
in units of the standard deviation between the previous polynomial fit and the retained data. 
ASPCAP uses a fourth-order polynomial and ten iterations. The algorithm looks for the spectrum's 
upper envelope, which is reached by employing small lower ($0.1\times\sigma$) and moderate upper ($3.0\times\sigma$) 
clipping thresholds. The first iteration fits the spectrum and replaces the rejected pixels with the fitted values.

Observations can suffer from systematic errors and depart from the 
synthetic spectra. These errors may be associated with
 minor typical instrument defects, the instrumental response, 
 or the contribution of the Earth's atmosphere and the interstellar medium. A good instrument characterization provides information on detector cosmetics 
(bad and/or saturated pixels), which can be used to avoid problematic pixels. 
The continuum fitting process and the 
{\ctwo} evaluation ignore bad pixels and those affected by cosmic rays. 
The data reduction pipeline produces flux-calibrated spectra from which sky emission 
has been subtracted and telluric absorption has been removed, along with uncertainties 
for the spectra. However, the sky subtraction is imperfect, especially for the bright OH lines, so the 
uncertainties in regions around such lines are inflated so that they are effectively masked out. Early ASPCAP analyses of some APOGEE data 
 showed little sensitivity of the parameter derivations to the masking of few spectral windows. The 
 chosen windows simulated those potentially affected by sky emission contamination.

  \subsection{Jobs Management and Data Organization}

The wrapper takes care of writing the {\ferre} input files, 
submitting the {\ferre} jobs to the execution queue, organizing the output, setting quality flags, and 
doing calibrations \citep[see][for more details on flags and calibration]{Holtzman15}. 

Output results are packed into FITS files \citep{Pence10}, with a structure\footnote{The data format for all files is described in the online documentation at 
https://data.sdss.org/datamodel/index-files.html, as well
as in Holtzman et al. (2015).}
that resembles that of the files containing APOGEE spectra.
The spectra themselves are included in output ASPCAP files (aspcapStar files).
These spectra are exactly as input to {\ferre} but they differ from those in the apStar files 
in two respects: they have been
continuum normalized (see Section \ref{dataprep}), and their
spectral range is slightly reduced to ensure the 
same spectral lines are used in the analysis of all stars,
regardless of their radial velocities. The best-fitting 
model spectra are also included in the files. The 
calculated parameters, abundances and their covariance matrices can be found 
in the allStar summary files. 

\begin{table*}
\tablenum{4}
\caption{Tests on Synthetic Spectra with Different ASPCAP Settings}
\centering
\begin{tabular}{ccrrrrrrrrrrrrrl}
\hline
\hline
Descrip. & Test & \multicolumn{2}{c}{[M/H]} & \multicolumn{2}{c}{[C/M]} & \multicolumn{2}{c}{[N/M]} & \multicolumn{2}{c}{[$\alpha$/M]} & \multicolumn{2}{c}{$\log{\xi_{\rm{t}}}$} & \multicolumn{2}{c}{$T_{\rm{eff}}$} & \multicolumn{2}{c}{$\log{g}$}  \\
&& \multicolumn{1}{c}{${\Delta}$} &\multicolumn{1}{c} {$\sigma$} & \multicolumn{1}{c} {${\Delta}$} & \multicolumn{1}{c}{$\sigma$} & \multicolumn{1}{c}{${\Delta}$} & \multicolumn{1}{c}{$\sigma$} & \multicolumn{1}{c}{${\Delta}$} & \multicolumn{1}{c}{$\sigma$} &\multicolumn{1}{c}{${\Delta}$} & \multicolumn{1}{c}{$\sigma$} &\multicolumn{1}{c}{${\Delta}$} & \multicolumn{1}{c}{$\sigma$} &\multicolumn{1}{c}{${\Delta}$} & \multicolumn{1}{c}{$\sigma$} \\
&& \multicolumn{2}{c}{(dex)} &\multicolumn{2}{c}{(dex)}&\multicolumn{2}{c}{(dex)}&\multicolumn{2}{c}{(dex)}&\multicolumn{2}{c}{(dex)} & \multicolumn{2}{c}{(K)} &\multicolumn{2}{c}{(dex)}\\

\hline
\multicolumn{16}{c}{On-nodes sample}\\

Linear Interpolation & A &  $+   0.000$  &   $   0.004$ &  $+   0.000$  &   $   0.020$ &  $+   0.000$  &   $   0.275$ &  $+   0.000$  &   $   0.006$ &  $+   0.001$  &   $   0.015$ &  $+   0.015$  &   $  10.414$ &  $+   0.000$  &   $   0.021$  \\
Quadratic Interpolation &  &  $    -0.000$  &   $   0.002$ &  $ -0.000$  &   $   0.006$ &  $+   0.000$  &   $   0.187$ &  $+   0.000$  &   $   0.002$ &  $+   0.001$  &   $   0.005$ &  $+   0.016$  &   $   4.005$ &  $+   0.000$  &   $   0.008$  \\
 &&  $+0.000$  &   $   0.003$&  $+0.000$  &   $   0.013$&  $-0.000$  &   $   0.313$&  $+0.000$  &   $   0.003$ &  $+0.001$  &   $   0.013$&  $+0.028$  &   $ 7.382$&  $+0.000$  &$0.016$\tablenote{Results listed are for the sample in common with that used in Test A, cubic interpolation case.}  \\
Cubic Interpolation &  &  $+ 0.000$  &   $   0.002$ &  $+   0.000$  &   $   0.010$ &  $+  0.000$  &   $   0.309$ &  $+   0.000$  &   $   0.002$ &  $+   0.001$  &   $   0.008$ &  $+   0.028$  &   $   5.706$ &  $+   0.000$  &   $   0.011$  \\   
Linear with PCA & B  &$  -0.003$  &   $   0.051$ &  $+   0.000$  &   $   0.202$ &  $+   0.044$  &   $   0.440$ &  $+   0.007$  &   $   0.048$ &  $+   0.006$  &   $   0.115$ &  $+   7.542$  &   $  69.880$ &  $+   0.012$  &   $   0.146$  \\
Quadratic with PCA &  &$+   0.001$  &   $   0.051$ &  $+   0.009$  &   $   0.196$ &  $+   0.085$  &   $   0.452$ &  $+   0.006$  &   $   0.050$ &  $+   0.005$  &   $   0.110$ &  $+   7.735$  &   $  69.958$ &  $+   0.016$  &   $   0.148$  \\
Cubic with PCA & &  $  -0.000$  &   $   0.050$ &  $+   0.009$  &   $   0.196$ &  $+   0.087$  &   $   0.454$ &  $+   0.006$  &   $   0.051$ &  $+   0.005$  &   $   0.108$ &  $+   7.233$  &   $  68.366$ &  $+   0.015$  &   $   0.144$  \\

\feh$ > -1$ \& \teff$ < 5500$ &  &  $+   0.002$  &   $   0.028$&   $+   0.002$ &   $   0.100$& $+   0.029$  &   $ 0.233$&  $+   0.004$  &   $   0.038$&  $+ 0.004$  &$   0.033$&  $+   5.045$  &   $  51.641$& $+   0.010$  &   $   0.106$  \\

\multicolumn{16}{c}{Off-nodes sample}\\

$S/N=\inf $ & C &   $  -0.002$  &   $   0.013$ &   $  -0.009$  &   $   0.039$ &  $+   0.019$  &   $   0.173$ &   $  -0.000$  &   $   0.011$ &   $  -0.002$  &   $   0.024$ &   $  -0.700$  &   $  20.950$ &   $  -0.000$  &   $   0.031$  \\
$S/N$=25 &  &  $+   0.035$  &   $   0.045$ &   $  -0.018$  &   $   0.095$ &  $+   0.072$  &   $   0.276$ &   $  -0.002$  &   $   0.037$ &   $  -0.002$  &   $   0.065$ &  $+  13.400$  &   $  61.650$ &  $+   0.003$  &   $   0.090$  \\
$S/N$=50 & &  $+   0.006$  &   $   0.022$ &   $  -0.010$  &   $   0.060$ &  $+   0.028$  &   $   0.200$ &   $  -0.001$  &   $   0.022$ &   $  -0.002$  &   $   0.036$ &  $+   2.400$  &   $  31.500$ &  $+   0.001$  &   $   0.053$  \\
$S/N$=100 & &   $  -0.001$  &   $   0.016$ &   $  -0.010$  &   $   0.045$ &  $+   0.021$  &   $   0.168$ &   $  -0.001$  &   $   0.015$ &   $  -0.003$  &   $   0.029$ &  $+   0.000$  &   $  23.250$ &  $+   0.000$  &   $   0.037$  \\
$S/N$=200 &  &   $  -0.002$  &   $   0.015$ &   $  -0.007$  &   $   0.041$ &  $+   0.024$  &   $   0.188$ &   $  -0.000$  &   $   0.012$ &   $  -0.002$  &   $   0.025$ &   $  -0.800$  &   $  21.600$ &  $+   0.000$  &   $   0.033$  \\
Blocking & D &  $  -0.002$  &   $   0.016$&   $  -0.010$ &   $   0.045$&  $+   0.022$  &   $ 0.196$&  $  -0.000$  &   $   0.014$&   $ -0.004$  &   $0.030$&   $  -0.700$  &   $  22.750$&   $ -0.001$  &   $   0.036$  \\

\multicolumn{16}{c}{Reduced on-nodes sample}\\

Ref.   &  E &  $+ 0.000$  &   $   0.017$  &  $+ 0.002$  &   $   0.042$& $+ 0.021$  &   $   0.201$&  $+0.002$  &   $   0.015$&  $  -0.000$  &   $   0.048$& $+2.600$  &   $  17.700$ & $+0.008$  &   $   0.036$  \\
Lower Res. &  &$+ 0.021$  &   $0.033$ &    $  -0.003$ &   $   0.076$&  $+   0.029$  &   $ 0.246$&  $+   0.009$  &   $   0.021 $ & $+   0.020$  &   $   0.101$ &   $ -15.800$  &   $  43.800$  &   $  -0.082$  &   $   0.099$\\
DR12 LSF &    &  $+ 0.024$  &   $0.032$ & $  -0.014$ &   $   0.055$&  $+   0.041$  &   $ 0.211$&  $+   0.013$  &   $   0.030$& $+   0.034$  &   $   0.095$ &  $ -14.300$  &   $  40.050$ &  $  -0.088$  &   $   0.093$ \\
\hline
\end{tabular}
\tablecomments{The impact of interpolation order (A), PCA compression (B), $S/N$ of spectra (C), masking windows (D), or incorrect LSF modeling (E) on
recovery of parameters by ASPCAP. All tests are performed on synthetic spectra generated similarly to the libraries that ASPCAP uses for fitting. In each column, $\Delta$ indicates the median difference between the parameters of input spectra and the best-fit ASPCAP results, and $\sigma$
is a robust measure of the dispersion of these differences. See text for description of the samples.}

\label{tests}
\end{table*}

\section{Tests on simulated data}
\label{validation}

\begin{figure} 
\figurenum{4}

\begin{center}
 \includegraphics[page=2,angle=0,trim=1.0cm 2.cm 1.5cm 9.5cm, clip,scale=0.70]{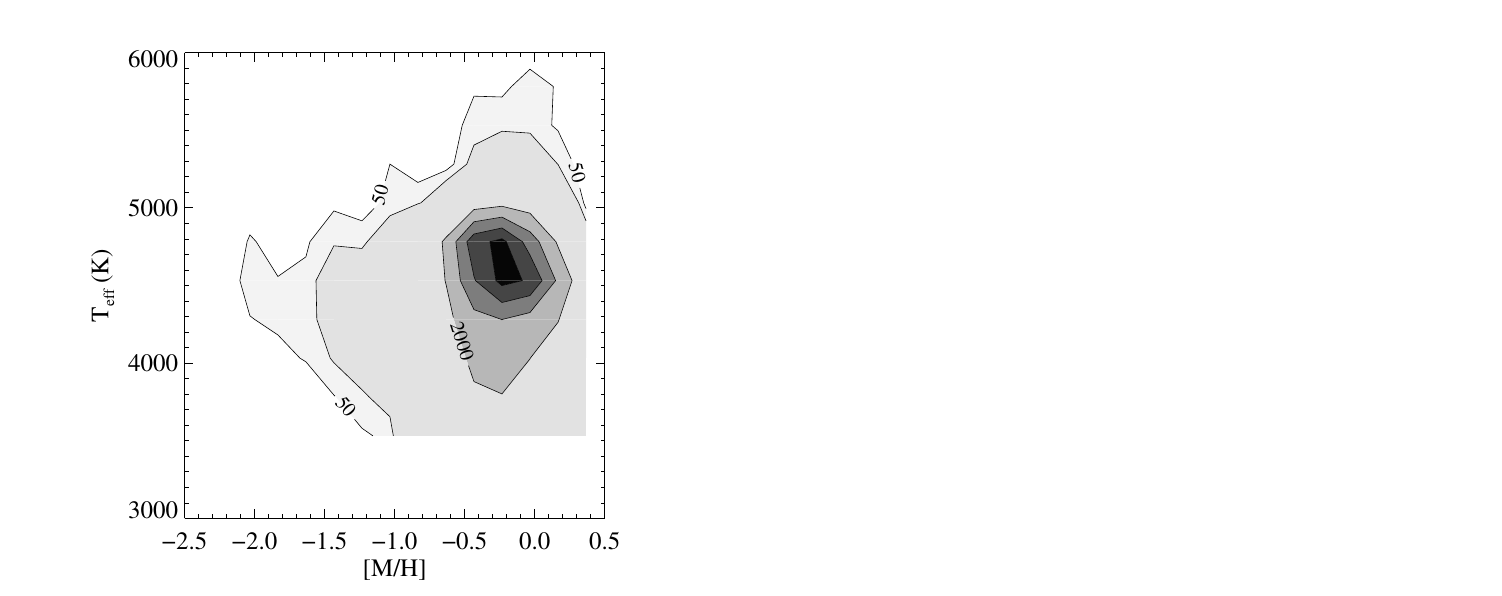}
\end{center}
\caption{The number of DR12 stars (not flagged as bad) in the \teff-\feh\ 
plane, using ASPCAP parameter values without calibration corrections. The 
contour lines are for levels of 50, 100, 2000, 4000, 6000, and 8000 stars.  
}
\label{teffmetal}
\end{figure}

The performance of ASPCAP is, naturally, dependent on stellar properties. For reference, Figure~\ref{teffmetal} plots the distribution
of DR12 stars in \teff\ versus \feh. Only stars on the main survey with reliable
parameters are shown; i.e., stars with neither of the bits set in the 
EXTRATARG bitmaks and with no BAD\_STAR bit set in the ASPCAPFLAG bitmask). 
While APOGEE stars span a wide range of {\teff} and {\feh} parameters, the great
majority of good main survey stars (83\%) lie in the range
$-0.7 \leq {\rm [M/H]} \leq 0.2$ and $3500\,{\rm K} \leq T_{\rm eff} \leq 5100\,{\rm K}$.

We evaluate the performance of our analysis methodology using 
two sets of simulated data and a 7D analysis (i.e., with microturbulence as a free parameter). First we use the 
very same model spectra in one of our libraries as simulated observations, to 
check whether there
are degeneracies among the many parameters involved in our analysis, 
and to test the effect of using PCA compression. We carry out these tests using linear, quadratic, and 
cubic interpolation in the grid of model spectra during the search. 
Second, 
we use a sample of model spectra computed for parameters off the
grid nodes. These are created in the same way as the grid synthetic spectra, 
computing model atmospheres and spectra for the
parameters. This data set is used to quantify the impact in the ASPCAP parameter results 
of interpolation errors, noise
in the spectra, and the effect of the information removal or degradation caused
by sky lines or telluric absorption.

The first data set includes 17,640 spectra extracted from a 
library with the same synthetic spectra used in the cool library ($3500\le T_{\rm eff} \le 6000$ K)
for DR12, but smoothed to a resolving power of R$=22,500$ with a Gaussian kernel. 
Since the spectra will be analyzed with the same library from which they come, the
details of the
broadening function do not matter. The parameters for this ``on-nodes" data set 
are uniformly distributed 
in the parameter space, leaving out the boundaries of the grid, where the search algorithm
runs into problems. 
In some tests, we used a subset of these spectra to reduce the computing
time (``reduced on-nodes sample''), with 194 spectra sampled uniformly from the larger sample. 
The second data set (``off-nodes sample'') is made of 1,000 
synthesized spectra with randomly distributed parameter values
and less extreme abundances (more typical of the APOGEE sample)---e.g., Figure~\ref{teffmetal}. 

The on-nodes sample covers $3750\le$ \teff\ (K) $\le 5750$, 
 $0.5\le$ {\logg} (cgs) $\le4.5$, $-2.0\le$ \feh\ $\le0.0$, $-0.75\le\ $\cfe\ and \afe$\ \le$ 0.75, 
 $-0.5\le[\rm{N/Fe}]\le0.5$ and $0.0\le \log{\xi_{\rm{t}}} $ (km~s$^{-1}$) $\le0.6$. The reduced 
on-nodes subset has a similar 
 coverage but sparser sampling. The coverage 
 of the off-nodes sample is restricted to a single value for the microturbulence, 2~km s$^{-1}$,
and $4000\le$ \teff(K) $\le6000$, $0.5\le$ \logg (cgs) $\le5.0$, $-2.0\le$\feh\ $\le 0.5$ , and 
 $-0.5\le\ $[X/M]$\ \le0.5$ for [C/M], [N/M], and [$\alpha$/M]. 
We quantify the results of our tests by comparing the true parameters of the test spectra with
those recovered by {\ferre} in terms of median offsets ($\Delta$) in Table~\ref{tests}. We use a robust measure of the dispersion in the offsets ($\sigma$)
to avoid outliers: we calculate 
the difference between the maximum and the minimum offsets after excluding the
largest 15.85\% of the sample and the smallest 15.85\%, and divide it by two, which would
correspond to the standard deviation in a normal distribution. 
The values of the dispersion we find in the different tests are in general larger than the values we would find 
by fitting a Gaussian curve to the distributions, but
lower than a straight calculation of the standard deviation, and we
think they are a more solid metric to compare the results from different
tests. These are the figures we report as $\sigma$ in 
Table~\ref{tests}.

\begin{figure} 
\figurenum{5}
\begin{center}

\includegraphics[page=6,angle=0,trim=1.5cm 0.cm 3.0cm 2cm, clip,scale=1,width=4.55cm]{figure5.pdf}\\
\includegraphics[page=7,angle=0,trim=0.0cm 4.0cm 2.5cm 0.6cm, clip,scale=1,width=4.65cm]{figure5.pdf}
\includegraphics[page=5,angle=0,trim=5.5cm 4.0cm 2.5cm 0.6cm, clip,scale=1,width=3.70cm]{figure5.pdf}\\
\includegraphics[page=1,angle=0,trim=0.0cm 1cm 2.5cm 2.4cm, clip,scale=1,width=4.65cm]{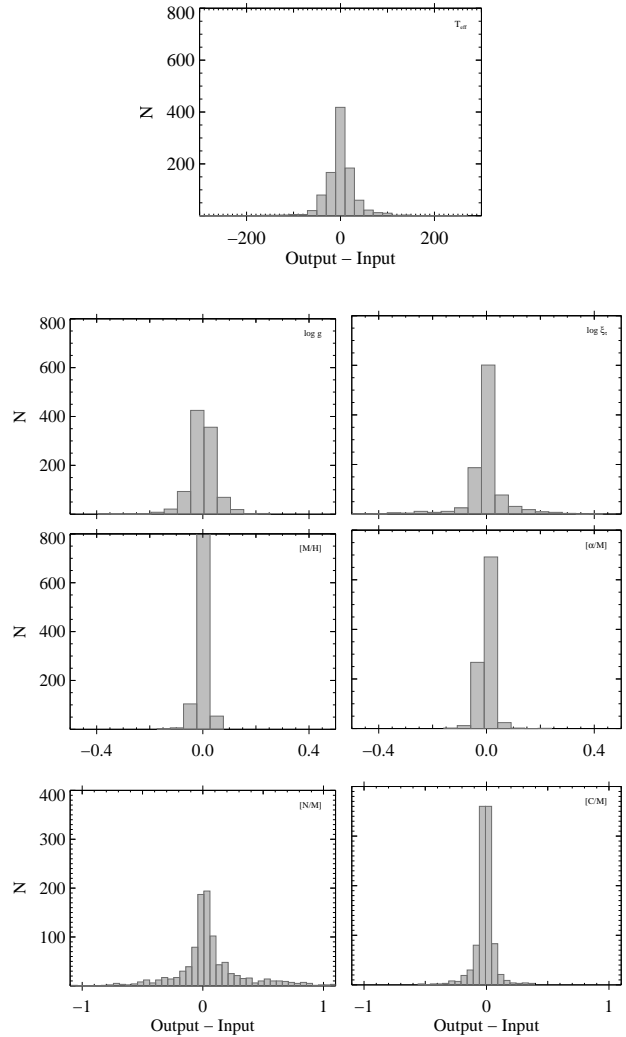}
\includegraphics[page=4,angle=0,trim=5.5cm 1cm 2.5cm 2.4cm, clip,scale=1,width=3.70cm]{figure5.pdf}\\
\includegraphics[page=3,angle=0,trim=0.0cm 0cm 2.5cm 2.4cm, clip,scale=1,width=4.65cm]{figure5.pdf}
\includegraphics[page=2,angle=0,trim=5.5cm 0cm 2.5cm 2.4cm, clip,scale=1,width=3.70cm]{figure5.pdf}\\
\end{center}

\caption{Histograms of the differences (output$\ -\ $input) for the seven global fit parameters for the off-node synthetic spectra with $S/N$=100.
}

\label{figstat}
\end{figure}

\begin{figure} 
\figurenum{6}

\begin{center}

 \includegraphics[page=13,angle=0,trim=0.0cm 0cm 2.4cm 0.0cm, clip,scale=1,width=4.4cm]{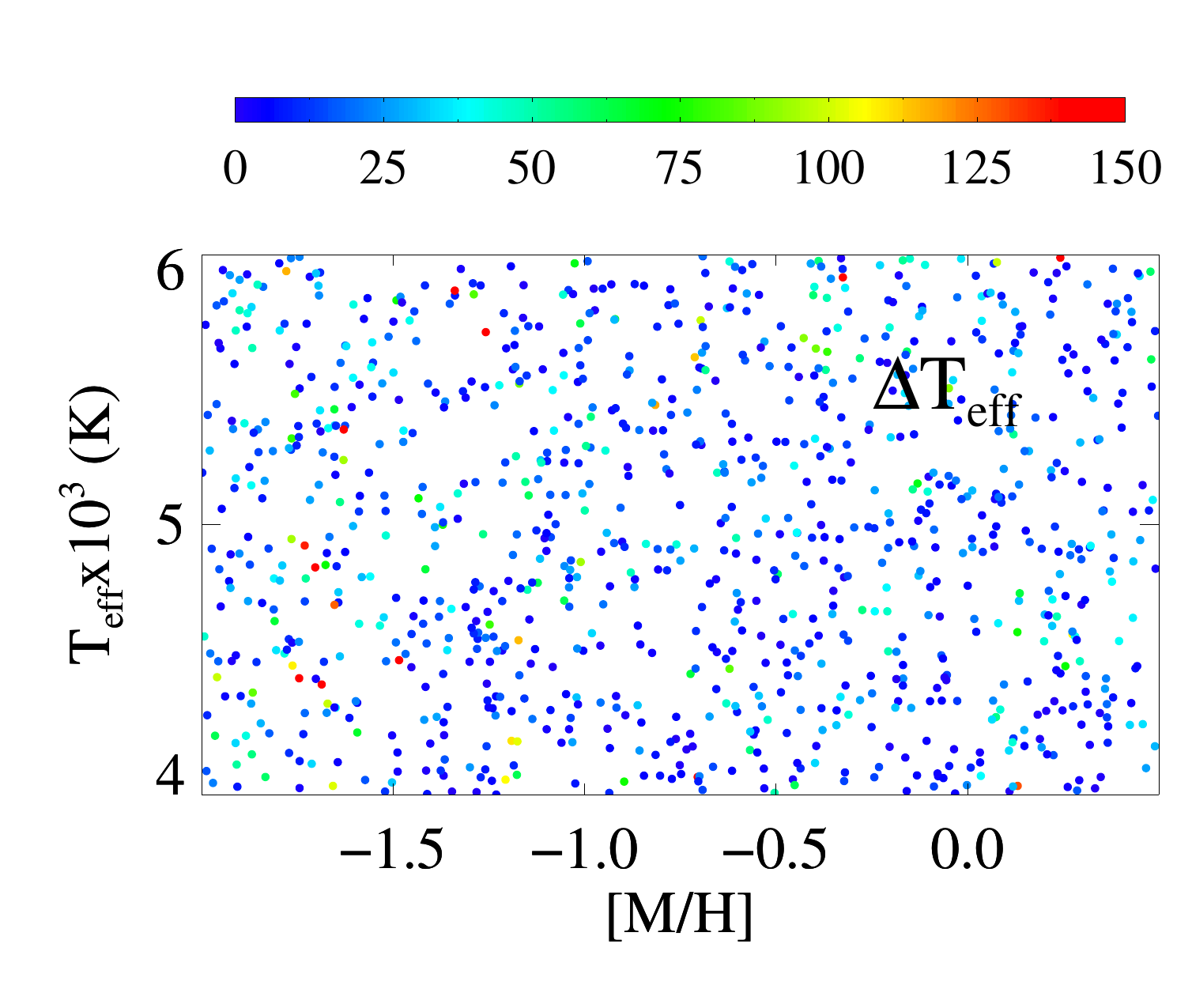}\\
\includegraphics[page=14,angle=0,trim=0.cm 4cm 2.6cm 0cm, clip,scale=1,width=4.55cm]{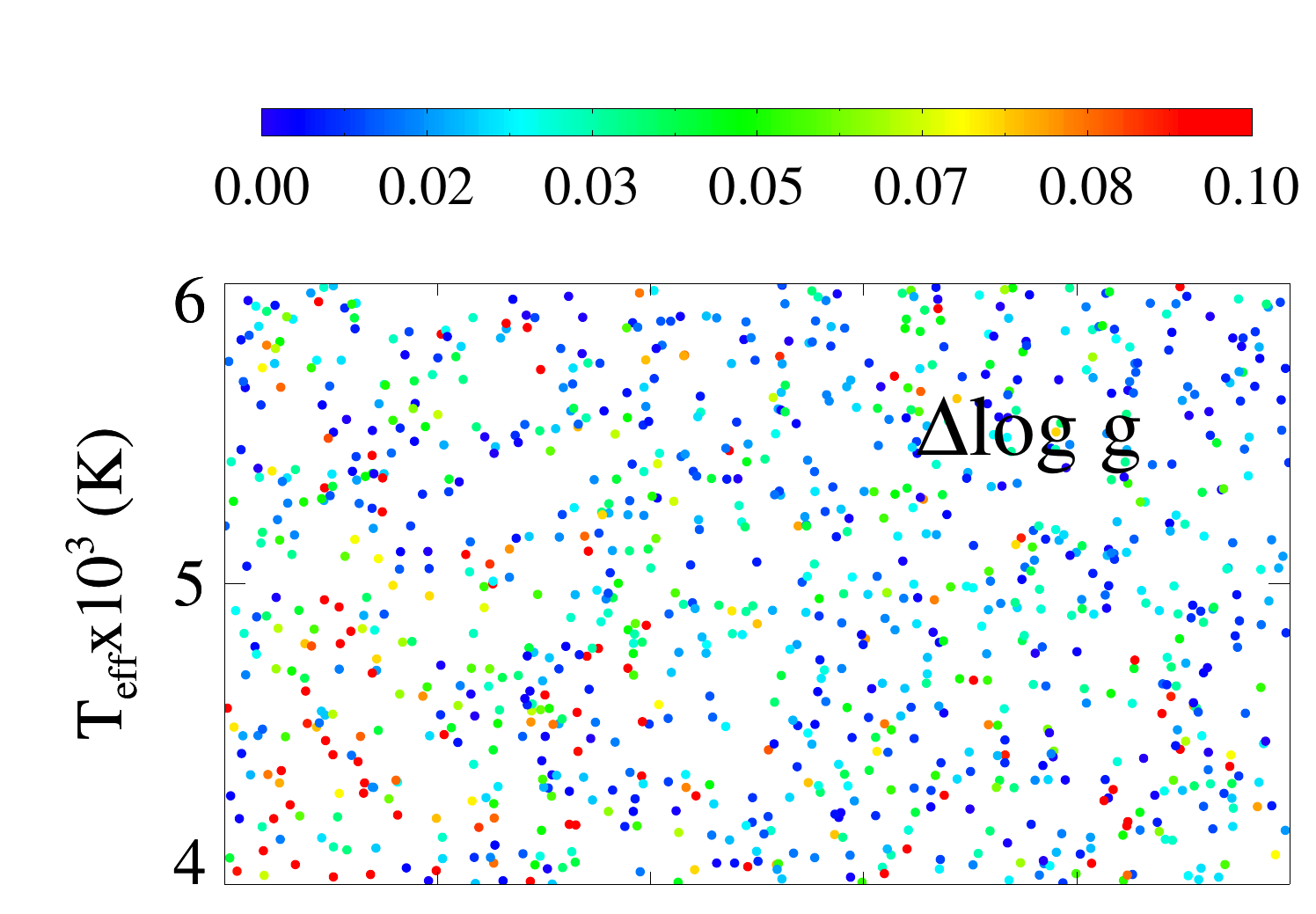}
 \includegraphics[page=12,angle=0,trim=3.8cm 4cm 2.6cm 0.0cm, clip,scale=1,width= 3.85cm]{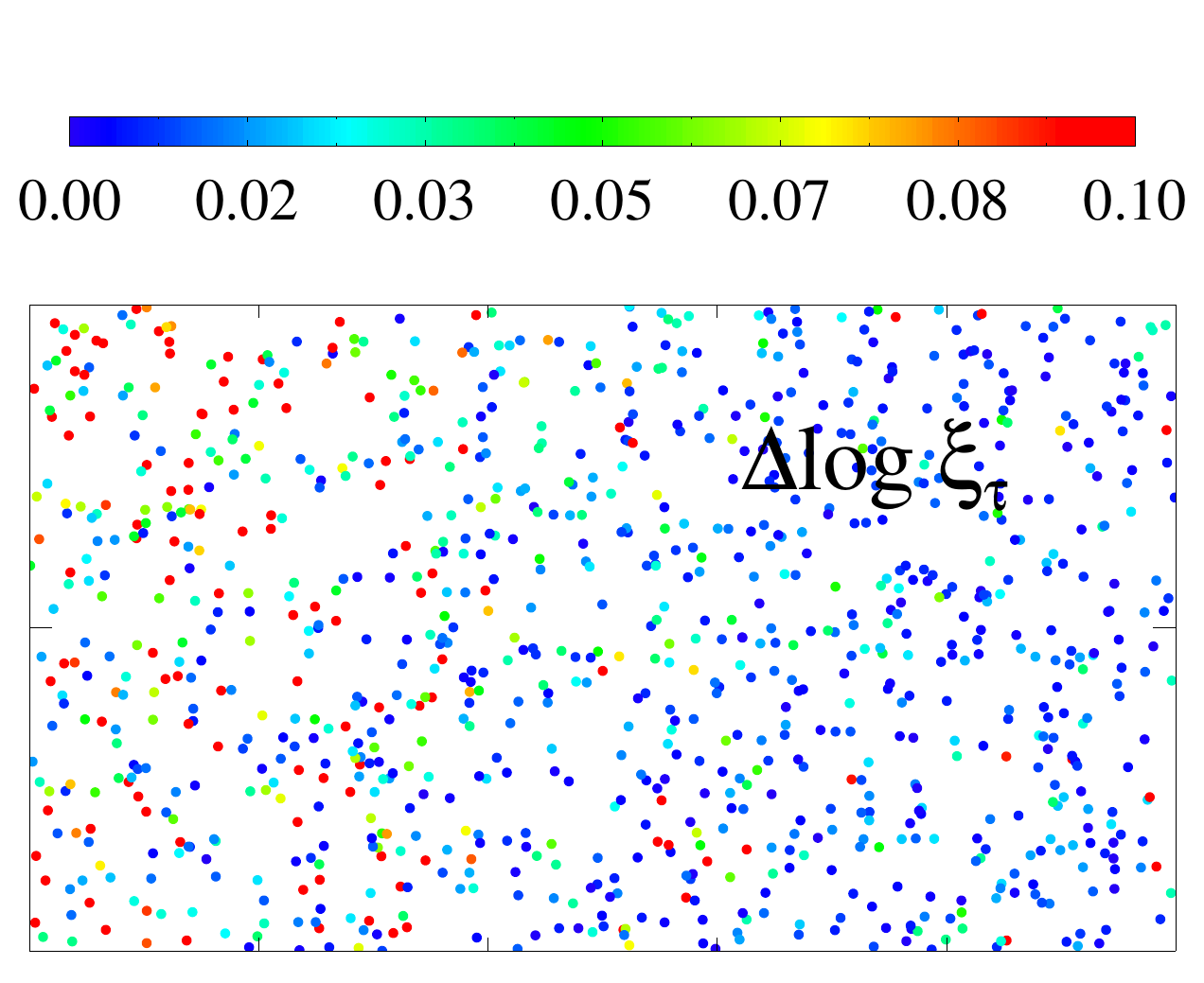}\\
\includegraphics[page=8,angle=0,trim=0cm 4cm 2.6cm 4.5cm, clip,scale=1,width= 4.55cm]{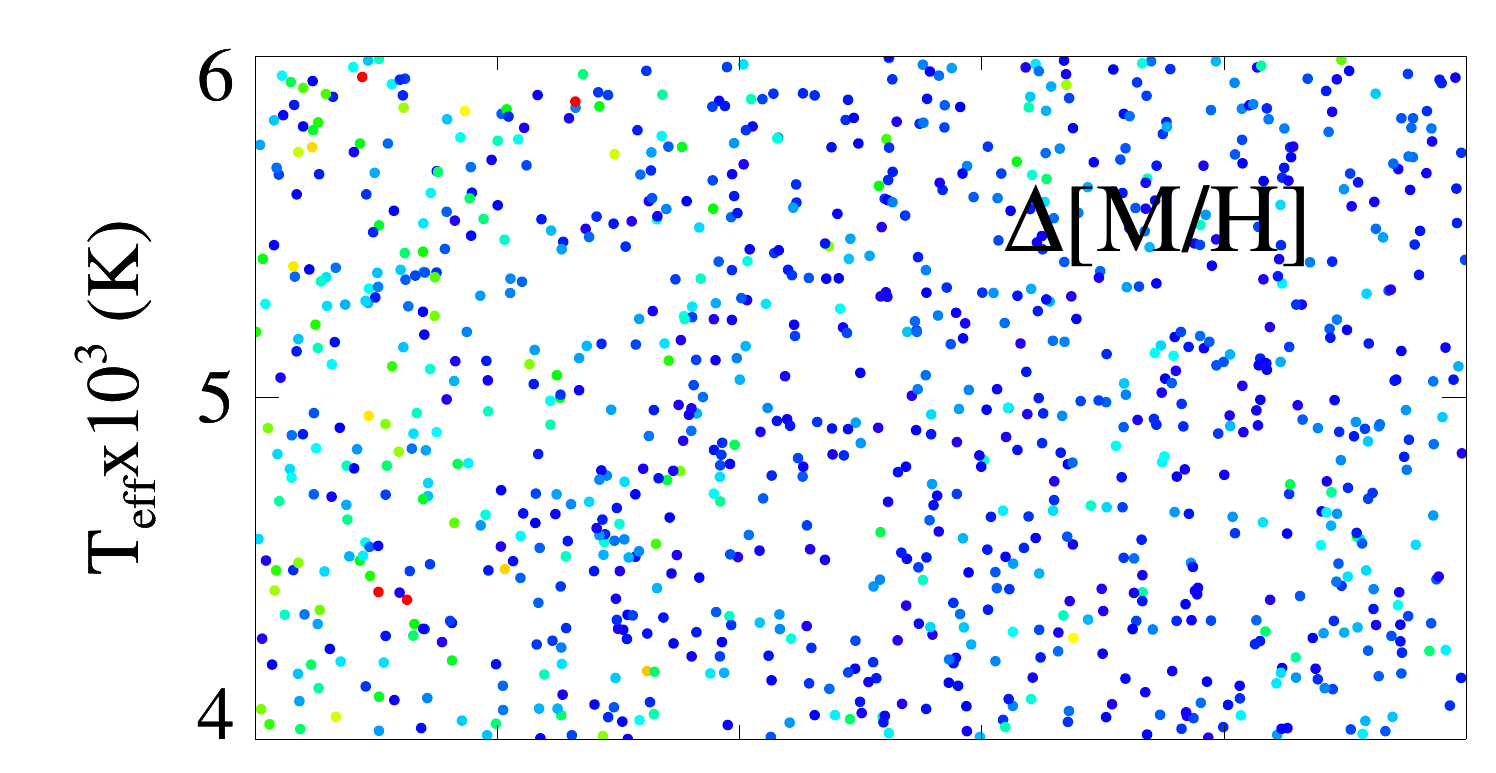}
 \includegraphics[page=11,angle=0,trim=3.8cm 4cm 2.6cm 4.5cm, clip,scale=1,width=3.85cm]{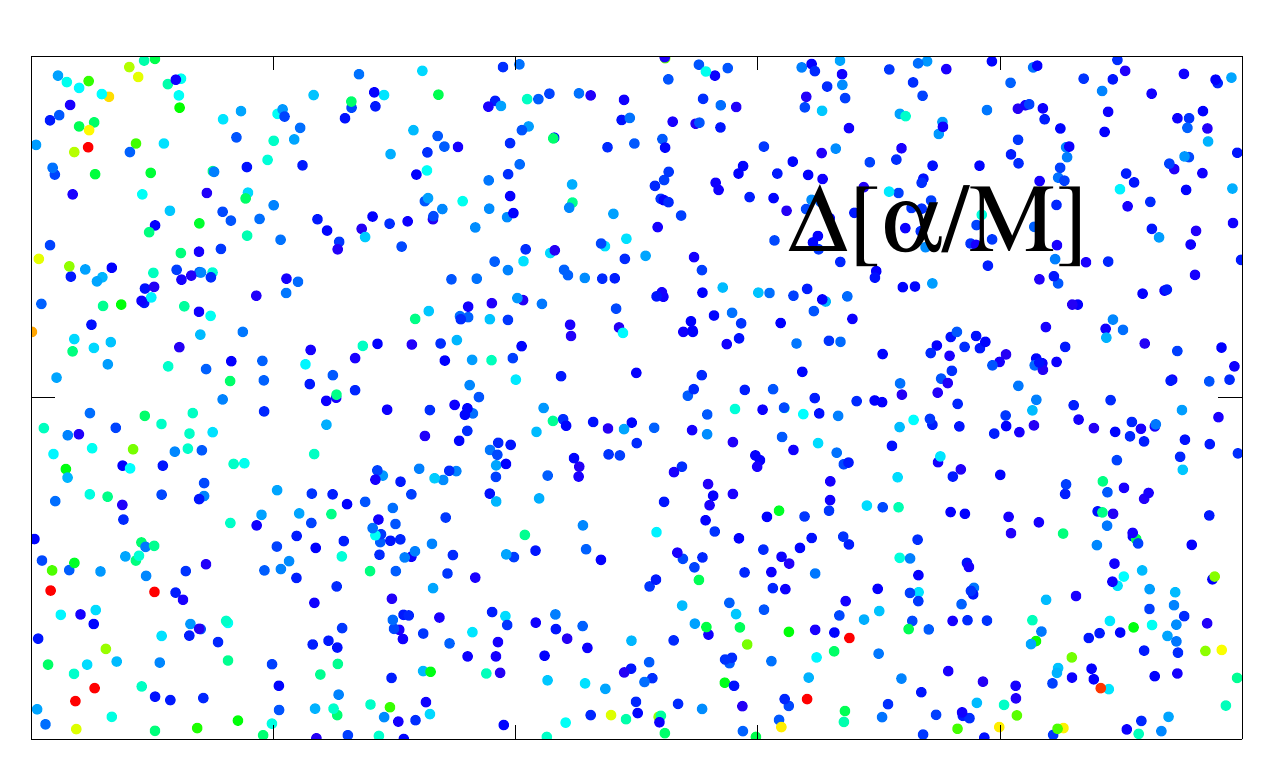} \\
 \includegraphics[page=10,angle=0,trim=0cm 0cm 2.6cm 4.5cm, clip,scale=1,width=4.55cm]{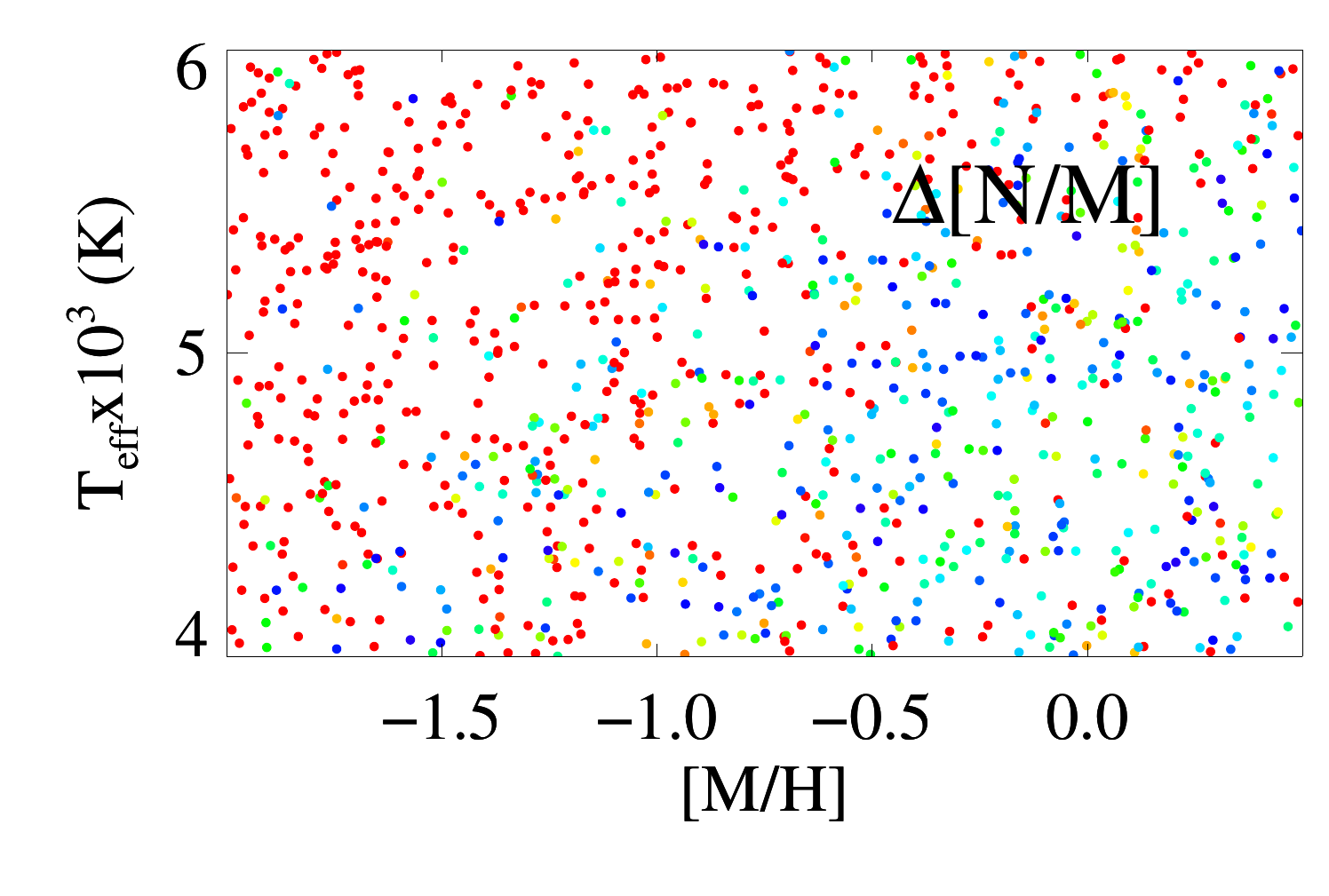}
 \includegraphics[page=9,angle=0,trim=3.8cm 0cm 2.6cm 4.5cm, clip,scale=1,width= 3.85cm]{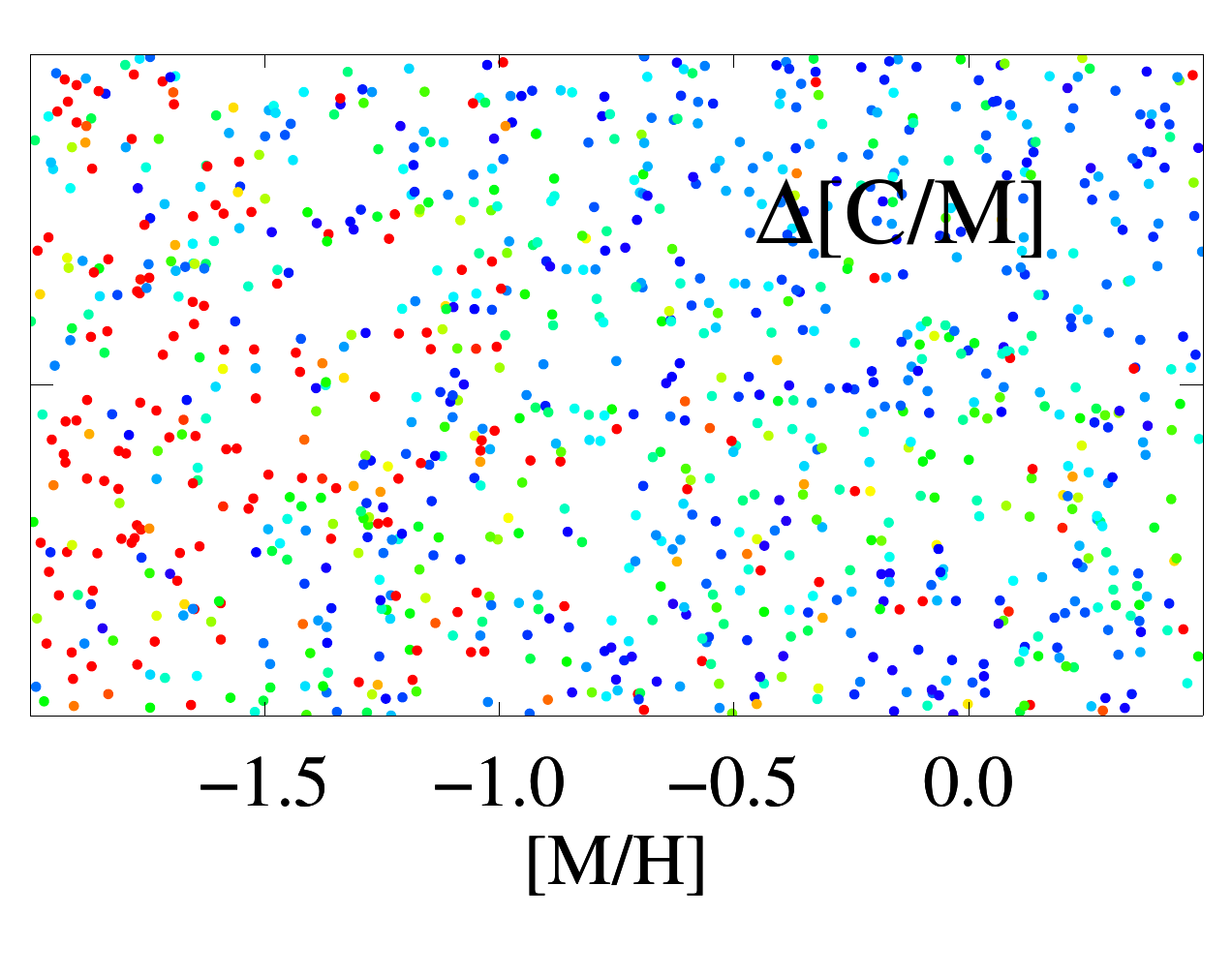}

\end{center}
\caption{The distribution of the input data in the \teff-\feh\ plane for the test with off-node synthetic spectra of $S/N$=100. Data are colored according to their parameter uncertainties, 
with each parameter displayed in a different panel. Data with larger differences than the maximum value of the color bar are scaled to that maximum value.} 
\label{figstat2}
\end{figure}

 Our tests with cubic interpolation produced the best overall results. 
  At the typical APOGEE metallicities (\feh$\ \gtrsim-1.0$), the 
 stellar parameters and the abundances of C, N, and $\alpha$-elements 
 were well recovered, even in the tests with PCA. 
 Of some concern is the compression at lower metallicities, especially for C and N. 
 The introduction of noise in the tests, at the level of $S/N=100$ per pixel, did not 
 compromise the quality of the results. The surface gravity and the microturbulent 
 velocity showed sensitivity to the LSF adopted, hence a detailed characterization 
  of the LSF was done in DR12. We present below the results 
  of our tests in more detail.

\subsection{Test A. Order of Interpolations}
\label{intsub}

As described above, {\ferre} searches for the model parameters that
best match the APOGEE data, and the evaluation of the model spectra
as a function of stellar parameters is performed by interpolation
in a pre-computed grid. 
We evaluated the performance of different interpolation 
schemes available in the code (linear, and B\`ezier quadratic and 
cubic polynomials) in M\'esz\'aros \& Allende Prieto (2013); currently 
{\ferre} uses the same order for all the parameters. Here we
test the effect of each scheme in the 7D analysis of the on-nodes sample.
Of course, the result of a model spectrum evaluation occurring on a node is exact (to numerical round-off), 
independent of the interpolation scheme, but the convergence of the search method 
(the Nelder-Mead algorithm) will be affected by the accuracy of the 
model spectrum evaluation off the nodes at each step in the search process. 

Our results are presented in Table~\ref{tests} as case A for linear, quadratic, and cubic interpolation. 
The cubic test was performed in a slightly smaller sample to speed-up the analysis. For comparison purposes, 
results of the quadratic test for that sub-sample are also listed in Table~\ref{tests}. In the three interpolation cases we recover the input parameter values of the simulations well. 
The best performance is quadratic and cubic interpolation, for which the dispersion is about two times lower for most parameters 
than for the linear case. 
ASPCAP uses cubic interpolation for also reducing uncertainties associated with the clustering of solutions 
around the nodes of the spectral libraries. All the tests show small $\sigma$ values for most parameters. 
Nitrogen is the exception, with a \nfe\ dispersion of 0.2--0.3~dex, as 
a result of the weakness of the CN bands in metal-poor spectra and 
the weak response of the CN lines to changes in the N abundance. 
For cubic interpolation, the dispersion of differences is $\leq 0.01$ dex for \logg,
the other abundance parameters, and $\xi_{\rm t}$, and $6\,{\rm K}$ for \teff.

\subsection{Test B. PCA Compression on the Nodes}
\label{PCAsubtest}

ASPCAP uses PCA-compressed libraries to reduce 
execution time and memory requirements. 
Tests show that the derivation of the carbon abundances for some stars 
could suffer significant uncertainties at this step, especially under extreme 
conditions (e.g., low metallicity). Figure~\ref{teffmetal} displays the distribution 
of the DR12 results in the {\teff}-{\feh} plane.

We repeated the interpolation Test A described in the previous section 
using the PCA compressed version of the very same library. The evaluation 
of the model fluxes on the nodes is no longer exact, and the interpolations
off the nodes are performed in PCA space---interpolating PCA coefficients
rather than fluxes, but the {\ctwo} evaluation is still done using 
fluxes, after the interpolated synthetic spectrum is uncompressed.

The results for linear, quadratic, 
and B\`ezier cubic interpolation are identified in Table~\ref{tests} as
Test B. The use of PCA introduces some distortion in the parameter
recovery, with larger offsets and dispersion in the parameter differences. 
The values of these statistics 
for the metallicity are still insignificant compared to other
sources of uncertainty.
That is also the case 
of the median offset values for \teff\ and \logg, but not of the dispersion, 
$\sim 70$~K and $0.15$~dex, respectively. More significant is the dispersion in \cfe\ 
($\sigma\sim0.2$~dex) and in \nfe\ ($\sigma=0.45$~dex) differences, which is of concern. 
The three polynomial orders lead to similar performance. While our evaluation 
of the performance on PCA compression with the chosen parameters was initially 
more optimistic, these tests suggest that we may have been overly aggressive compressing 
the synthetic spectra with PCA. However, additional tests performed in spectra off-nodes show 
less impact of the PCA than the apparent for the on-node sample.

However, as mentioned earlier, the on-nodes simulation is not representative 
of the APOGEE data, since the simulation uniformly samples the whole
parameter space in the grids, which is far broader than that spanned by 
the APOGEE stellar sample.
The uncertainties are significantly higher for low metallicity stars, 
especially those with low {\afe} and high \cfe\ (or low \cfe\ for the {\cfe} parameter uncertainties), than for the rest of the sample. The results for 
the samples restricted to \feh$\ >-1$ and without the \afe$\ \le0$ spectra are 
better than for the entire sample. 

In Table \ref{tests}, we also report the statistics for the analysis with
cubic interpolation restricted to metallicities\ $\ge-1.0$ and \teff\ $\le$
5500~K. In general, the derived offsets and dispersion become smaller 
for all parameters.

\subsection{Test C. PCA Compression with Noise for the Off-nodes Sample}
\label{errsub}

The off-nodes sample is more representative of APOGEE data. 
We analyzed this data set (Test C in Table~\ref{tests}) with and without added noise to 
test how sensitive our results are to the 
 $S/N$ of the data ($S/N$-values are given per pixel). 
 Our uncertainty estimates, which are based on 
 the recovery of the input parameter values, include both systematic 
 and random contributions. All of these tests use cubic interpolation and PCA 
 compression.

The first case we tested corresponds to a run with noiseless spectra, as
in tests A and B above. With this test, 
we estimate pure systematic uncertainties. 
The input parameters are very well recovered with small uncertainties: 
21~K in \teff\ and $<0.04$~dex for the other parameters (see case $S/N=$\ inf 
in Table~\ref{tests}). Nitrogen remains uncertain at a level of $\sigma$(\nfe) 
$\sim0.17$~dex.

The performance of {\ferre} is better in the off-nodes than the on-nodes 
associated test, which at first may seem contradictory, since interpolation 
errors are expected to be larger for the former. This might be due to the differences in the 
number of initial searches performed in the analysis, and to the differences in 
the sample regarding the size and parameter space coverage. The off-node sample has a fixed 
microturbulence of 2.0~km~s$^{-1}$ (although this parameter is searched for), 
and less extreme C, N and $\alpha$ abundances relative to their iron content. The use of 
a common subsample delivers a similar {\ferre} performance.
 
The noise injection in the tested spectra introduces changes in the quality of the recovered
parameter values, because random errors are now contributing to the total uncertainty.
The quality of the results is already acceptable 
at a $S/N$ of 25 (compared to our goal of $<0.1$~dex abundance errors),
except for \nfe, which has a dispersion of $0.28$~dex. 
For other quantities,
the most significant dispersions at this $S/N$ are 0.095~dex for \cfe\
and $\sim 62$~K for \teff.

As expected, the higher the $S/N$, the better the performance. 
The results for the median offset and robust dispersion increase about a
factor of two between the tests at $S/N= 50$ and 25. At $S/N=50$, 
the results are already of high quality, with $\sigma$ smaller than
 30 K for \teff\ and $<0.1$ dex for the rest 
of parameters (except N). At $S/N=100$, the results show a smaller improvement, and the benefits of working at a higher $S/N$ of 200 are marginal. 
In fact, the quality of the results for $S/N=100$ or 200 are similar 
to those of the associated noiseless test. We recall that APOGEE 
combined spectra typically enjoy a $S/N>70$ per half-resolution element,
and $87\%$ of DR12 stars have $S/N > 100$ per half-resolution element. 
The S/N requirement for APOGEE is of $100$ (per half a resolution element), which 
is set by the goal of getting precise abundances at the level of 0.1 dex or better---see 
Majewski et al. (2015) for more details.

Figure~\ref{figstat} shows the results for tests with the off-nodes
sample degraded to $S/N=100$: the offset 
(output$\ -\ $input parameter values) distributions. 
The distributions for \cfe\ and \nfe\ present significant wings (see second right and bottom left panels in 
Figure~\ref{figstat}), which suggests that the parameters of some 
spectra are not well 
recovered. 

A closer look at the distribution of errors as a function of
 {\teff} and \feh\ reveals a dependence on metallicity for \microt, \cfe\ and \nfe. 
 Figure~\ref{figstat2} shows our input data points colored according to the result quality (measured as the offsets). The \feh\ $\lesssim-1.0$ spectra 
present the highest \cfe, \nfe, and $\log{\xi_{\rm{t}}}$ uncertainties, which can reach values larger than 
0.1~dex. The large uncertainty in \nfe\ extends to solar metallicities 
in warm spectra (\teff$\ \gtrsim 5000$~K). Main-sequence gravities show 
larger than average \cfe\ and \nfe\ uncertainties, as well. The other parameter uncertainties are less dependent on metallicity.
 
The large uncertainties at low metallicity and at dwarf gravities 
are not a concern for the bulk of APOGEE data. 
As seen previously in Figure~\ref{teffmetal},
most stars lie at \feh\ $\gtrsim -1.0$. Nonetheless, it is important
to realize that the spectroscopic information content decreases significantly at low
metallicity (or warm temperatures), and our analysis strategy is less able 
to discern parameters with as high accuracy/precision. 

{\ferre} has two main 
options for estimating the random errors in derived parameters: 
inverting the curvature matrix or carrying
out multiple searches after adding Gaussian noise to the
spectrum (see \S\ref{errors}).
The second option is quite time consuming, so it is
valuable for tests but not well suited for large data samples.

\begin{figure} 
\figurenum{7}
\begin{center}

\includegraphics[page=19,angle=0,trim=0cm 2.4cm 2.4cm 0cm, clip,scale=1,width=4.75cm]{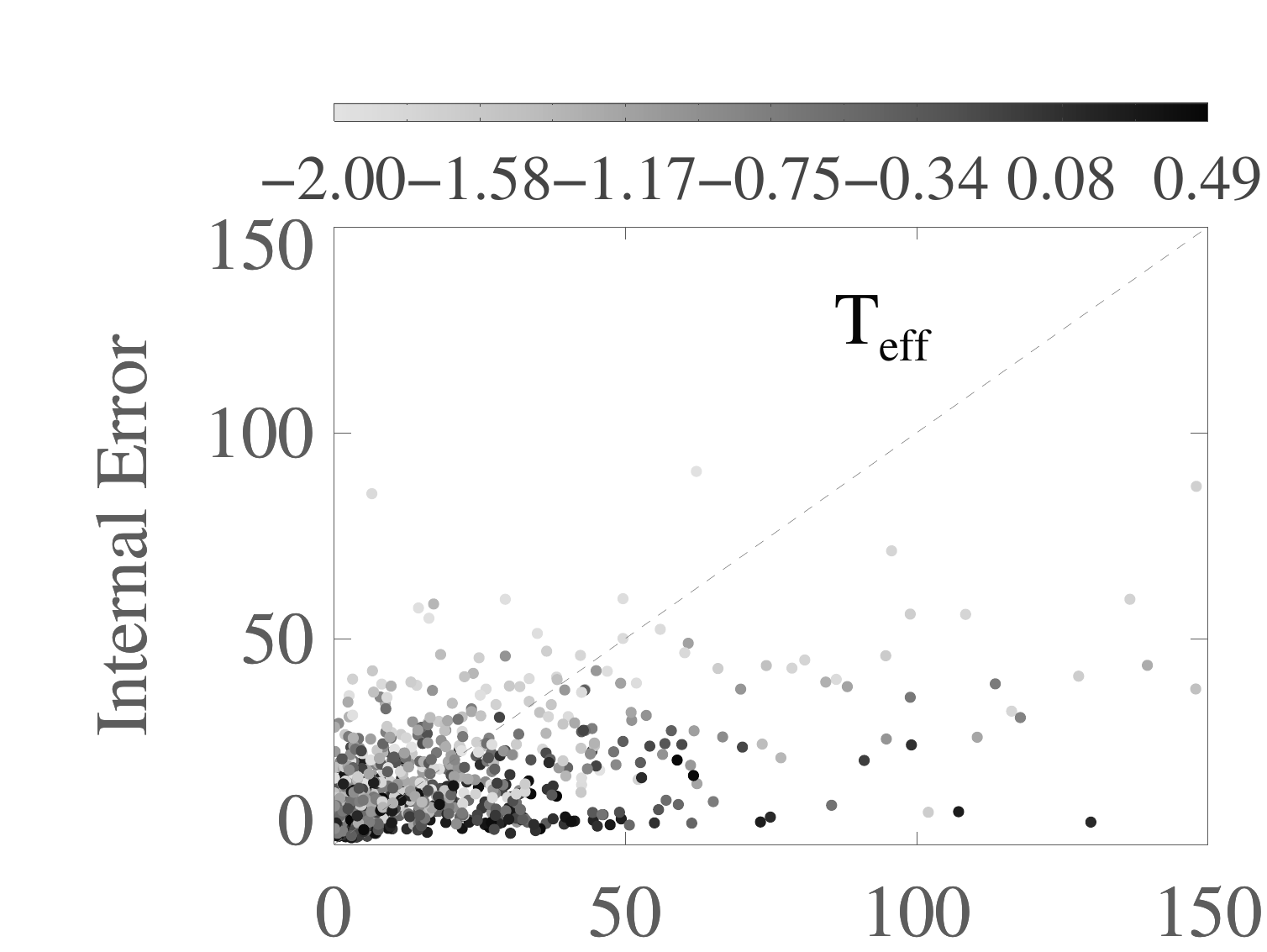}
\includegraphics[page=19,angle=0,trim=5.5cm 2.4cm 2.4cm 0cm, clip,scale=1,width=3.75cm]{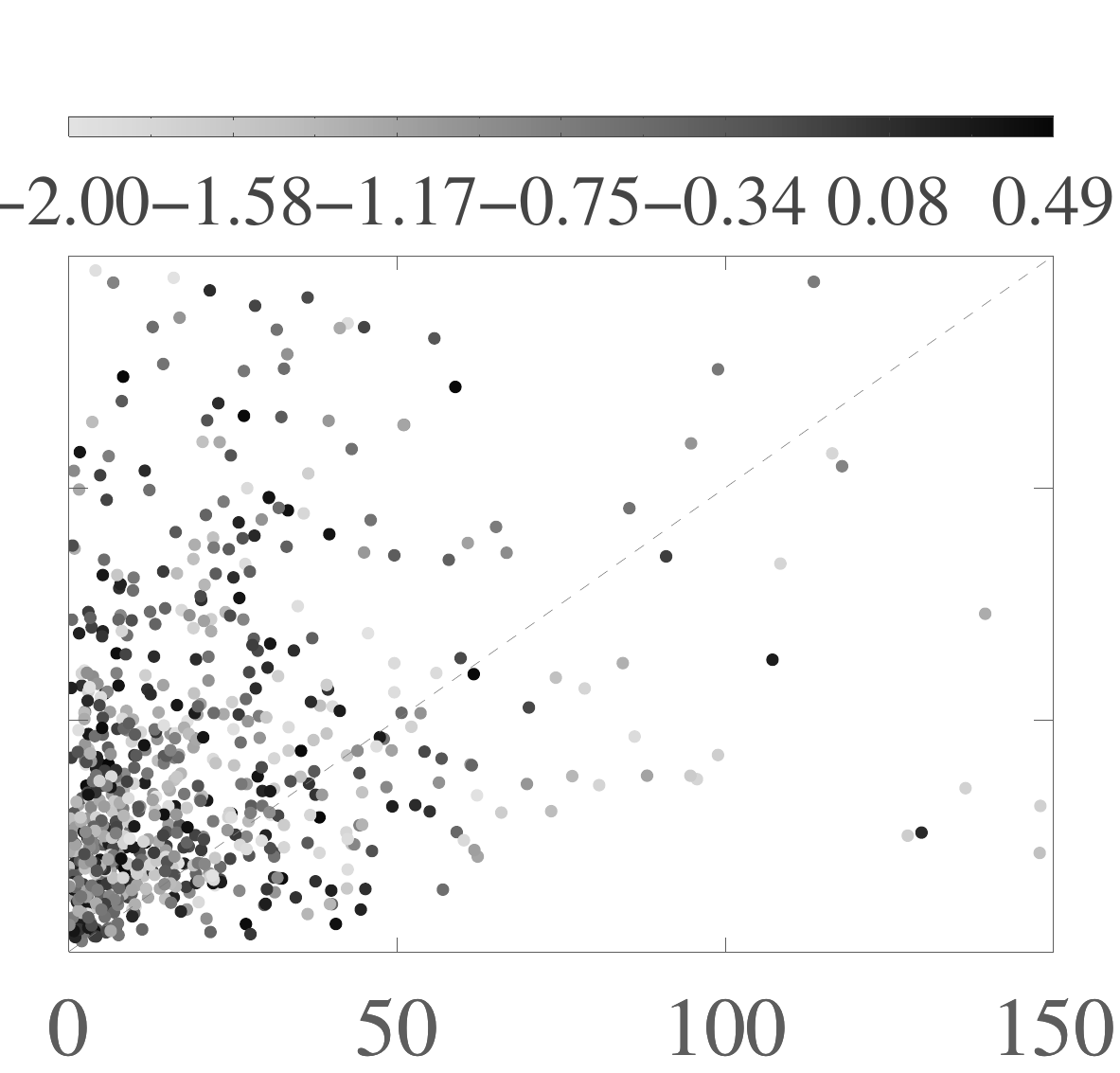}\\
\includegraphics[page=21,angle=0,trim=0cm 2.4cm 2.4cm 4.2cm, clip,scale=1,width=4.75cm]{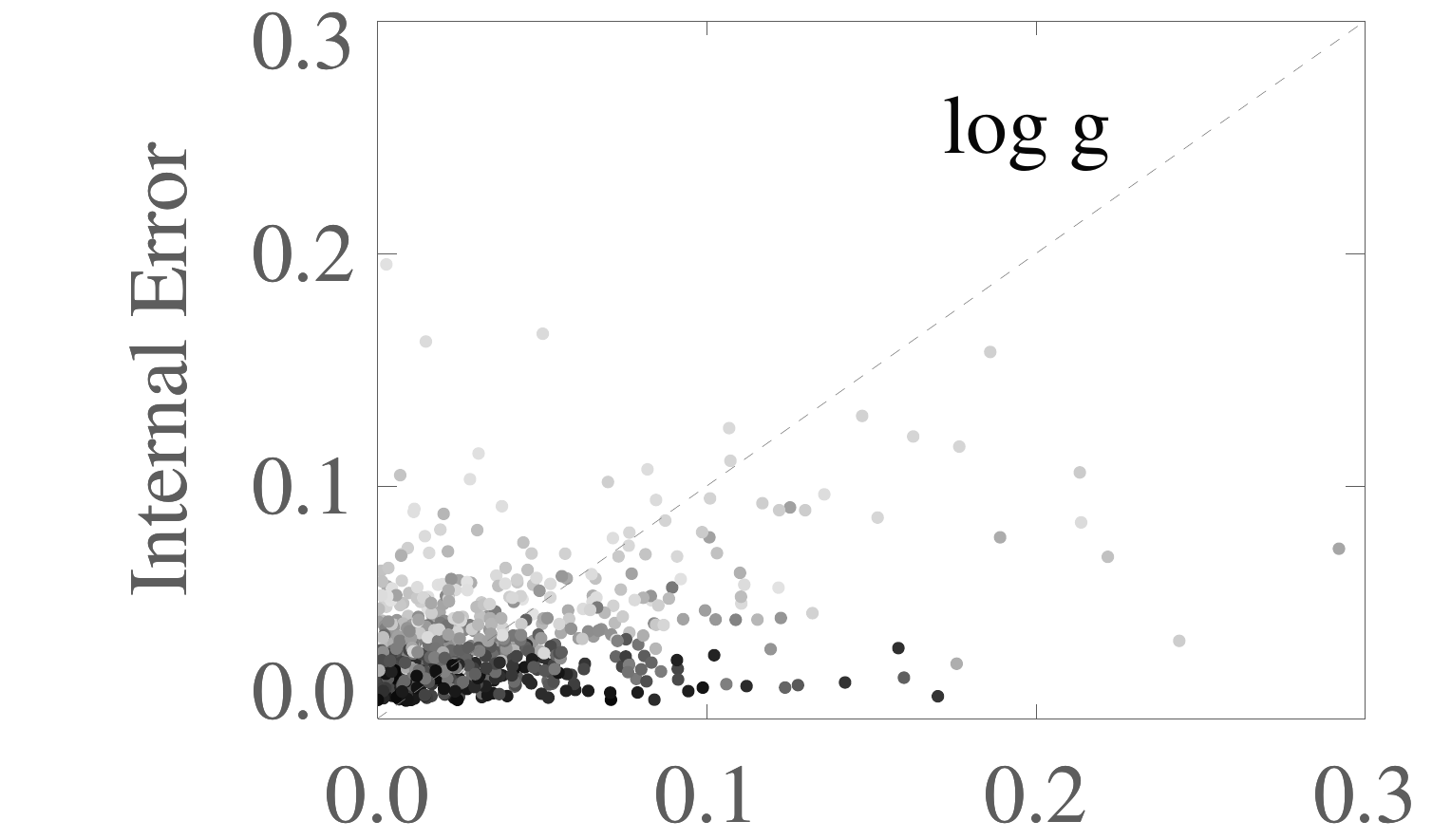}
\includegraphics[page=21,angle=0,trim=5.5cm 2.4cm 2.4cm 4.2cm, clip,scale=1,width=3.75cm]{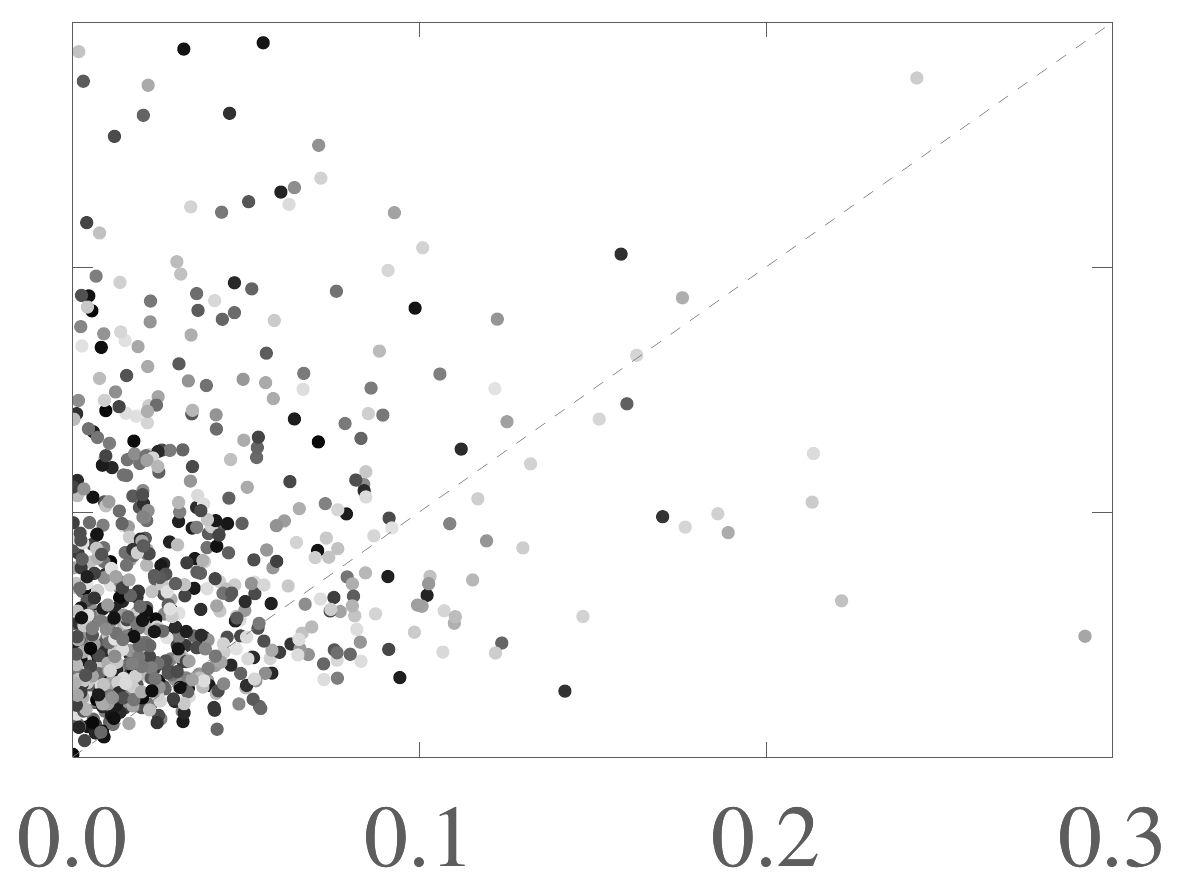}\\
\includegraphics[page=17,angle=0,trim=0cm 2.4cm 2.4cm 4.2cm, clip,scale=1,width=4.75cm]{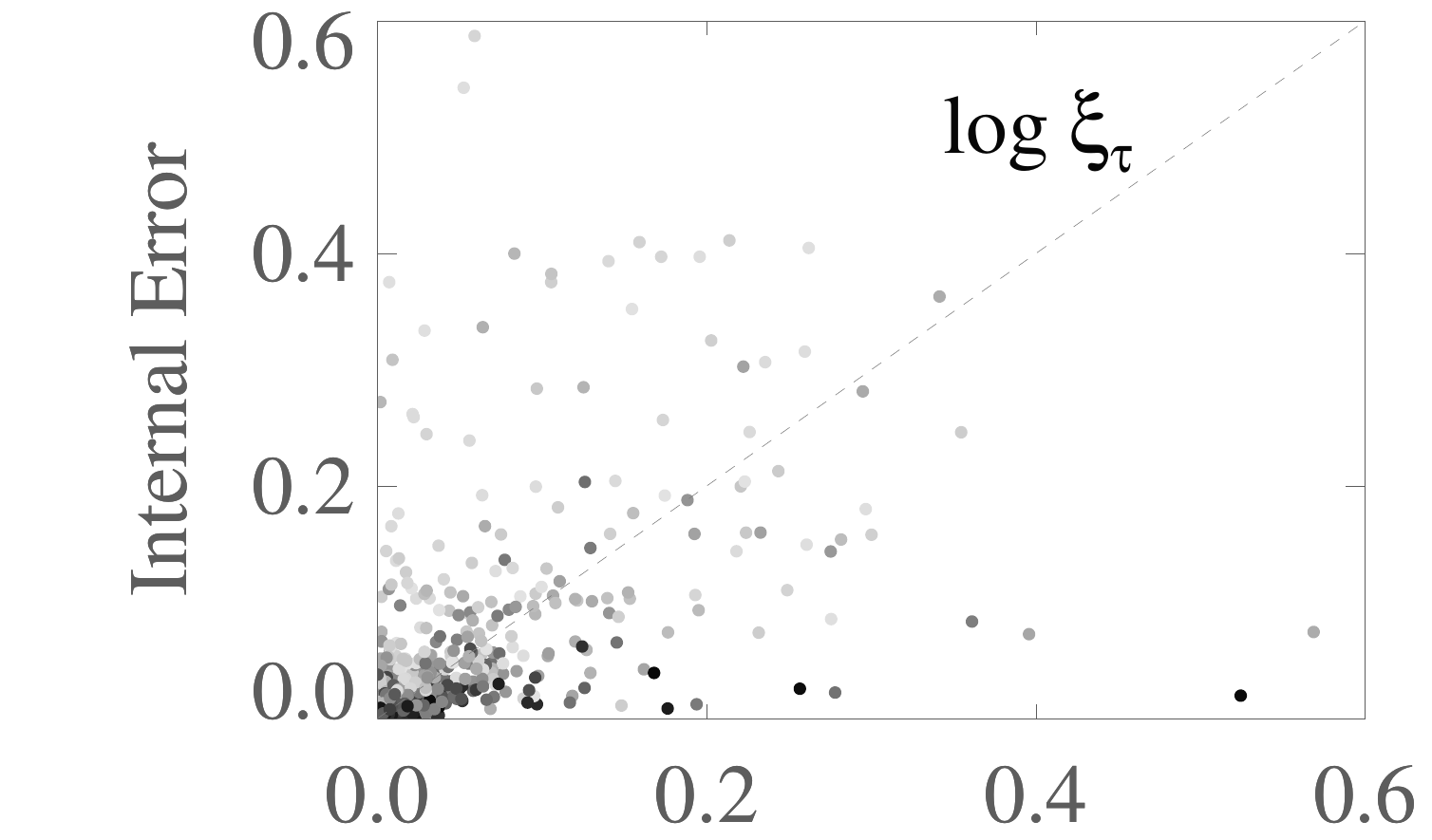}
\includegraphics[page=17,angle=0,trim=5.5cm 2.4cm 2.4cm 4.2cm, clip,scale=1,width=3.75cm]{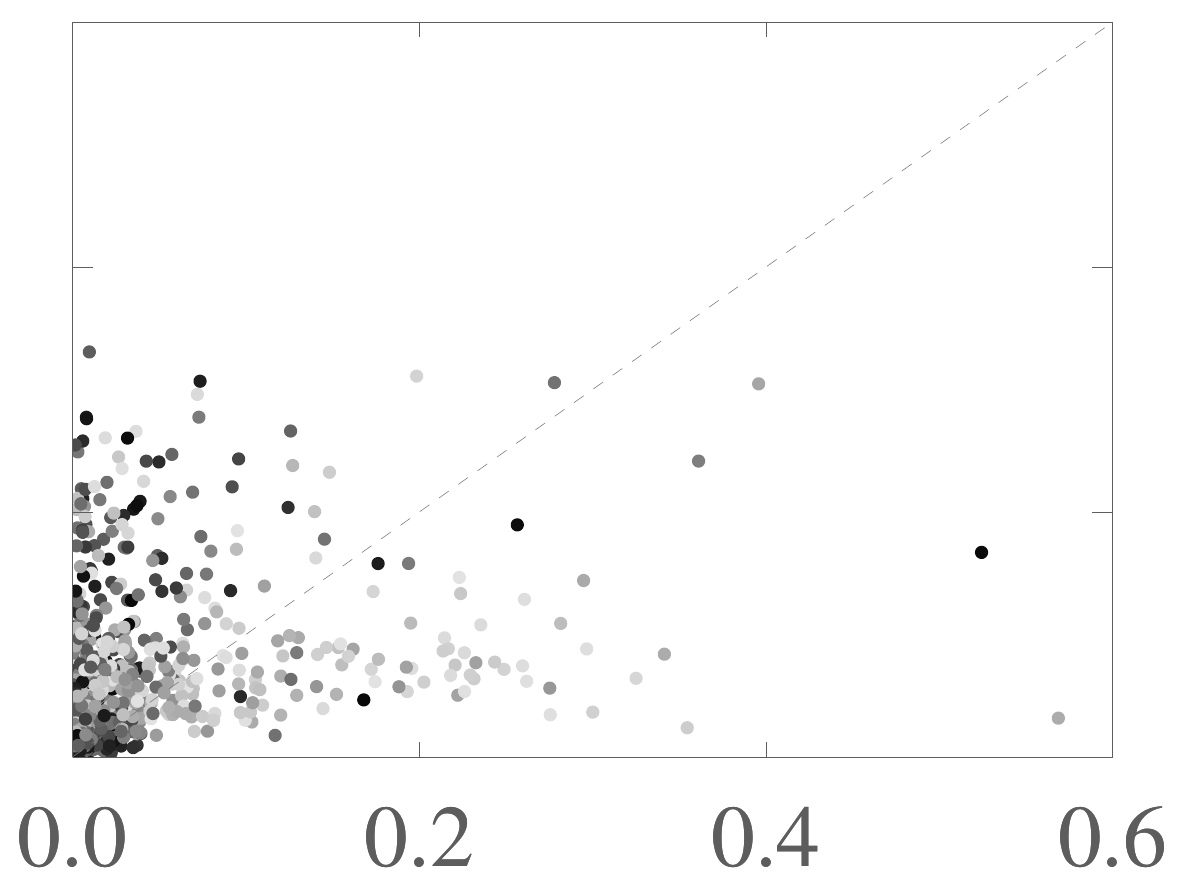}\\
\includegraphics[page=9,angle=0,trim=0cm 2.4cm 2.4cm 4.2cm, clip,scale=1,width=4.75cm]{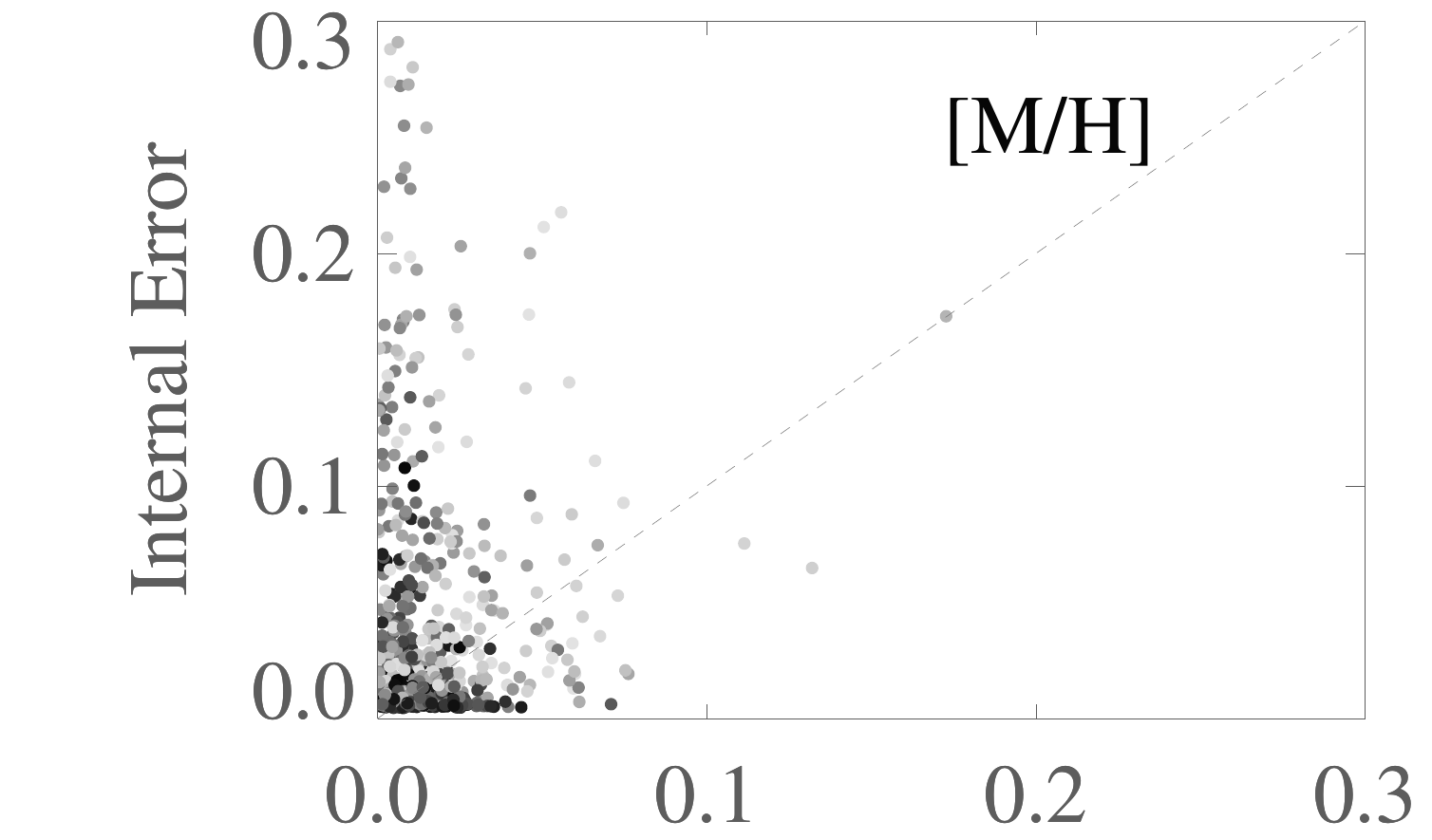}
\includegraphics[page=9,angle=0,trim=5.5cm 2.4cm 2.4cm 4.2cm, clip,scale=1,width=3.75cm]{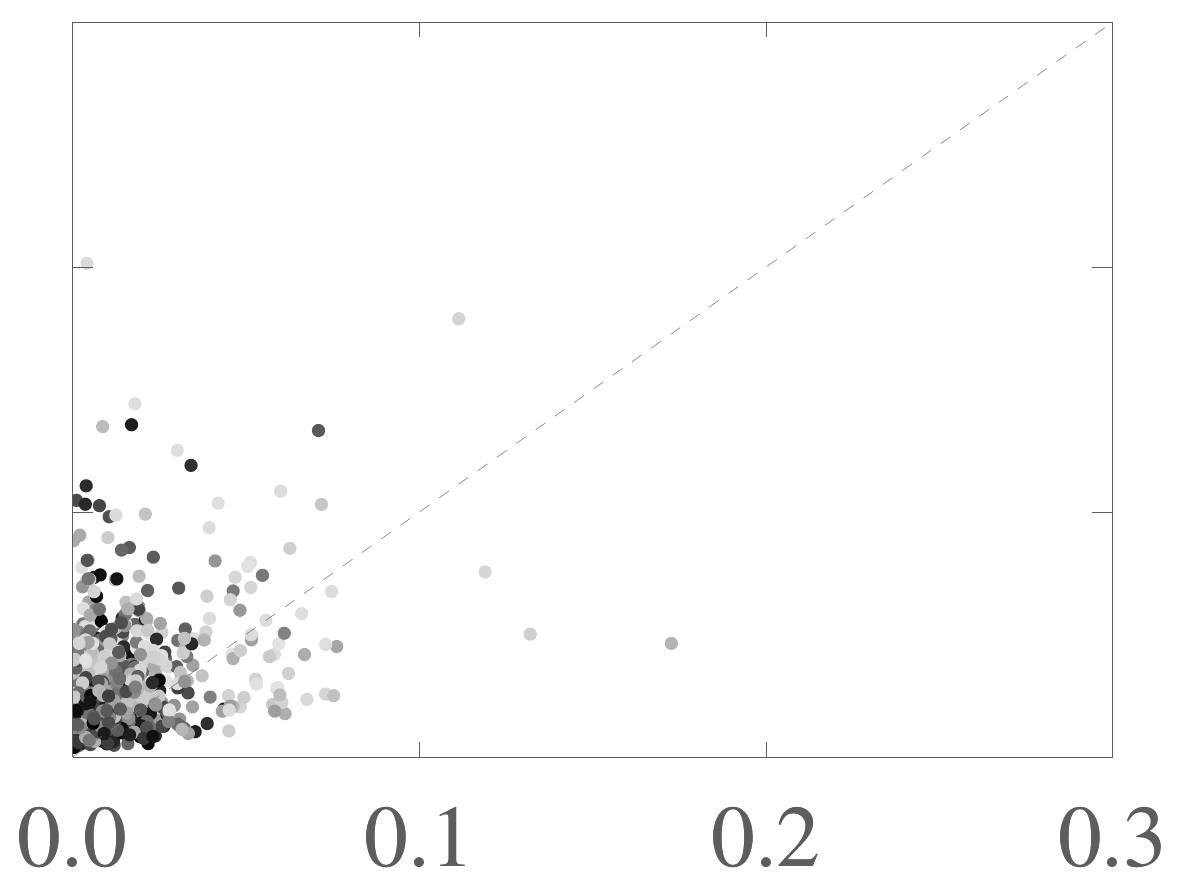}\\
\includegraphics[page=15,angle=0,trim=0cm 2.4cm 2.4cm 4.2cm, clip,scale=1,width=4.75cm]{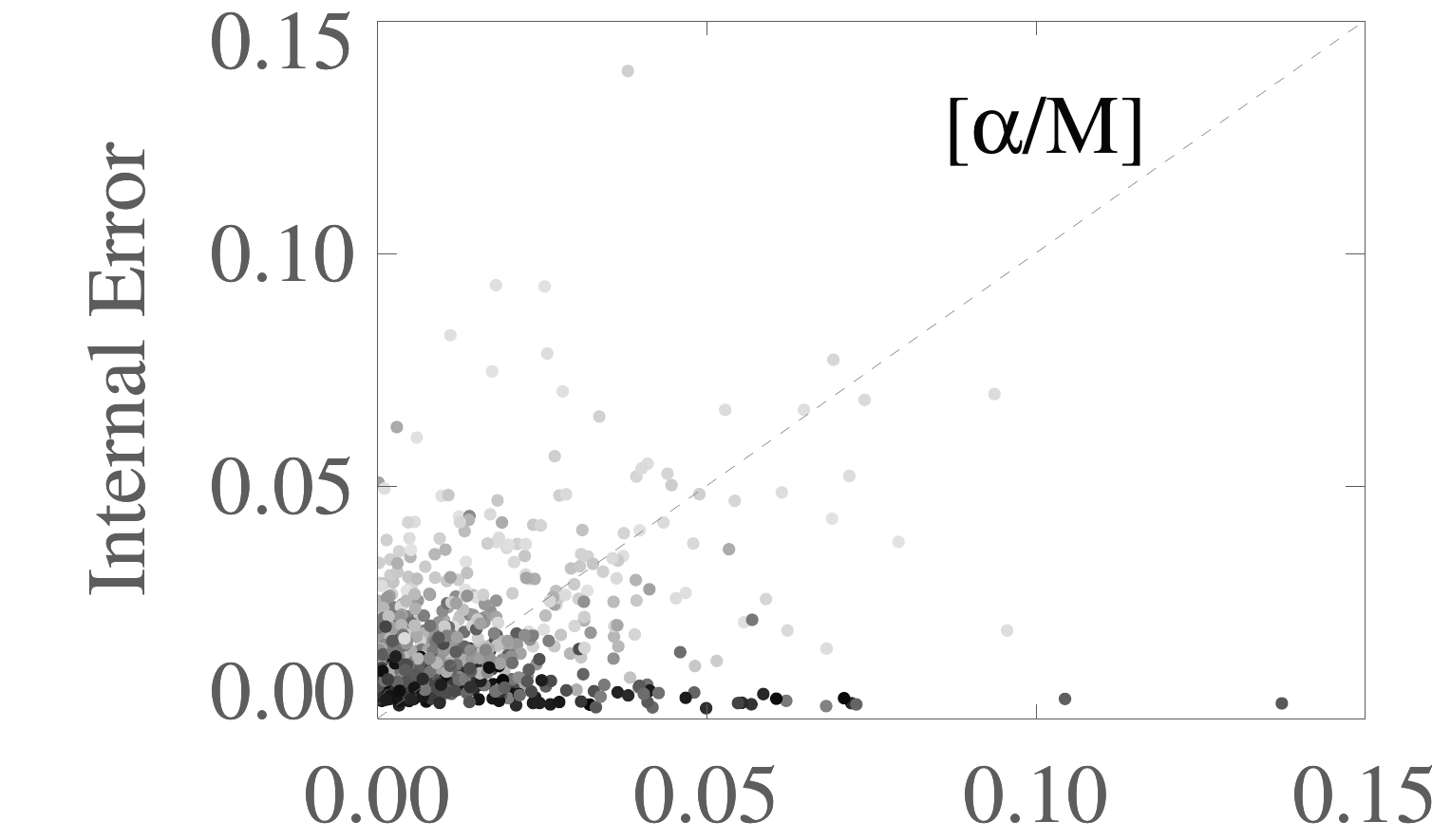}
\includegraphics[page=15,angle=0,trim=5.5cm 2.4cm 2.4cm 4.2cm, clip,scale=1,width=3.75cm]{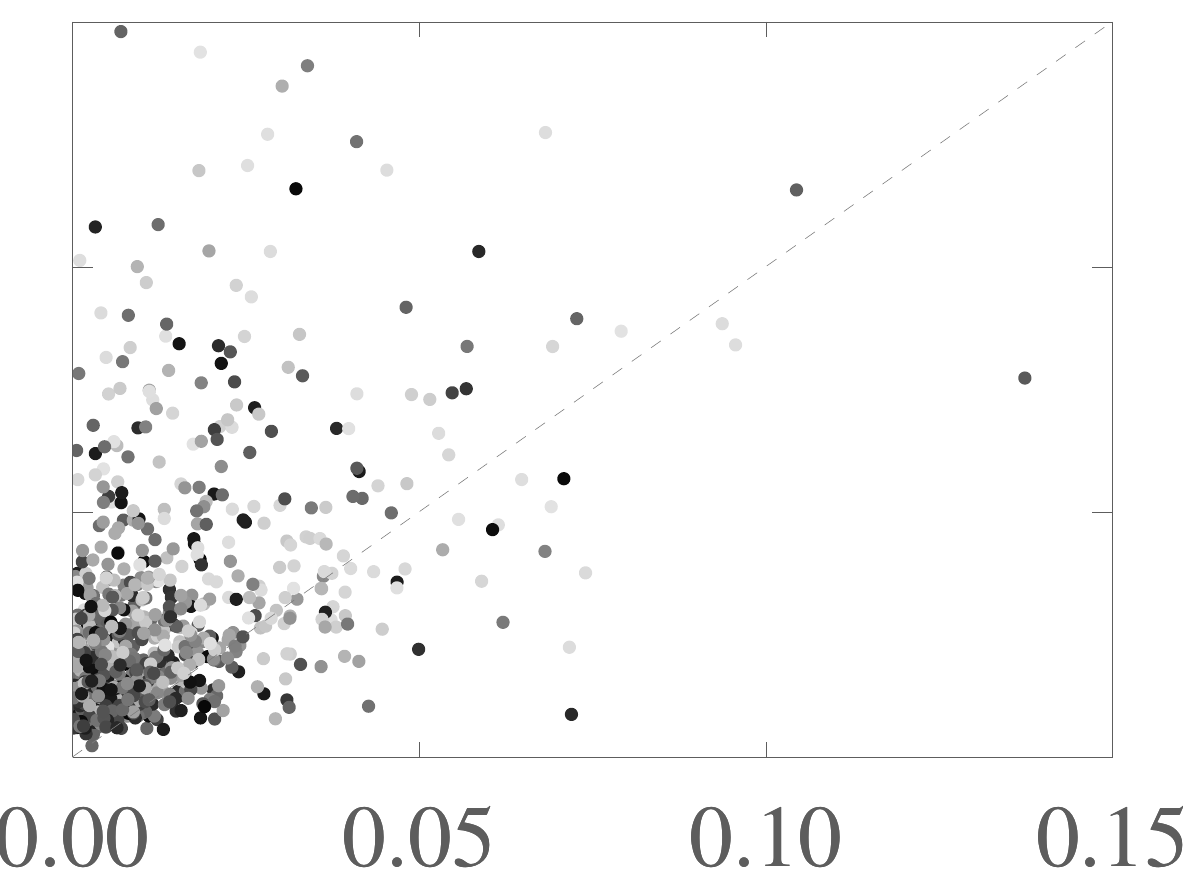}\\
\includegraphics[page=13,angle=0,trim=0cm 2.4cm 2.4cm 4.2cm, clip,scale=1,width=4.75cm]{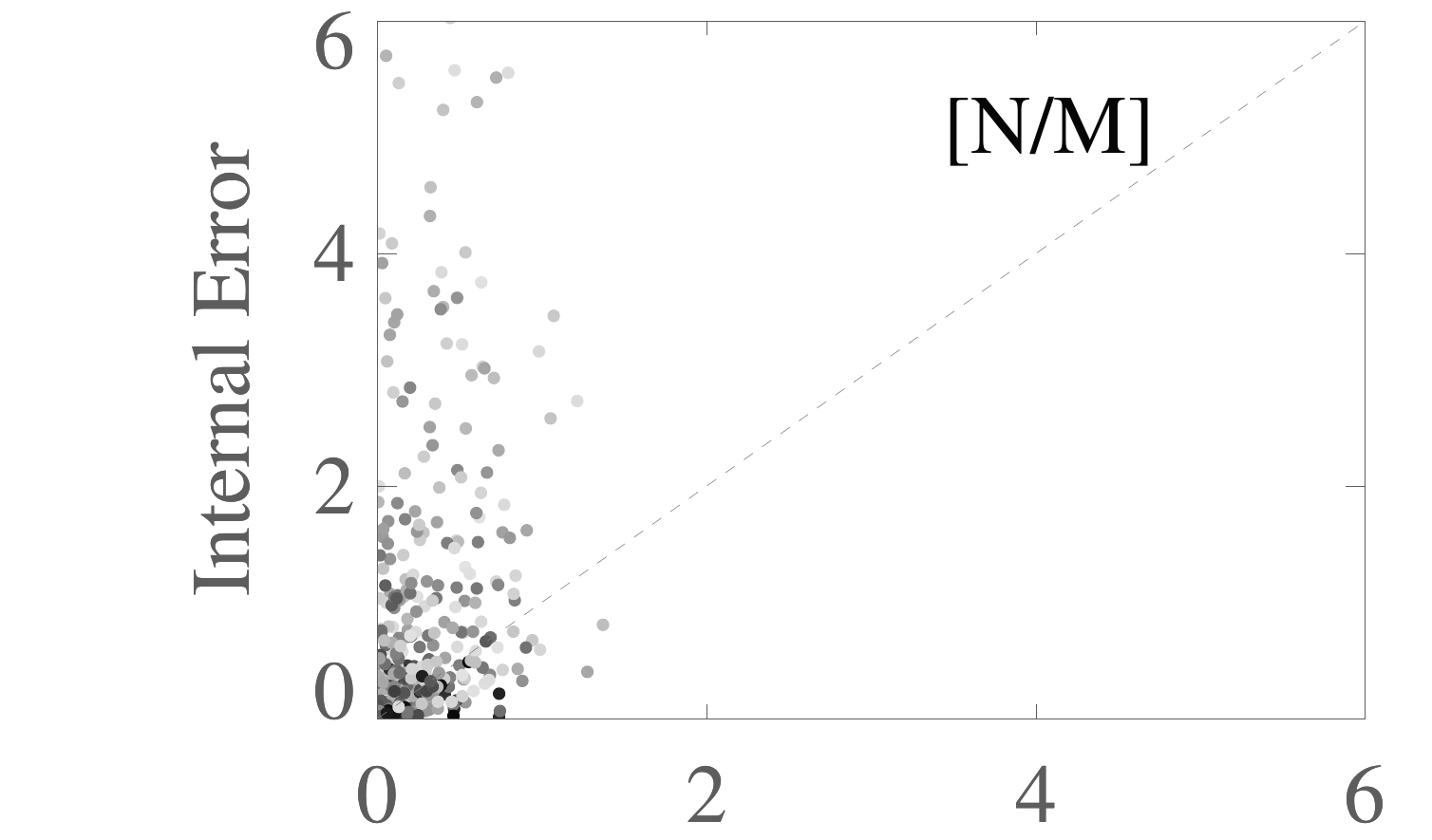}
\includegraphics[page=13,angle=0,trim=5.5cm 2.4cm 2.4cm 4.2cm, clip,scale=1,width=3.75cm]{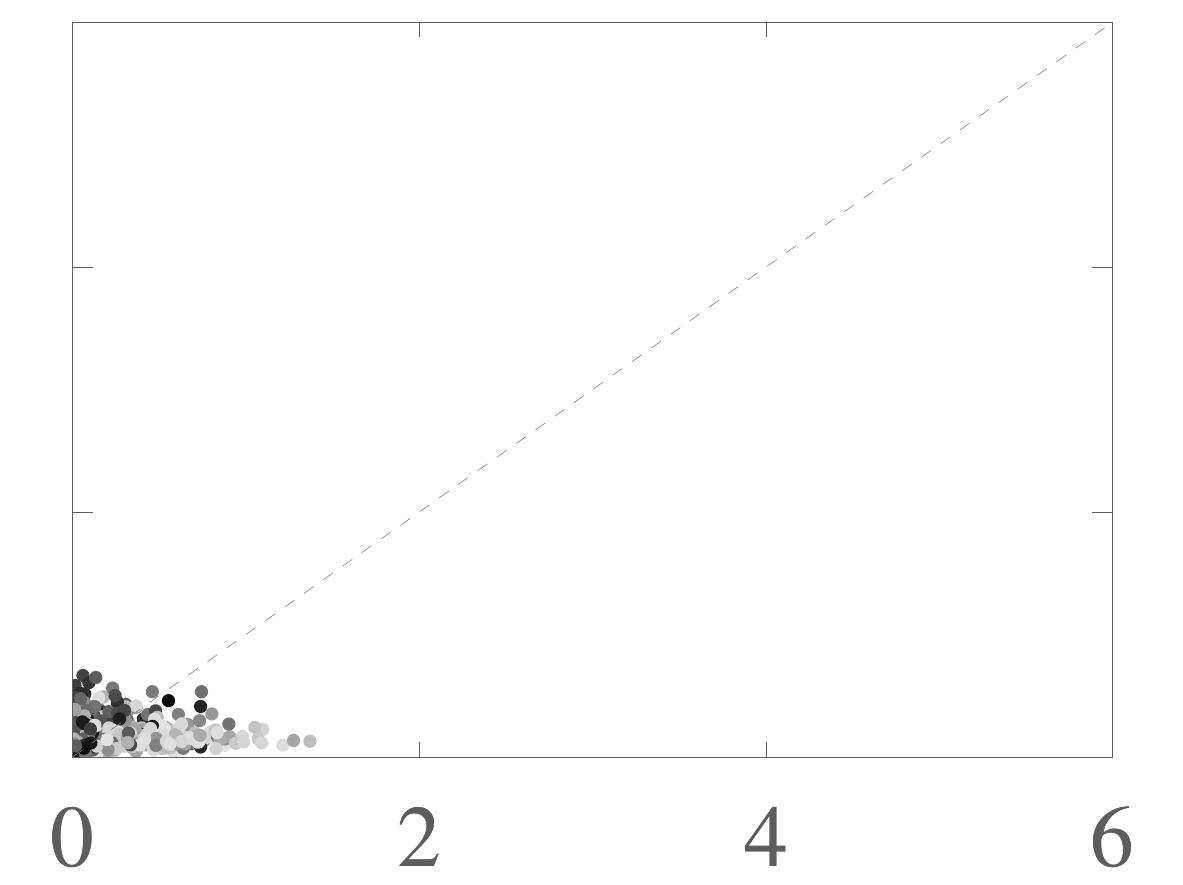}\\
\includegraphics[page=11,angle=0,trim=0cm 0cm 2.4cm 4.2cm, clip,scale=1,width=4.75cm]{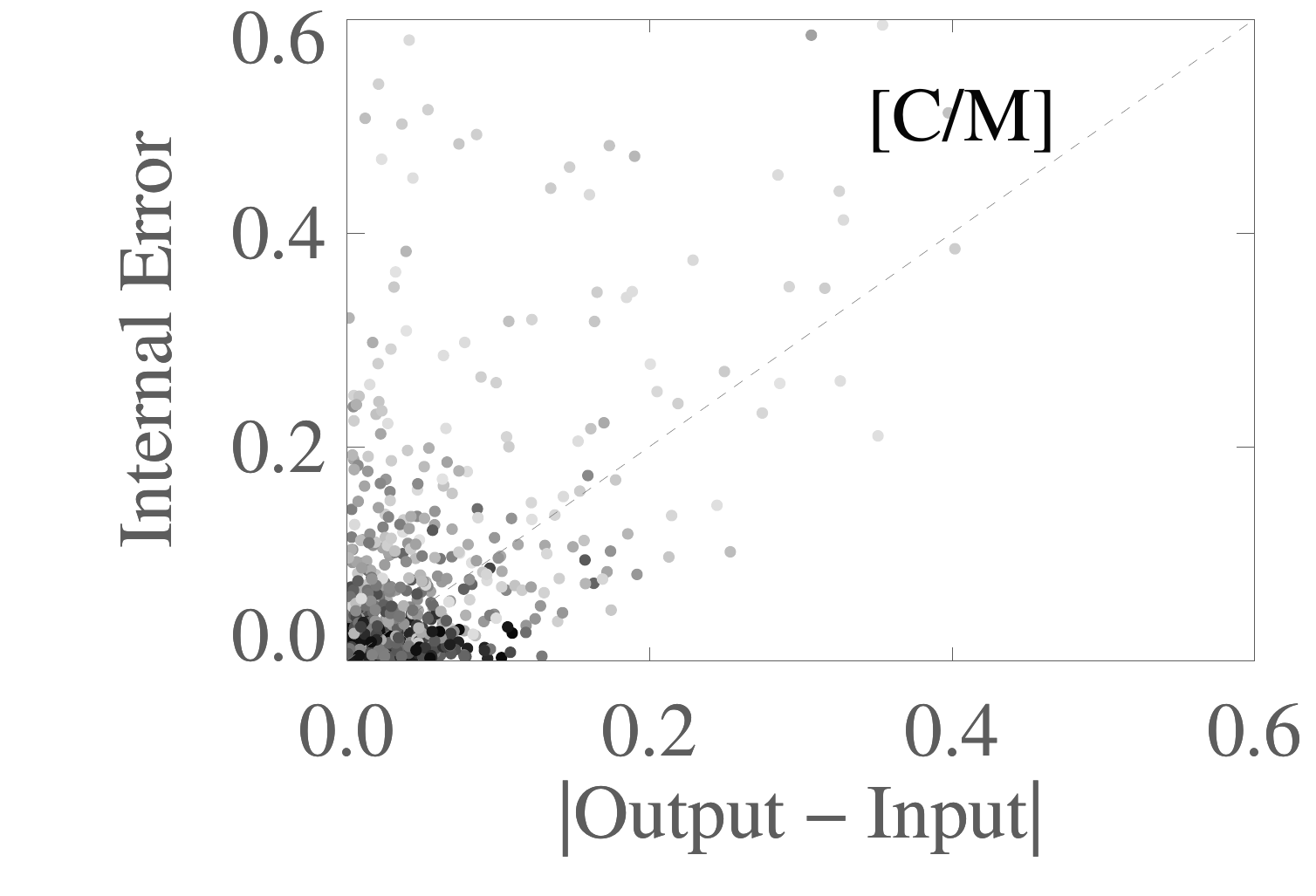}
\includegraphics[page=11,angle=0,trim=5.5cm 0cm 2.4cm 4.2cm, clip,scale=1,width=3.75cm]{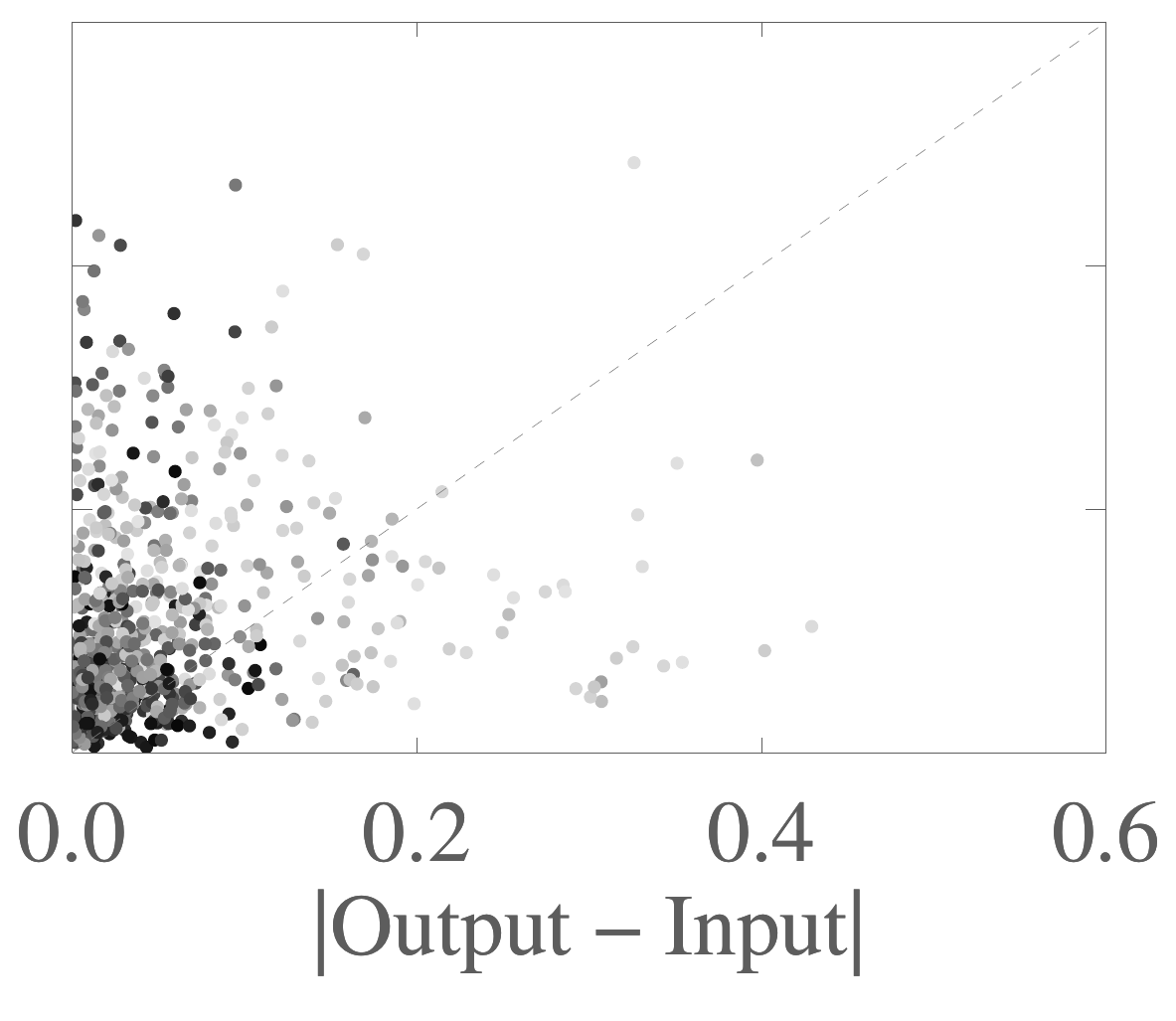}\\

\end{center}

\caption{{\ferre} internal parameter errors versus error estimates from differences between
input and output parameters of synthetic spectra. Results are presented for 
the off-node ($S/N$=100) test with the curvature matrix (left panels) and the multiple search error (right panels) options. Shading indicates
metallicity and dashed lines show the one-to-one relation.}
\label{figerr}
\end{figure}

Figure~\ref{figerr} plots these two internal error estimates 
against the difference between input and output parameter values for
the $S/N=100$ off-nodes sample test. Comparing the left and right columns
shows that the curvature matrix errors are, in an average sense, comparable or better 
to those found by multiple searches with Gaussian added noise. Furthermore,
while there is obviously scatter between these internal errors and the
$|{\rm output}-{\rm input}|$ differences (the error should only predict
the difference in an rms sense), the general magnitude of these internal
error estimates appears to be correct in this synthetic spectrum test.
Some low metallicity spectra (\feh$\  < -1$) show excessive internal
error estimates in the abundance parameters and microturbulence, especially
at low $S/N$; a better agreement is reached with higher $S/N$. For the case 
of S/N=100, and when curvature-matrix errors are considered, the values can exceed the 
maximum range in Figure~\ref{figerr}. The ranges displayed correspond to the 
majority of the sample. Cases with large $|{\rm output}-{\rm input}|$ in Figure~\ref{figerr} 
also tend to have large internal error estimates, indicating that these
will generally flag parameter values that have large uncertainty. 
However, not all large internal errors correspond to large differences.

Both the curvature matrix and multiple search methods yield much smaller
error estimates than those found empirically from scatter in open clusters 
\citep{Holtzman15}. This difference suggests that the
actual errors (``random'' as well as systematic) are typically dominated by mismatch between the model spectra and the true spectra.
This mismatch can be a consequence of imperfect theoretical modeling
or of imperfectly representing details of the data such as LSF variations
or telluric subtraction errors. Given the high $S/N$ ratio of APOGEE
spectra, it is not surprising that modeling errors dominate over photon
noise in many circumstances, though photon noise may still be the limiting
factor for individual elements especially at low metallicity. Unfortunately,
this class of errors is difficult to quantify based on internal properties
of the fits. A positive implication is that improved modeling could reduce
the parameter errors for APOGEE DR12 by a substantial factor, with
no changes to the data themselves.

\subsection{Test D. Effect of Masking Windows in the Global Fit}

$H$-band spectra from the ground suffer substantial degradation from 
 Earth's atmosphere, which imprints OH emission lines and 
absorption by O$_2$, CH$_4$, H$_2$O, and CO$_2$. Depending on the strength of
these features, their presence and removal increase the uncertainties in the observed fluxes, in some cases to the point that 
data become useless at particular wavelengths. The fluxes at wavelengths significantly affected by these features of CH$_4$, 
H$_2$O, and CO$_2$ are weighted according to the uncertainties in the telluric-corrected fluxes.

We have evaluated their impact (Test D in Table~\ref{tests}) by 
using the actual error bars associated with the APOGEE spectrum of the star 2M18161497-1738507 (APOGEE field 4339, $l=14\degr$, $b=0\degr$) to identify the spectral windows 
to mask in our synthetic spectra. The star was selected arbitrarily among those with only one APOGEE visit to avoid uncertainties associated 
with the combination of multiple visits. Some 17\% of the total number of pixels are rendered unusable in this particular spectrum, which is a typical figure
for the APOGEE data.
The analysis results for the $S/N=100$ realization for the off-nodes 
sample with blocked spectral windows do not show a significant degradation compared to the analysis of the same data set without blocked
windows, nor do they show significantly larger errors.

\subsection{Test E. Uncertainties Associated with Modeling the Line Spread Function}
\label{lsfsub}

\begin{figure} 
\figurenum{8}
\begin{center}

\includegraphics[page=8,angle=0,trim=0cm 1.0cm 3.5cm 4cm, clip,scale=1,width=4.68cm]{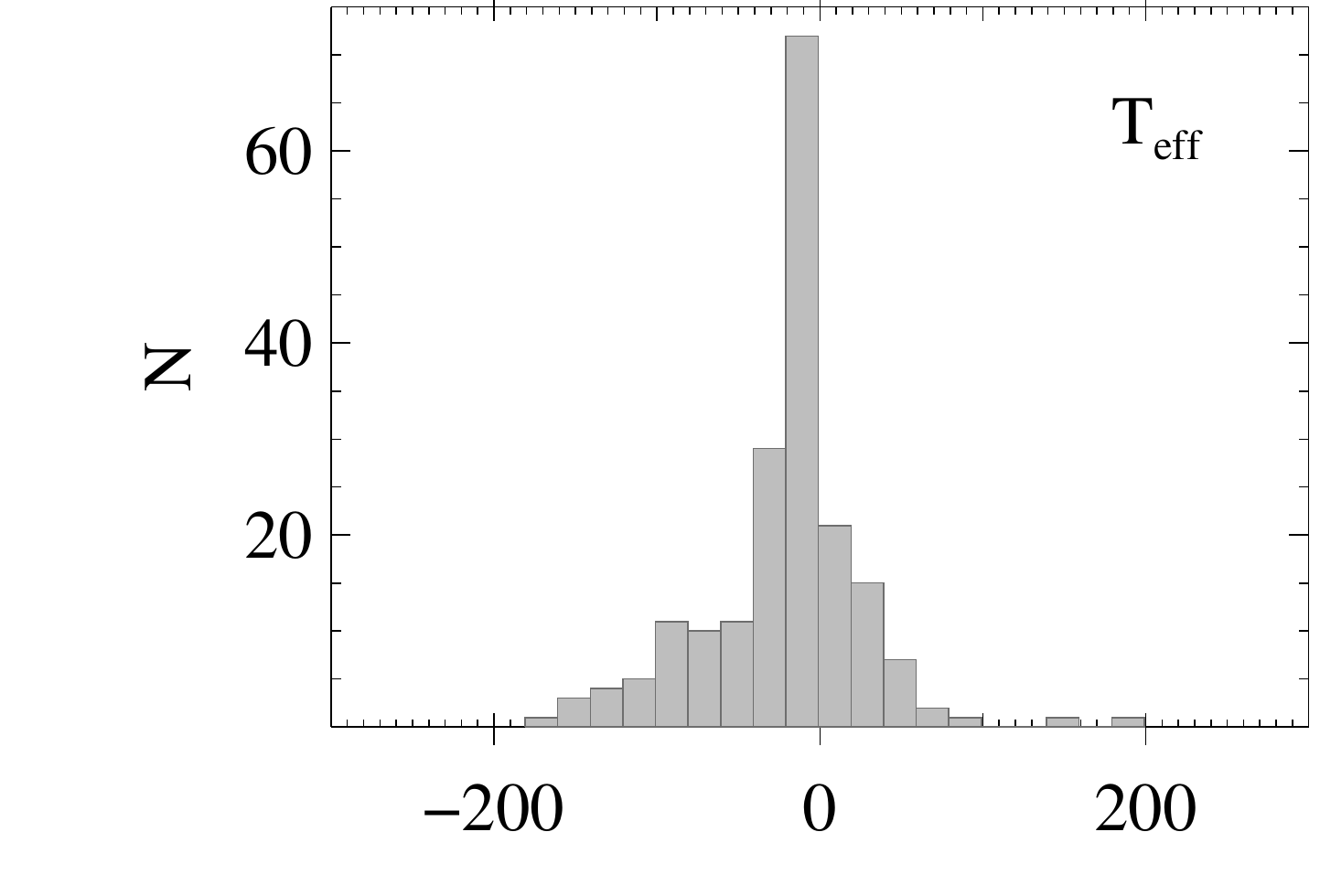}
\includegraphics[page=8,angle=0,trim=6cm 1.0cm 3.5cm 4cm, clip,scale=1,width=3.55cm]{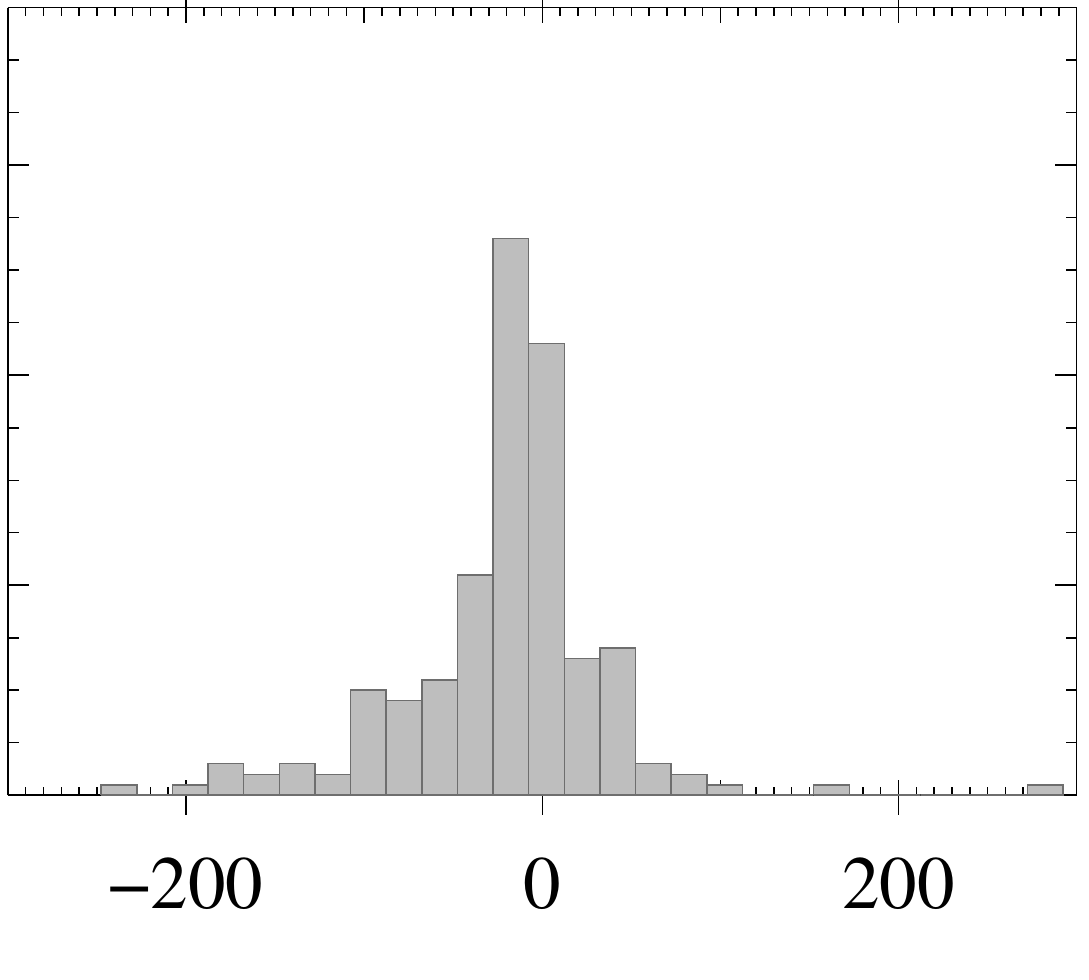}\\
\includegraphics[page=7,angle=0,trim=0cm 4.1cm 3.5cm 4cm, clip,scale=1,width=4.68cm]{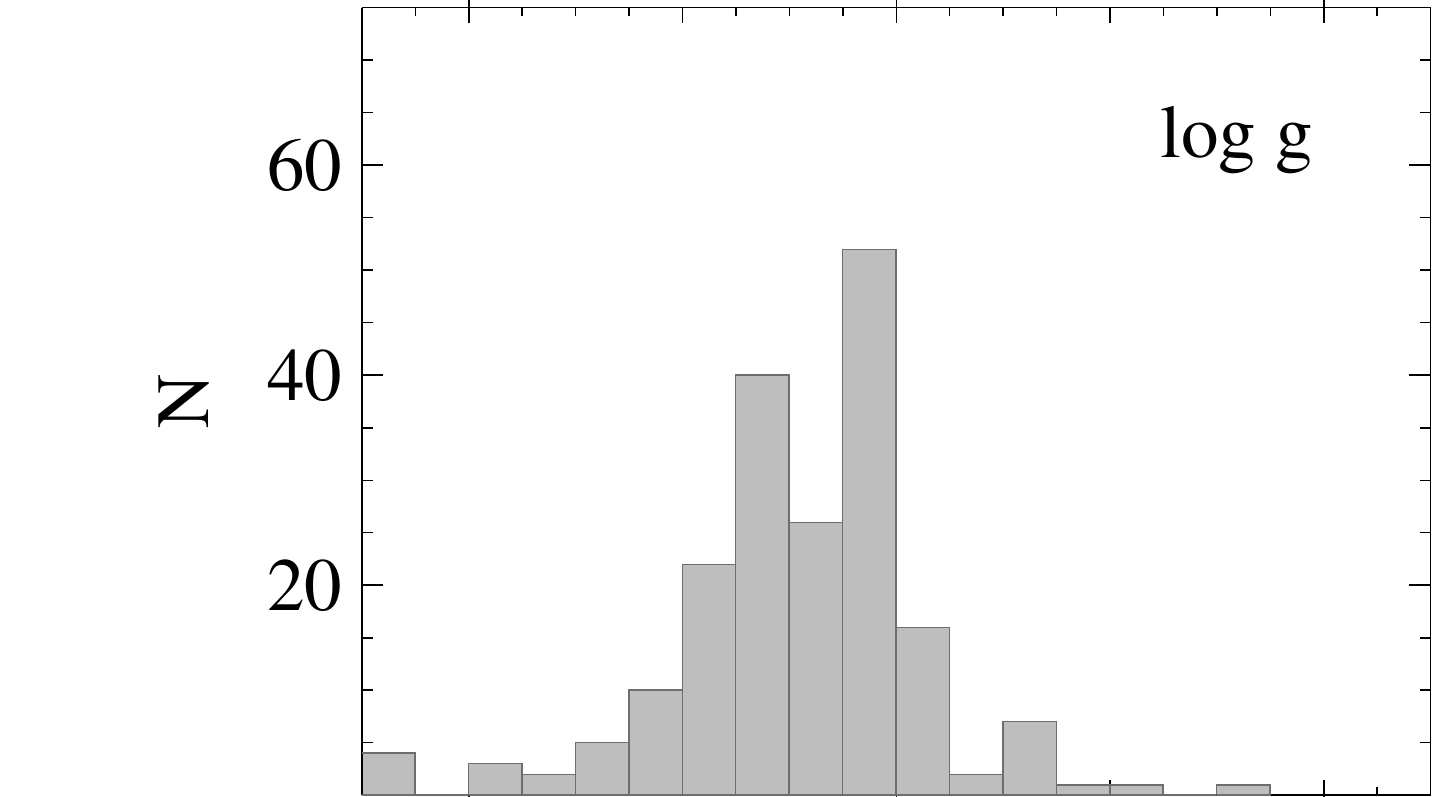}
\includegraphics[page=7,angle=0,trim=6cm 4.1cm 3.5cm 4cm, clip,scale=1,width=3.55cm]{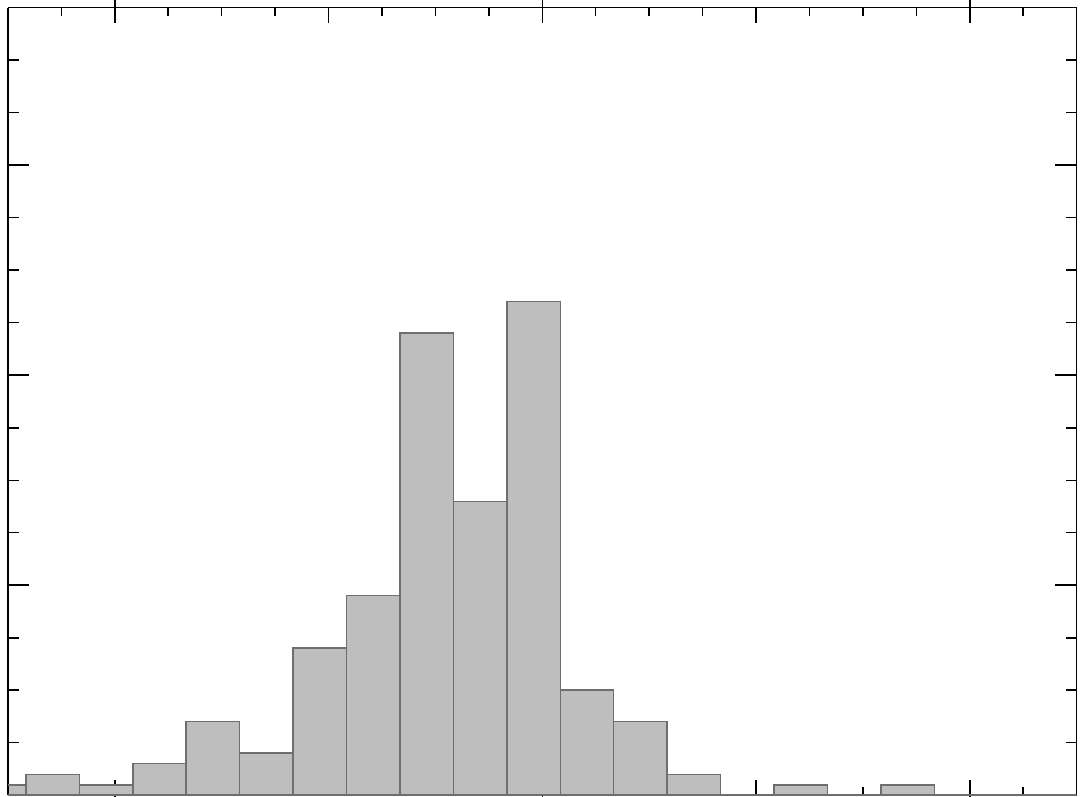}\\
\includegraphics[page=2,angle=0,trim=0cm 4.1cm 3.5cm 4cm, clip,scale=1,width=4.68cm]{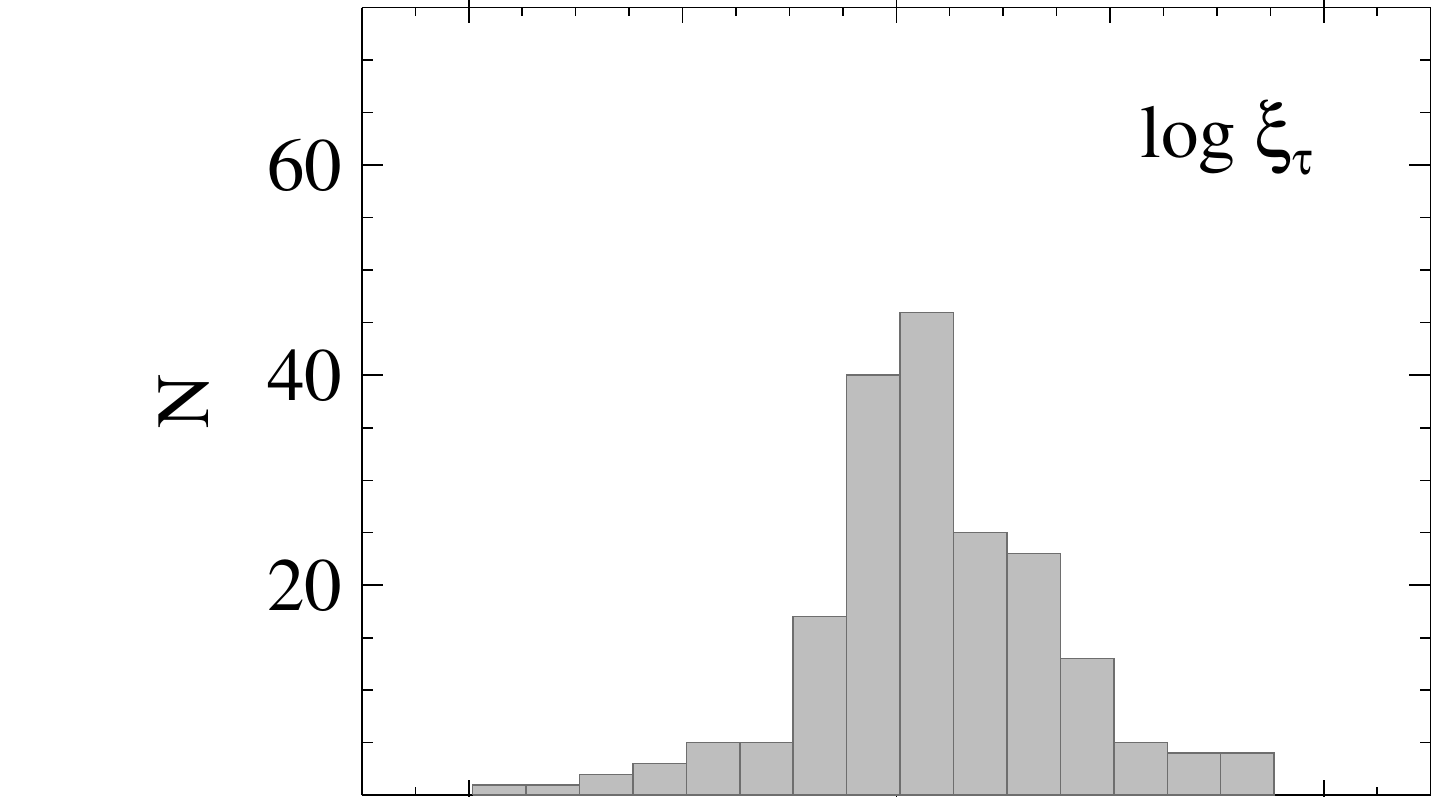}
\includegraphics[page=2,angle=0,trim=6cm 4.1cm 3.5cm 4cm, clip,scale=1,width=3.55cm]{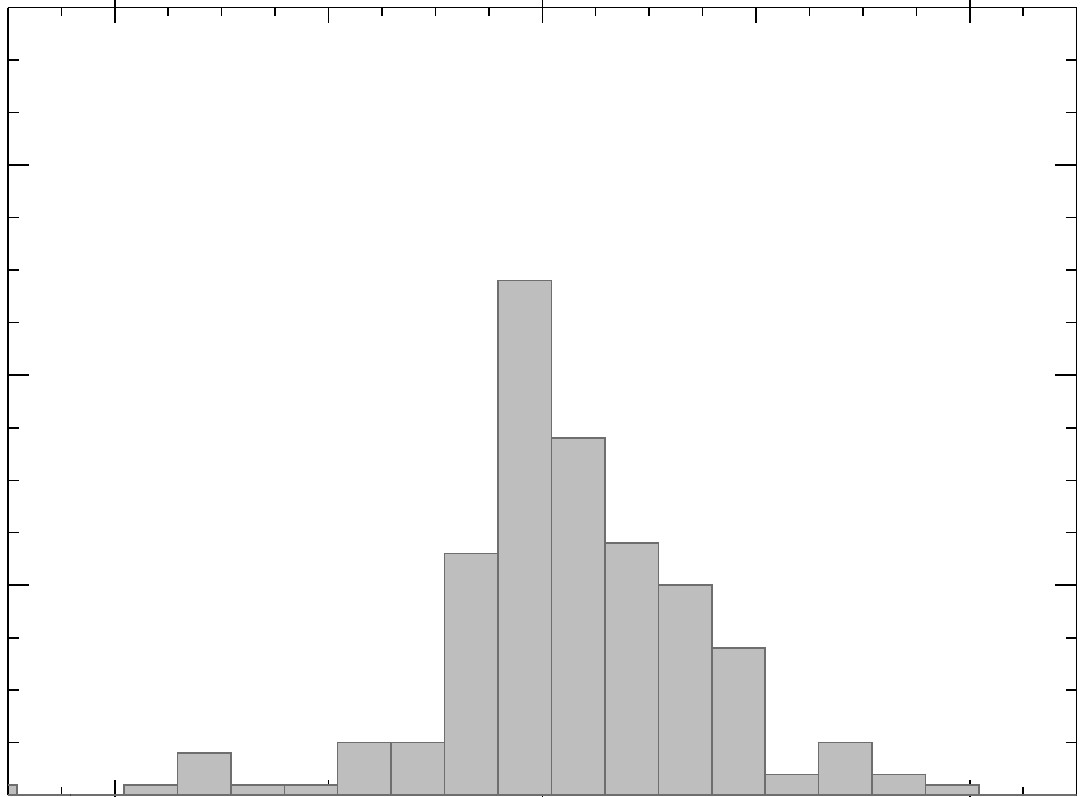}\\
\includegraphics[page=6,angle=0,trim=0cm 4.1cm 3.5cm 4cm, clip,scale=1,width=4.68cm]{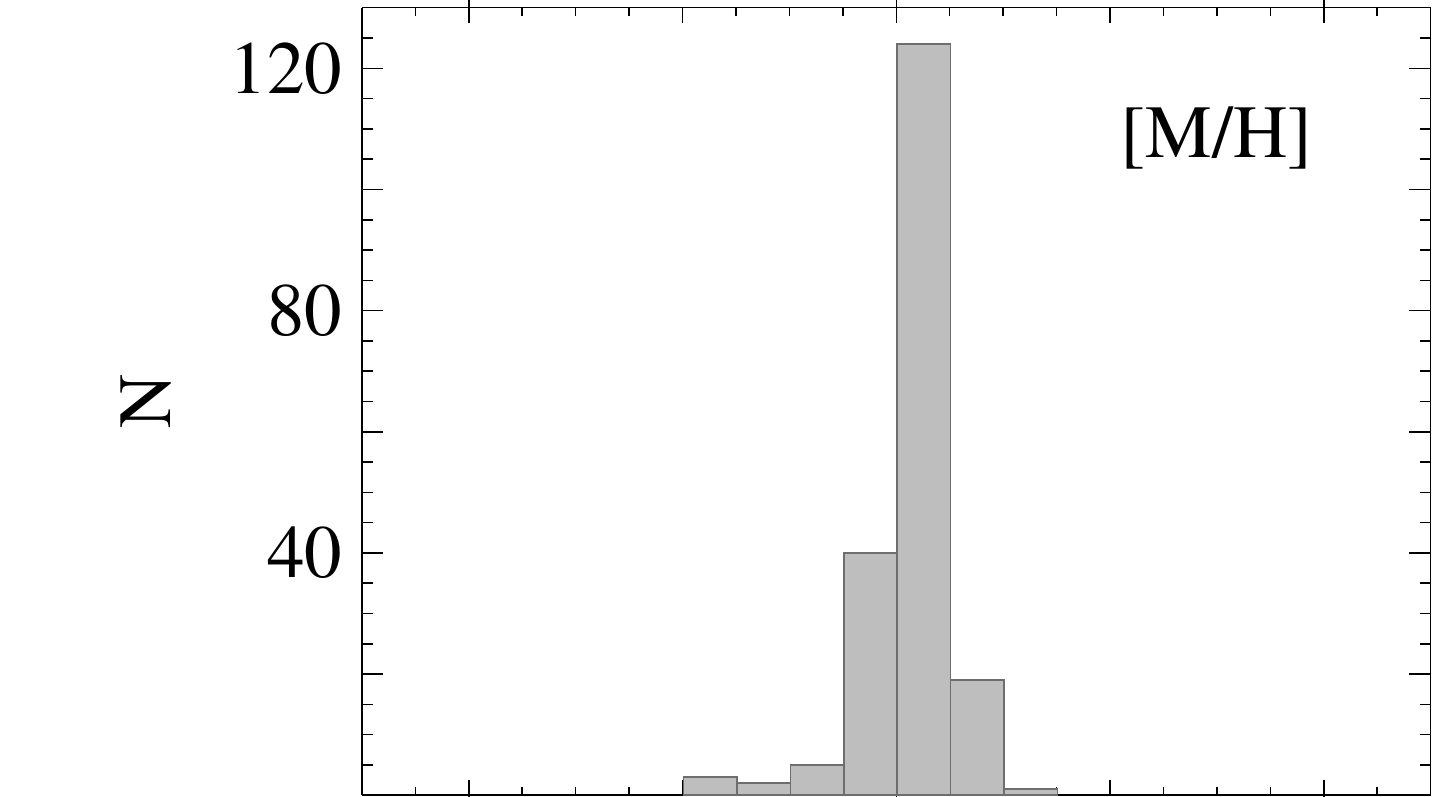}
\includegraphics[page=6,angle=0,trim=6cm 4.1cm 3.5cm 4cm, clip,scale=1,width=3.55cm]{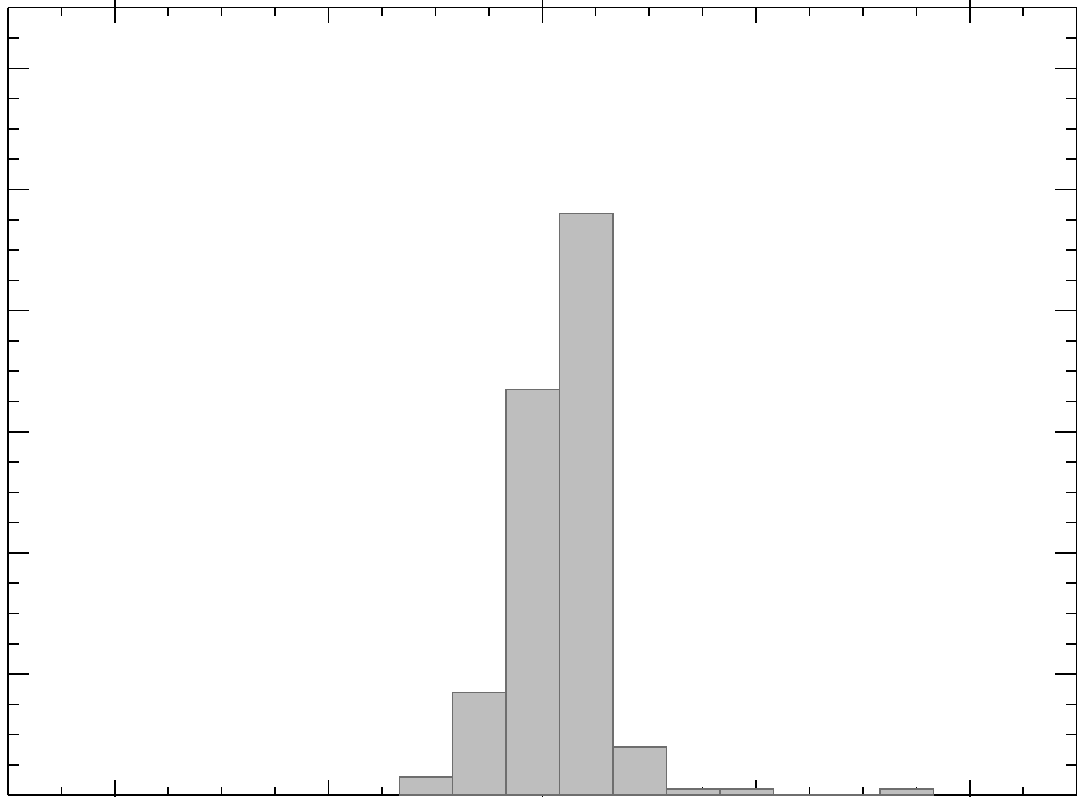}\\
\includegraphics[page=5,angle=0,trim=0cm 1.1cm 3.5cm 4cm, clip,scale=1,width=4.68cm]{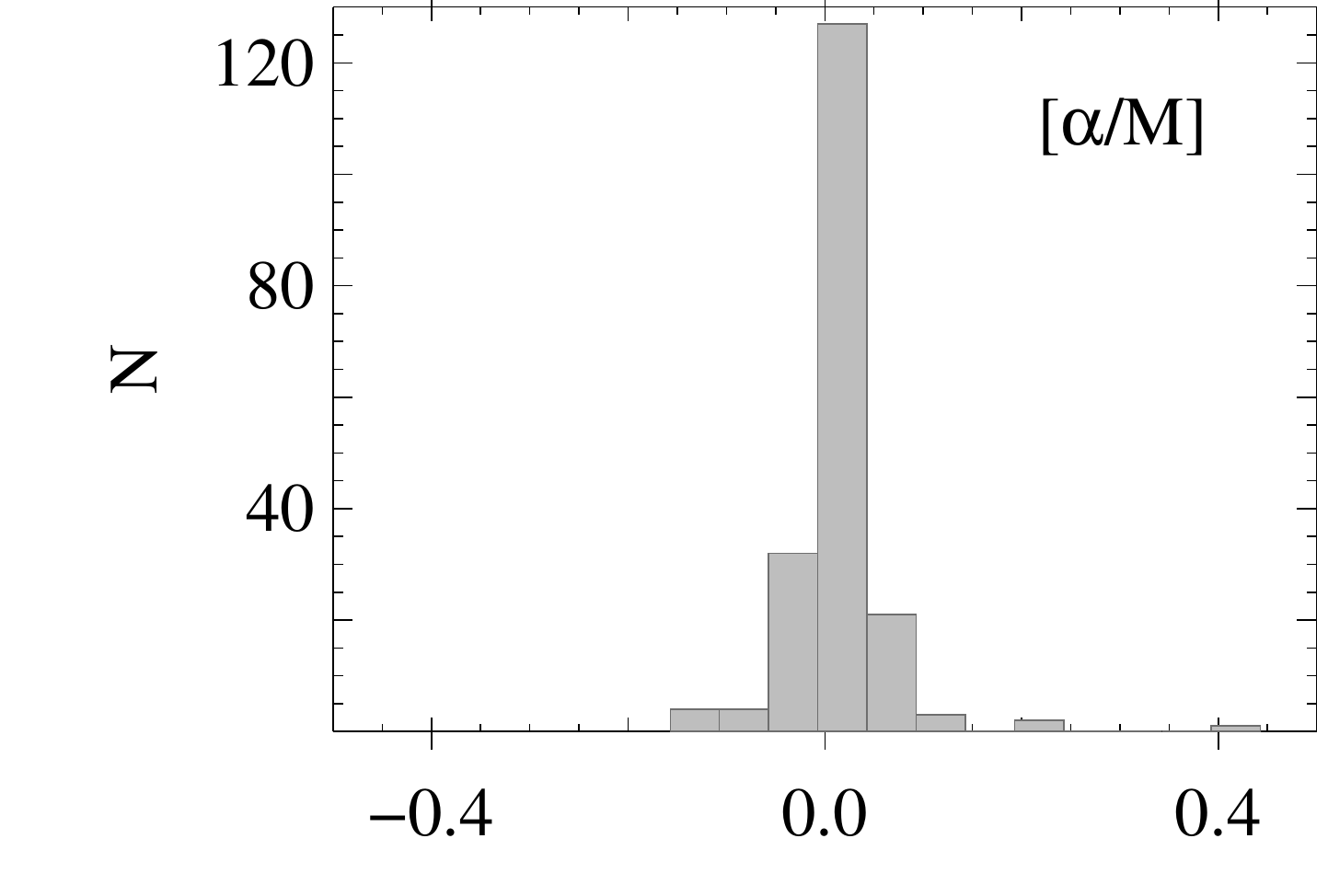}
\includegraphics[page=5,angle=0,trim=6cm 1.1cm 3.5cm 4cm, clip,scale=1,width=3.55cm]{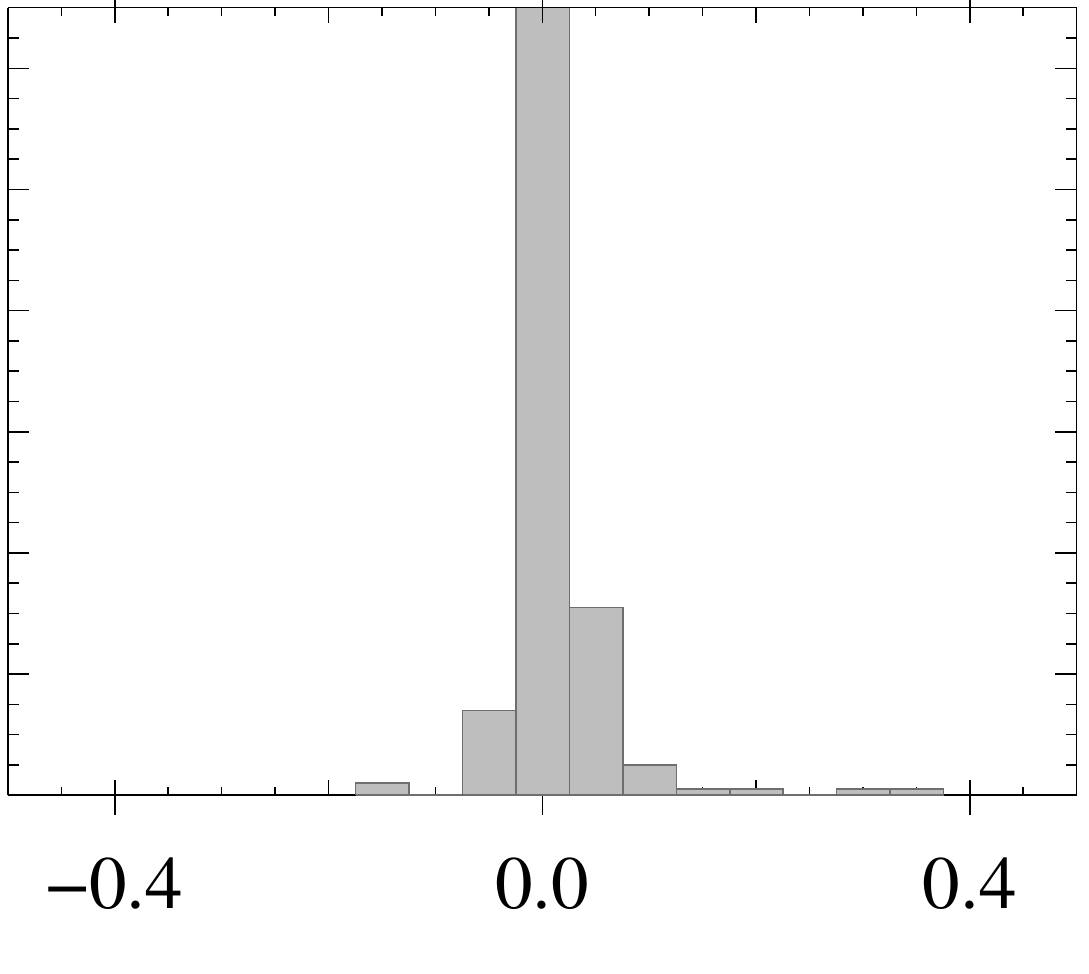}\\
\includegraphics[page=4,angle=0,trim=0cm 4.1cm 3.5cm 4cm, clip,scale=1,width=4.68cm]{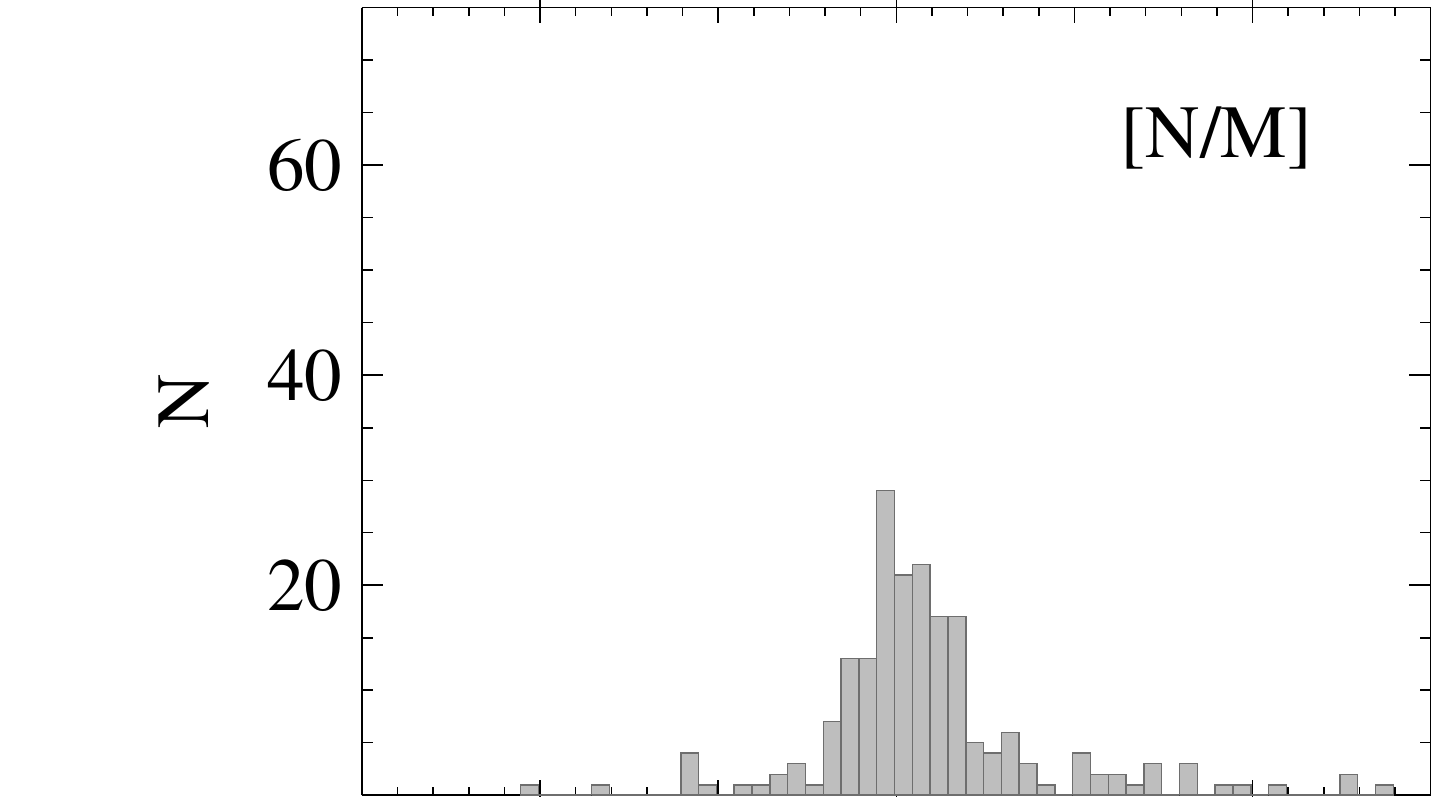}
\includegraphics[page=4,angle=0,trim=6cm 4.1cm 3.5cm 4cm, clip,scale=1,width=3.55cm]{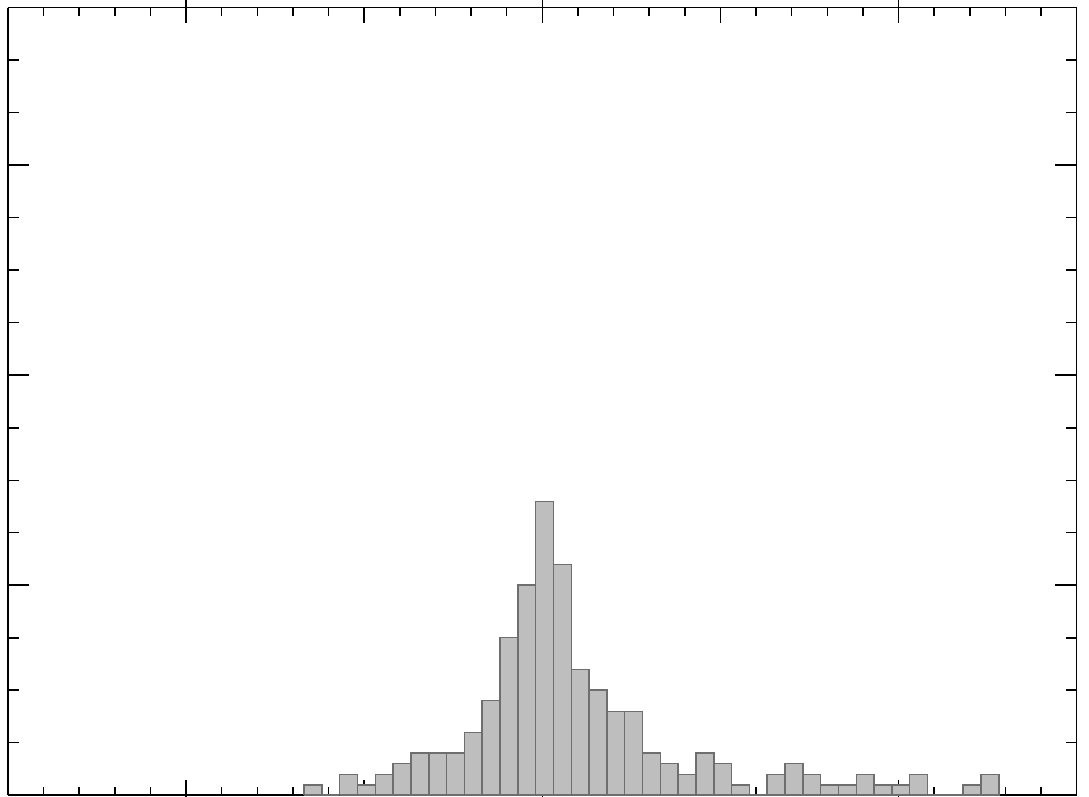}\\
\includegraphics[page=3,angle=0,trim=0cm 0.0cm 3.5cm 4cm, clip,scale=1,width=4.68cm]{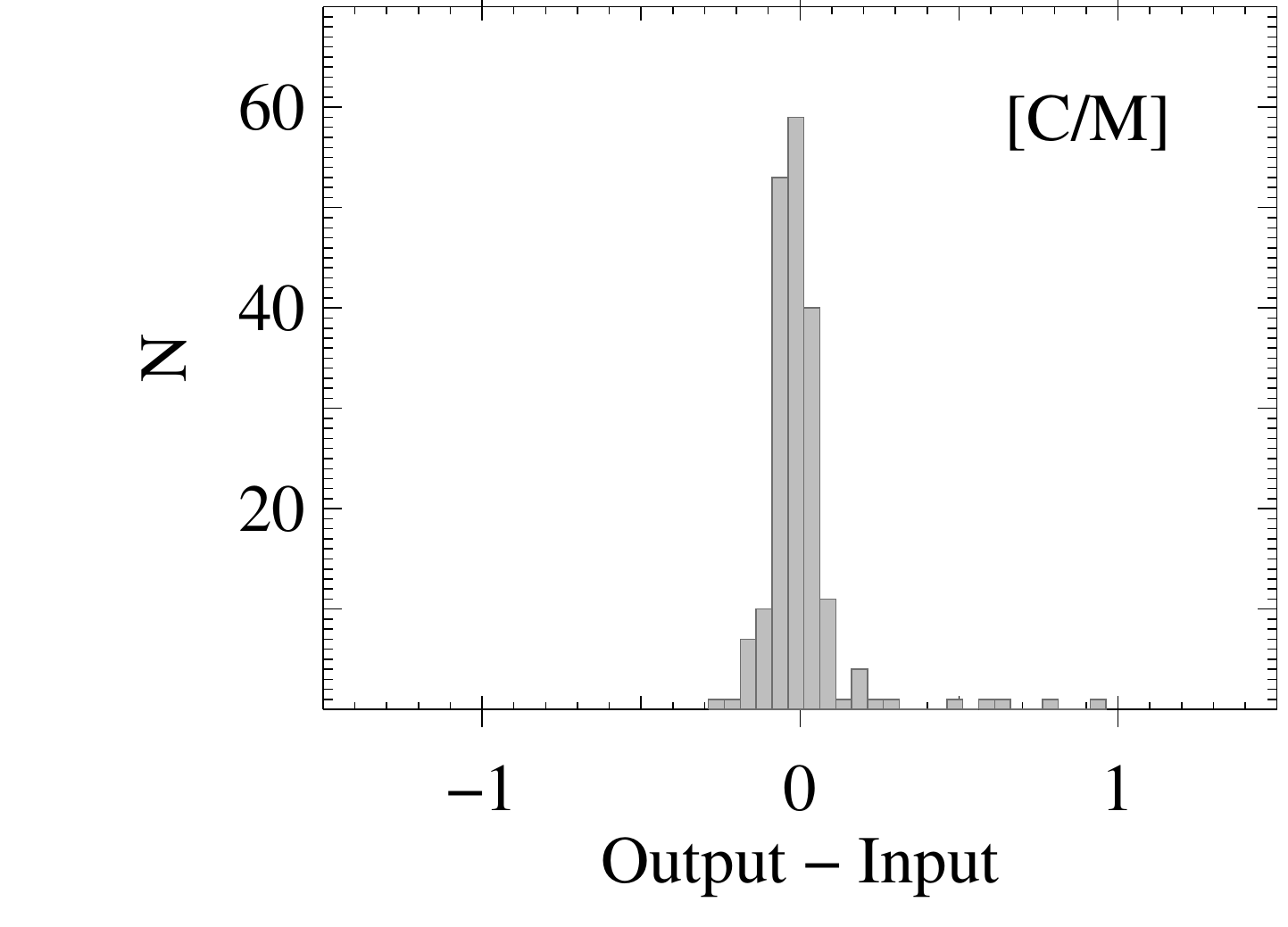}
\includegraphics[page=3,angle=0,trim=6cm 0.0cm 3.5cm 4cm, clip,scale=1,width=3.55cm]{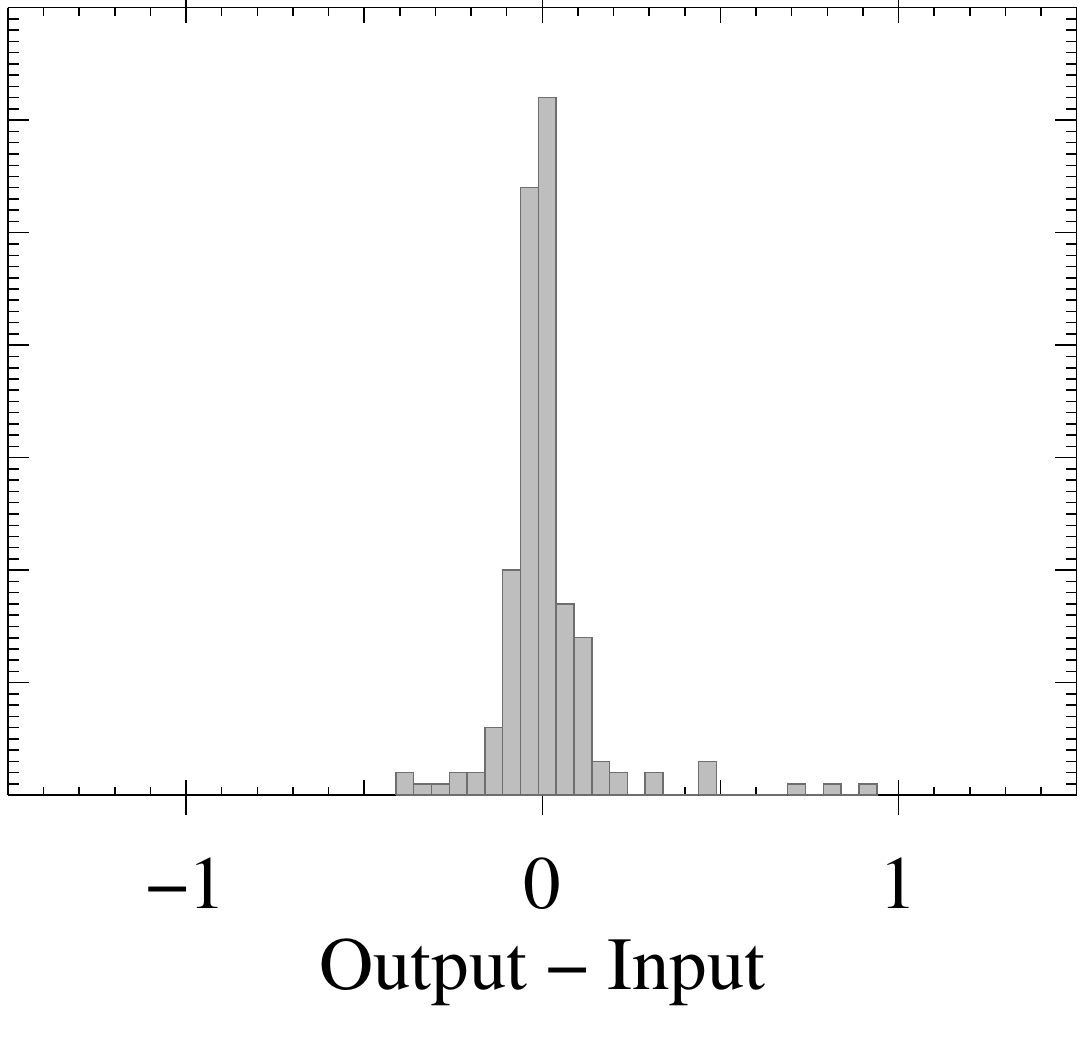}\\

\end{center}
\caption{The distribution of the parameter (output$\ -\ $input) differences for the LSF test that 
uses a $R=18,500$ (left) and a DR12 LSF (right) library to analyze synthetic spectra of 
$R=22,500$. Each row is for a different parameter.}
\label{lsffig}
\end{figure}

 The APOGEE LSF is not a Gaussian and changes 
 significantly along the pseudoslit (i.e., with fiber) and wavelength (Nidever et al. 2015, Wilson et al., in preparation). 
 Inaccurate LSF modeling in the spectral analysis can introduce 
uncertainties in ASPCAP parameter determinations, especially for \teff, \logg, and 
\microt. ASPCAP went from employing a Gaussian LSF kernel of constant 
resolving  power $R=22,500$ in DR10 to a more realistic LSF shape in DR12.
The DR12 LSF was varied as a function of wavelength, based on the average
LSF for five different fibers spread across the slit, but the same 
LSF was adopted for all spectra.

We have evaluated the effect of the LSF approximation we have been using by 
carrying out two experiments: (1) analyzing spectra from a library 
convolved with a Gaussian LSF equivalent to $R=22,500$ with a library for $R=18,000$, and (2) analyzing the same spectra
using the DR12 equivalent 7D library. In both tests we used the reduced on-node sample and noiseless
spectra. The choice of the low resolution Gaussian was to investigate the effect of assuming a wrong spectral resolution in the analysis 
(a possible case for some fibers in DR10/DR12). We used a rather
drastically mismatched resolution to test an extreme case. The test with 
the DR12 library helped to study the effect of adopting a wrong LSF shape+resolution for the ASPCAP parameter determinations. 
In this second test we used a Gauss-Hermite LSF of variable $R$ to analyze Gaussian convolved spectra of 
constant $R$. A similar magnitude of effect would be 
expected for the inverse case, which roughly accords with the
APOGEE DR10 analysis. 

Figure~\ref{lsffig} and the final three lines of Table~\ref{tests} (Test E) summarize the
results of the test with the correct LSF (for a comparison reference), and of the other two tests. 
The analysis of the spectra adopting an erroneously low 
spectral resolution shows that the impact of this systematic can be significant for some 
parameters. 
Overall the impacts of assuming an
incorrect spectral resolution or assuming an incorrect LSF shape and wavelength dependence, 
on top of the PCA compression and the cubic interpolation, are comparable in magnitude. Reassuringly, the median offsets are small in both tests:
below 0.025~dex in {\feh} and about $-16$~K in {\teff} and $-0.09$~dex in {\logg}. 
The largest median offset is for {\nfe}, which rises
by 0.03--0.04~dex. However, the distribution of offsets is fairly
broad for {\teff}, {\cfe}, and {\nfe}, and {\logg}, with dispersions of
$\approx 40$~K, 0.07~dex, 0.24~dex, and 0.09~dex, respectively, 
and some extreme outliers, not significant different from the equivalent test 
with the right LSF modeling. The values of {\teff}, {\logg}, and {\microt} are 
affected most by the LSF accuracy, severely at high surface gravity for the 
case of {\teff} and {\logg}.

The errors in DR10 parameters associated with inaccurate LSF modeling
should be comparable to those shown in our second test.
The corresponding errors in DR12 should be smaller, because these
ASPCAP analyses incorporate the non-Gaussian and wavelength-dependent LSF,
and their main omission is the fiber-to-fiber variation. Comparing to the results of Test C, the effects of LSF modeling errors are
larger (typically by a factor of 1.5--2.0) than those of interpolation
errors and noise at the level of $S/N=100$. (Note that the
reduced on-nodes sample used in Test E is five times smaller than
the off-nodes sample used in Test C, 194 versus \ 1000 spectra.)
While median offsets remain small compared to our accuracy goals, the
dispersions suggest that imperfect LSF modeling may still contribute non-negligibly 
to the ASPCAP error budget, particularly for the surface gravity, and 
for some classes of stars. 

\begin{figure*} 
\figurenum{9}
\begin{center}
{\includegraphics[page=1,angle=0,trim=3.0cm 2.5cm 9cm 1.5cm, clip,scale=0.39]{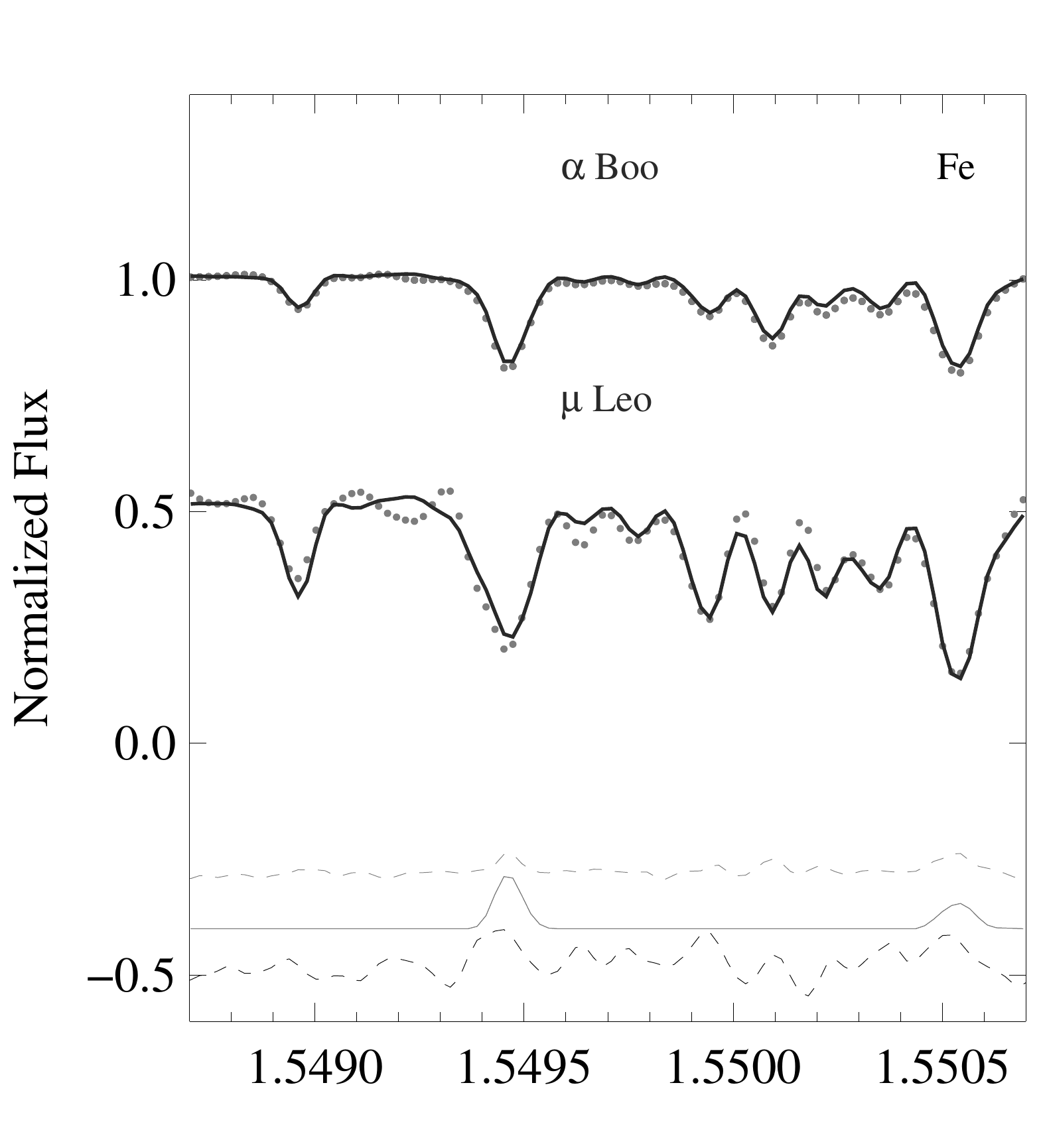}}
{\includegraphics[page=2,angle=0,trim=4.8cm 2.5cm 8cm 1.5cm, clip,scale=0.39]{figure11.pdf}}
{\includegraphics[page=1,angle=0,trim=5.8cm 2.5cm 9cm 1.5cm, clip,scale=0.39]{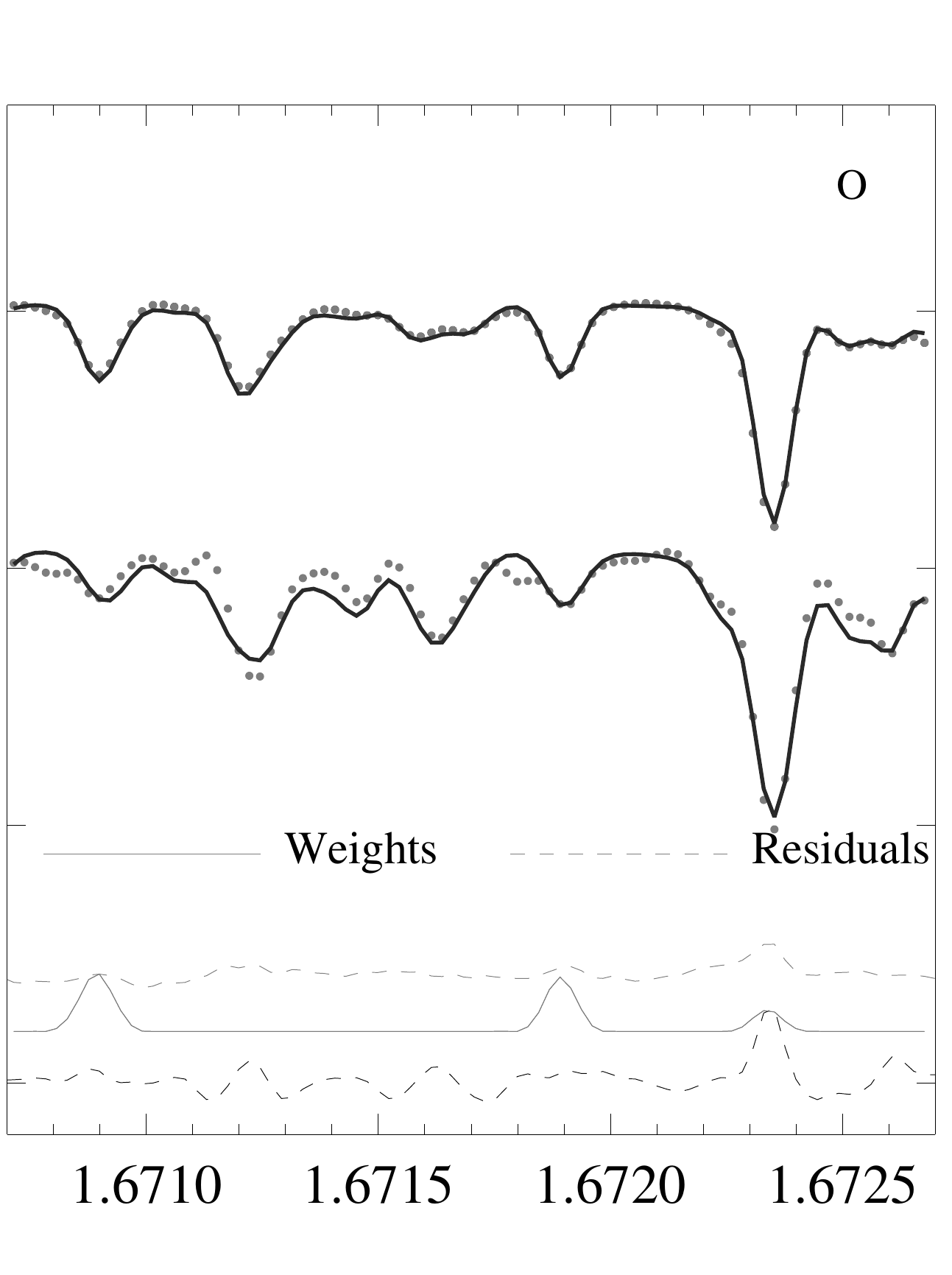}}\\
\end{center}
\caption{FTS spectrum (circles) and best spectral synthesis (black solid line) for Arcturus ($\alpha$ Boo) and $\mu$~Leo, for
small spectral windows targeted to different chemical elements. Wavelengths are vacuum values. Fits for iron (left), 
silicon (middle), and silicon (right) are presented, along with the abundance weights (grey solid line) described in Section~\ref{abndet}. 
Residuals between the 1-m- and the FTS-spectrum are also presented (dashed lines, top: Arcturus, bottom: $\mu$~Leo). Spectra, residuals and weights are shifted to fit the 
figure and the weights are also scaled for visibility.}

\label{abnsynfig}

\end{figure*}

\section{Examples with real data}

\begin{figure*} 
\figurenum{10}
\begin{center}

{\includegraphics[page=3,trim=0cm 0cm 0.0cm 0.5cm, clip,scale=1,width=6cm]{figure12.pdf}}\\
{\includegraphics[page=4,trim=0cm 3.6cm 3.0cm 5.2cm, clip,scale=1,width=5.2cm]{figure12.pdf}}
{\includegraphics[page=2,trim=5.5cm 3.6cm 3.0cm 5.2cm, clip,scale=1,width=4.1cm]{figure12.pdf}}
{\includegraphics[page=1,trim=5.5cm 3.6cm 3.0cm 5.2cm, clip,scale=1,width=4.1cm]{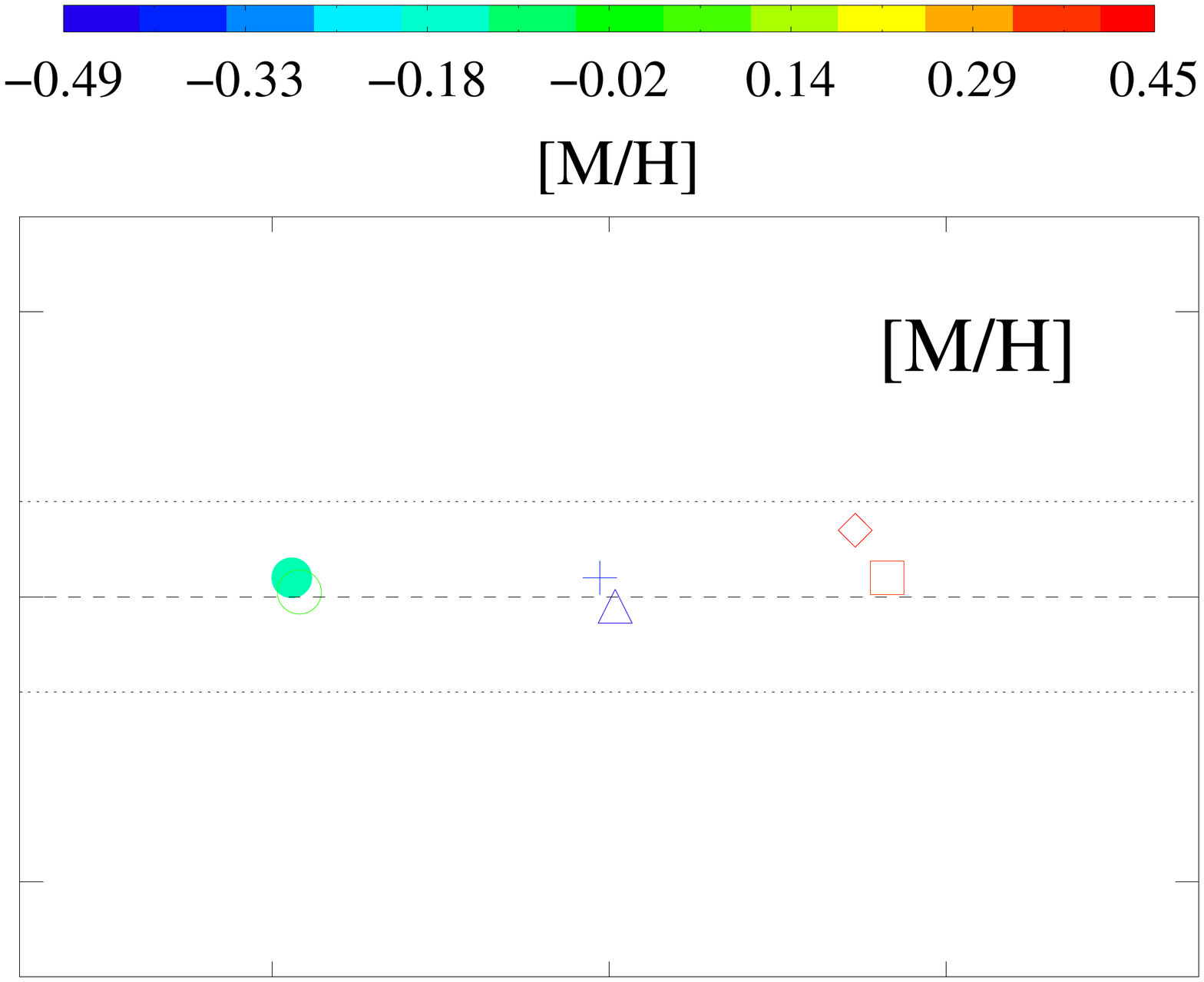}}
{\includegraphics[page=5,trim=5.5cm 3.6cm 3.0cm 5.2cm, clip,scale=1,width=4.1cm]{figure12.pdf}}\\
{\includegraphics[page=6,trim=0cm 3.6cm 3.0cm 5.2cm, clip,scale=1,width=5.2cm]{figure12.pdf}}
{\includegraphics[page=7,trim=5.5cm 3.6cm 3.0cm 5.2cm, clip,scale=1,width=4.1cm]{figure12.pdf}}
{\includegraphics[page=8,trim=5.5cm 3.6cm 3.0cm 5.2cm, clip,scale=1,width=4.1cm]{figure12.pdf}}
{\includegraphics[page=9,trim=5.5cm 3.6cm 3.0cm 5.2cm, clip,scale=1,width=4.1cm]{figure12.pdf}}\\
{\includegraphics[page=10,trim=0.0cm 3.6cm 3.0cm 5.2cm, clip,scale=1,width=5.2cm]{figure12.pdf}}
{\includegraphics[page=11,trim=5.5cm 3.6cm 3.0cm 5.2cm, clip,scale=1,width=4.1cm]{figure12.pdf}}
{\includegraphics[page=12,trim=5.5cm 3.6cm 3.0cm 5.2cm, clip,scale=1,width=4.1cm]{figure12.pdf}}
{\includegraphics[page=13,trim=5.5cm 3.6cm 3.0cm 5.2cm, clip,scale=1,width=4.1cm]{figure12.pdf}}\\
{\includegraphics[page=14,trim=0cm 0.cm 3.0cm 5.2cm, clip,scale=1,width=5.2cm]{figure12.pdf}}
{\includegraphics[page=15,trim=5.5cm 0.cm 3.0cm 5.2cm, clip,scale=1,width=4.1cm]{figure12.pdf}}
{\includegraphics[page=16,trim=5.5cm 0.cm 3.0cm 5.2cm, clip,scale=1,width=4.1cm]{figure12.pdf}}
{\includegraphics[page=17,trim=5.5cm 0.cm 3.0cm 5.2cm, clip,scale=1,width=4.1cm]{figure12.pdf}}\\

\end{center}

\caption{Stellar parameters and abundances differences ({\sc qascap}$\ -\ $reference) for four well studied stars.
For Arcturus and $\mu$-Leo we show both FTS and 1-m+APOGEE results, yielding six points in each test.
Data are colored according to their stellar metallicity derived from the global fit (\feh). The reference values for the 
comparison are from \cite{Smith13}. Differences for metallicities derived from both the 
global fit and dedicated Fe spectral windows are presented and seen to be nearly identical. The dashed and the dotted lines denote 0 and 
$\pm0.2$~dex differences, respectively. }
\label{ftsfig}
\end{figure*}

The pipeline was tested on real data using high quality ($S/N > 100$, $R\ge45,000$) 
$H$-band spectra for a set of bright field giants with previously derived abundances 
in the literature. The test data were obtained with the Fourier Transform Spectrograph (FTS) 
at the 4-m Mayall Telescope on Kitt Peak. The stars are 
the giants Arcturus ($\alpha$ Boo), $\beta$ And, $\delta$ Oph, and $\mu$ Leo. The stellar 
parameter coverage is $3825\le$ {\teff} $\le4550$~K, $0.90\le\ $\logg$\ \le1.70$ and $-0.47 \le$\ [Fe/H]\ $\le+0.31$. The 
spectra were smoothed with a Gaussian kernel to the APOGEE nominal spectral resolution 
of $R=22,500$. Observations of Arcturus and $\mu$ Leo taken with the APOGEE spectrograph linked to 
the New Mexico State University 1-m telescope at APO (Feuillet et al., in preparation) are also available and 
were used in our tests. All spectra were 
analyzed with seven free parameters, using a quick\footnote{Observed spectra were normalized and if needed also resampled, and run through \ferre.} version of ASPCAP 
({\sc qaspcap}). Gaussian and DR12 LSF libraries\footnote{ The 7D libraries are p\_apsKK-01-23k\_w123, p\_apsasGK\_131216\_lsfcombo5v6\_w123.} 
were adopted in the analysis of the non-APOGEE and 1m-spectra, respectively. Table~\ref{fts} gives the stellar parameters and chemical 
abundances\footnote{Abundances are given as $A(X)=\log{(N_X/N_\mathrm{H})}$+12, with $N_X$ the number density of atoms of element X.} 
from the analysis, along with reference values from \cite{Smith13}. Figure~\ref{abnsynfig} illustrates 
the quality of the spectral fits for the O, Si and Fe lines of Arcturus and {$\mu$}~Leo, along with the 
differences between the spectra from different sources.

 \subsection{Arcturus}

The Sun is a solid reference for spectroscopic studies of main-sequence stars, but Arcturus 
can be considered as a more appropriate one for studies of giant stars, which are the bulk 
of APOGEE targets. The APOGEE 
line list is based on a solar and Arcturus analysis (Shetrone et al. 2015). In this study, we analyzed two 
Arcturus spectra, the FTS atlas \citep[][$R=100,000$]{Hinkle95} 
degraded to the APOGEE spectral resolution, and the APOGEE 1m-spectrum.

Figure~\ref{ftsfig} shows a comparison of the stellar parameters and chemical abundances derived 
with {\sc qaspcap} with those in Smith et al. (2013). The latter study reported abundances 
from the manual analysis of the FTS spectra with a different line list and method than ASPCAP. 
Note that Smith el al.~(2013) derived {\teff} from photometric calibrations and \logg\ from luminosity 
along with stellar evolution models, while ASPCAP 
derives them purely spectroscopically. Differences in the derived stellar parameters for the studied stars
(e.g., {\teff} and  {\logg} values), as well as differences in the atomic data in the adopted line lists, 
can introduce abundance offsets with respect to their results. Also, the use of 
slightly different features for abundance determinations in this study and Smith et al. (2013; see Section 4.3) 
could be responsible for part of the differences in the abundances.

The use of the Arcturus atlas spectrum removes the parameter/abundance 
uncertainties associated with the modeling of the LSF. Our analysis of those data delivers effective 
temperature, metallicity, and microturbulent velocity values which are overall in good agreement with 
the results by Smith et al. (see blue crosses in Figure~\ref{ftsfig}). 
Compared to Smith et al., ASPCAP finds
Arcturus to be slightly cooler and more metal-rich, 
with offsets  of $-85$~K and $+0.04$~dex, and to have slightly higher {\microt} ($+0.06$~\kms)---see Table~\ref{fts}. 
The agreement for \logg\ is worse; ASCPAP infers a higher {\logg} by $\sim+0.4$~dex,
well outside the estimated 0.1~dex uncertainty reported by Smith et al. This 
discrepancy highlights the need to calibrate ASPCAP-derived 
values of {\logg} against empirical data.  
Both DR10 and DR12 release calibrated {\logg} values in addition to the direct
ASPCAP estimates (see \citealt{Meszaros13} and \citealt{Holtzman15}
for more details). 

We find good agreement with Smith et al. (2013) for the elemental abundances,
with differences that are typically smaller than or of the order of their estimated uncertainties.
The elements for which we find the best agreement ($< 0.05$~dex) 
are Fe, Ca, and Mn. They are followed by the abundance of Ni, with an offset less than $0.1$~dex. Offsets of about 
0.2~dex or larger are observed for  N, O, Si, K, and V. In the case of Si, the discrepancy between the results is larger than the 
estimated uncertainty, and it is probably related to our adoption of smaller $\log{gf}$ 
values for the Si~{\sc i} lines. Different ${\log{gf}}$s can also be responsible for the 
large abundance offset for K. For other elements, however, the differences in $\log{gf}$ in the adopted line lists
were typically smaller than 0.1~dex.

The APOGEE 1m-spectrum has high $S/N$ but suffers from distortions associated with the persistence in the 
detectors (see Majewski et al. 2015, Nidever et al. 2015), a complex LSF, and other factors that can degrade the performance of ASPCAP. Nonetheless, 
the results of our analysis are overall in good agreement with the reference values of Smith et al. (2013), and 
comparable to that obtained from the analysis of the Arcturus atlas spectrum (compare triangles versus crosses in Figure~\ref{ftsfig}). The exception is the microturbulent velocity, which has an offset 
of $-0.43$~{\kms} for the 1-m spectrum, significantly larger than that obtained for the Atlas spectrum ($+0.06$~\kms ). 
The DR12 analysis however fixes \microt\ based on \logg\ (see \S\ref{database}), yielding 
a value $\xi_{\rm t}=1.82\,$\kms\ that is only 0.04~\kms\ below Smith et al's value.
Other DR12 stellar parameters for Arcturus agree well with those found here for an analysis with the \microt\ as a free parameter:
\teff$\ = 4206$~K, \logg$\ = 2.01$~[cgs], and \feh$\ =-0.54$. The exception to this agreement is for the 1-m spectrum 
microturbulent velocity. Elemental abundance differences between the Atlas and 1-m analyses are generally
small, indicating little sensitivity to details of the data and
LSF.  The abundances showing the largest differences ($0.15-0.2$ dex)
are N, Al, Ti, and V. We note that the abundances of especially Ti, Si, and Al, along with {\logg} and {\microt}, are 
sensitive to the adopted LSF modeling (Gaussian versus a DR12 LSF) in the analysis of the 
1m-spectrum. 

 \subsection{Analysis of Other Stars}

We analyzed the FTS spectra of the super-solar metallicity red giant $\mu$~Leo 
(\fehw\ $=+0.31$) and the cool red giants $\beta$~And, and 
$\delta$~Oph (\teff\ $\lesssim 3900$~K), and the 1m-spectrum of $\mu$~Leo. 
The FTS spectra were retrieved from the Kitt Peak National Observatory archive \citep{Hall79}. 
This sample allows us to test ASPCAP results in a more metal-rich regime than the 
one probed in the analysis of Arcturus discussed above.

The differences between the results obtained for $\mu$~Leo (red diamond),  $\beta$~And (filled green circle), and $\delta$~Oph 
(empty green circle) using ASPCAP and those from the manual analysis in Smith et al. (2013)
are also shown as a function of {\teff} in Figure~\ref{ftsfig}.
At these higher metallicities the agreement between the stellar 
parameters from ASPCAP and Smith et al. (2013) remains good (see top panel of Figure~\ref{ftsfig}), 
similar or in some cases even better than for Arcturus. In general, and in particular for surface gravity 
and microturbulence, there seems to be an indication of a dependence on the effective temperature 
of the star.
\begin{itemize}

\item{ Surface gravity: overall there seems to be a systematic difference between the surface gravities derived 
by ASPCAP and Smith et al. (2013) in the sense that ASPCAP results are systematically larger.
The differences found for the surface gravities of $\beta$~And, and $\delta$~Oph are 
$\sim 0.2-0.3$~dex, but
for the most most metal rich and hottest star in our sample, $\mu$~Leo, 
the discrepancy is significant ($+0.68$~dex) and much larger than the expected uncertainties 
in Smith et al. (2013).}

\item{Microturbulence: as noted previously, there seems to be an overall trend with effective 
temperature. However, the microturbulent velocities generally agree with Smith et al. 
to within 0.4~\kms\ for the three stars.}

\item{Metallicities: the global metallicities agree with those of
Smith et al. (2013), to within 0.04, 0.01, and 0.14 dex
($\beta$~And, $\delta$~Oph, and $\mu$~Leo, respectively).
Agreement of Fe abundances is still better.}
\end{itemize}

Our individual chemical abundances match those presented in Smith et al. (2013), with differences 
typically smaller than $0.2$~dex. 
The best agreement is found for Ca. For some of the elements, there is a suggestive
trend of the abundance differences with the effective temperature of the stars,
e.g., for C and Ni, and perhaps a marginal trend for Fe, Mg, and Ca. 
Of course, with only four stars, two of similar temperature, the ability
to reliably identify such trends is limited.
For $\mu$~Leo, the \logg\ offset of +0.68 dex significantly affects the derived C 
abundance and may be responsible for the worse agreement on \cfe.
There is also some dependence of the abundance differences with Smith et al. 
on {\feh} for some of the elements, e.g., N, O and V. The trends of the abundance differences 
with {\teff} and {\feh} may result from the different stellar parameters adopted by Smith et al. 

Some of the elements with the most discrepant abundances for Arcturus (N, O, K, and Si), 
are also discrepant for some of the other test stars, differing by $\gtrsim 0.15$~dex from the 
values of \cite{Smith13}. 
However, these differences are still within the uncertainties estimated
by Smith et al. Only Si systematically exceeds the Smith et al.\ value in all the stars
by an amount larger than expected errors, further supporting the 
idea that atomic data are a major contributor to the differences between these
two studies. There 
is dispersion in the K abundance offset with Smith et al. (2013), which rules out the 
atomic data as the major source of discrepancy. Other elements that show 
larger-than-expected discrepancies are Mg ($\beta$~And) and Mn ($\delta$~Oph).

The result obtained from the APOGEE 1m-spectrum of $\mu$~Leo (red squares in Figure~\ref{ftsfig}) 
are similar to those from the FTS spectrum. The parameters that depend the most on which observations 
are adopted are, as in the case of Arcturus, the effective temperature and microturbulence. 

In summary, there is good overall agreement between our abundances and those of Smith et al. (2013). The dispersion 
(standard deviation) of abundance differences is $< 0.10$~dex. The abundances 
of Ca, Fe, and Ni (with the exception of $\mu$~Leo) generally agree quite well, while our most discrepant abundances 
are Si (not surprising given the differences in adopted atomic data). Abundance differences 
with the values in Smith et al. (2013) are due partly to the differences in the stellar parameters, 
in the atomic data, and/or in the analyzed spectra. The microturbulent velocity, in particular,
changes significantly depending on the source of the spectra.
 
 \begin{table*}
\tablenum{5}
\begin{center}
\setlength{\tabcolsep}{0.02in}
\caption{Stellar Parameter and Chemical Abundances}
\begin{tabular}{l|ccc|cc|cc|ccc|}
\hline
\hline
\multicolumn{1}{c}{}& \multicolumn{3}{c}{$\alpha$ Boo} & \multicolumn{2}{c}{$\beta$ And} & \multicolumn{2}{c}{$\delta$ Oph} & \multicolumn{3}{c}{$\mu$ Leo}\\

 & FTS & 1-m & Ref. &FTS & Ref. &FTS & Ref. &FTS & 1-m & Ref.\\ 
\hline

         $T_{\rm{eff}}$ (K)                &  $+ 4189$ &$+ 4207$ &$+  4275 $ & $+ 3823$ & $+  3825 $ & $+ 3832$ & $+  3850 $ & $+ 4492$  & $+ 4530$ &$+  4550$ \\
                $\log{g}$ (cgs)                 &  $+ 2.09$  &$+ 2.02$ &$+  1.70 $ & $+ 1.17$ & $+  0.90 $ & $+ 1.39$ & $+  1.20 $ & $+ 2.78$  & $+ 2.76$ &$+  2.10$ \\
            $\xi_{\rm{t}}$ (\kms)               &  $+ 1.92$ & $+ 1.43$ &$+  1.86 $ & $+ 2.32$ & $+  2.19 $ & $+ 2.25$ & $+  1.91 $ & $+ 1.95$  & $+ 1.28$ &$+  1.82$ \\
$[\mathrm{M}/\mathrm{H}]$  &  $-0.43 $ &$-0.49 $ &$ -0.47 $ &  $-0.18 $ & $ -0.22 $ & $+ 0.00 $ & $ -0.01 $ & $+ 0.45$   & $+ 0.35$ &$+  0.31$ \\
                                            A(Fe)  &  $+ 7.00$ &$+ 6.95$ & $+  6.98 $ & $+ 7.23$ & $+  7.23 $ & $+ 7.41$ & $+  7.44 $ & $+ 7.88$ & $+ 7.76$ &$+  7.76$ \\
                                            A(C)  &  $+ 8.09$ & $+ 7.99$ & $+  7.96 $ & $+ 8.05$ & $+  8.06 $ & $+ 8.27$ & $+  8.24 $ & $+ 8.80$ & $+ 8.69$ &$+  8.52$ \\
                                             A(N)  &  $+ 7.45$ & $+ 7.29$ & $+  7.64 $ & $+ 7.96$ & $+  8.05 $ & $+ 8.03$ & $+  8.20 $ & $+ 8.69$ & $+ 8.53$ &$+  8.71$ \\
                                              A(O)  &  $+ 8.45$ & $+ 8.39$ & $+  8.64 $ & $+ 8.58$ & $+  8.78 $ & $+ 8.71$ & $+  8.77 $ & $+ 9.16$ & $+ 9.04$ &$+  9.05$ \\
                                           A(Mg)  &  $+ 7.27$ & $+ 7.19$ & $+  7.15 $ & $+ 7.48$ & $+  7.26 $ & $+ 7.58$ & $+  7.54 $ & $+ 7.89$ & $+ 7.79$ &$+  7.85$ \\
                                            A(Al)  &  $+ 5.99$ & $+ 6.21$ & $+  6.16 $ & $+ 6.26$ & $+  6.12 $ & $+ 6.53$ & $+  6.45 $ & $+ 6.87$ & $+ 6.87$ &$+  6.90$ \\
                                            A(Si)  &  $+ 7.41$ & $+ 7.31$ & $+  7.12 $ & $+ 7.48$ & $+  7.18 $ & $+ 7.71$ & $+  7.53 $ & $+ 8.03$ & $+ 7.96$ &$+  7.76$ \\
                                             A(K)  &  $+ 4.63$ & $+ 4.58$ & $+  4.79 $ & $+ 4.75$ & $+  4.86 $ & $+ 5.32$ & $+  5.18 $ & $+ 5.58$ & $+ 5.33$ &$+  5.63$ \\
                                          A(Ca)  &  $+ 5.88$ & $+ 5.83$ & $+  5.84 $ & $+ 6.08$ & $+  6.02 $ & $+ 6.21$ & $+  6.24 $ & $+ 6.54$ & $+ 6.55$ &$+  6.62$ \\
                                            A(Ti)  &  $+ 4.46$ & $+ 4.62$ & $+  4.59 $ & $+ 4.90$ & $+  4.72 $ & $+ 5.02$ & $+  5.07 $ & $+ 5.52$ & $+ 5.44$ &$+  5.40$ \\
                                             A(V)  &  $+ 3.43$ & $+ 3.29$ & $+  3.61 $ & $+ 3.76$ & $+  3.66 $ & $+ 3.92$ & $+  3.86 $ & $+ 4.28$ & $+ 4.34$ &$+  4.18$ \\
                                          A(Mn)  &  $+ 4.90$ & $+ 4.88$ & $+  4.86 $ & $+ 5.27$ & $+  5.18 $ & $+ 5.57$ & $+  5.34 $ & $+ 5.89$ & $+ 5.89$ &$+  5.79$ \\
                                           A(Ni)  &  $+ 5.84$ & $+ 5.83$ & $+  5.77 $ & $+ 6.02$ & $+  6.01 $ & $+ 6.21$ & $+  6.18 $ & $+ 6.73$ & $+ 6.60$ &$+  6.60$ \\

\hline
\end{tabular}
\end{center}
\label{fts}
\end{table*}

\subsection{The Open Cluster M67}

\begin{figure} 
\figurenum{11}
\begin{center}
{\includegraphics[page=3,trim=2cm 2.2cm 16.2cm 10cm, clip,scale=1,width=4cm]{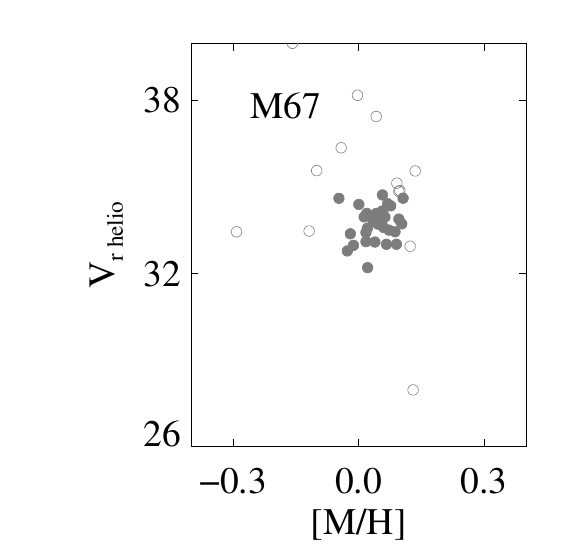}}\\
{\includegraphics[page=6,trim=2.5cm 3.9cm 17.0cm 10.7cm, clip,scale=1,width=3.5cm]{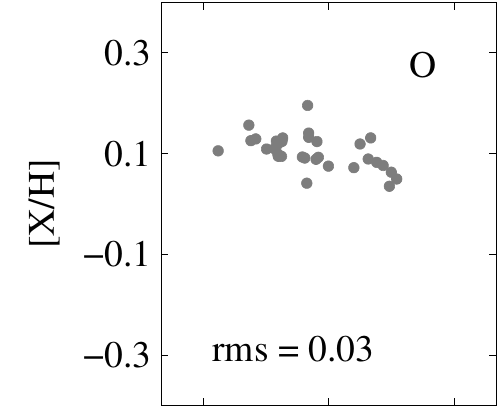}}
{\includegraphics[page=13,trim=5.2cm 3.9cm 17.0cm 10.7cm, clip,scale=1,width=2.4cm]{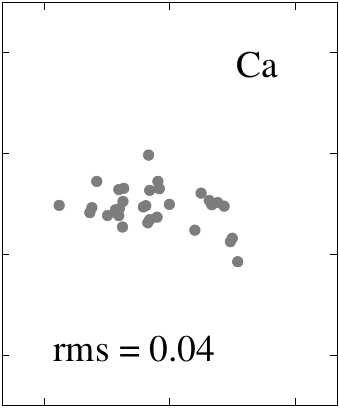}}
{\includegraphics[page=17,trim=5.2cm 3.9cm 17cm 10.7cm, clip,scale=1,width=2.4cm]{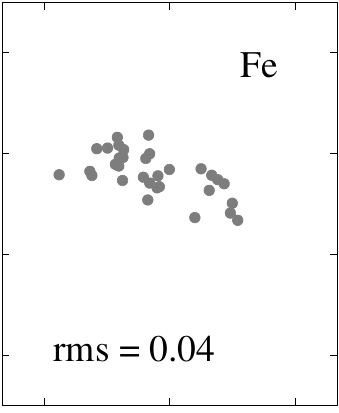}}\\
{\includegraphics[page=10,trim=2.5cm 2cm 17cm 10.7cm, clip,scale=1,width=3.5cm]{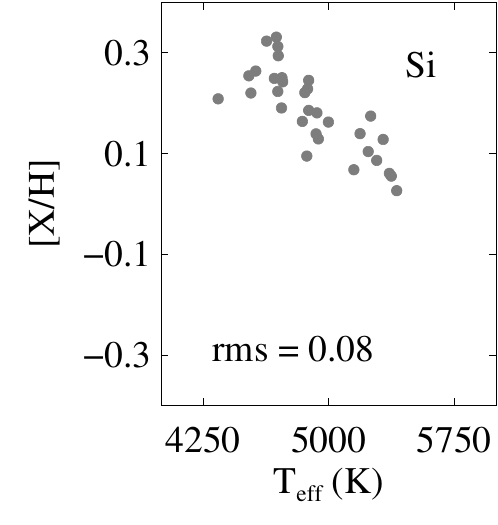}}
{\includegraphics[page=14,trim=5.2cm 2cm 17cm 10.7cm, clip,scale=1,width=2.4cm]{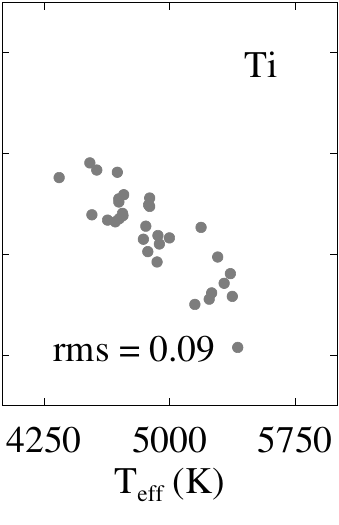}}
{\includegraphics[page=16,trim=5.2cm 2cm 17cm 10.7cm, clip,scale=1,width=2.4cm]{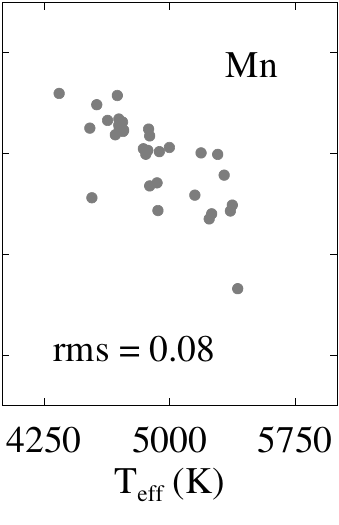}}\\

\end{center}
\caption{The DR12 heliocentric radial velocities and chemical abundances for a sample of $\sim30$ M67 
members. The filled circles in top-left panel denote velocity and \feh\ (from the global fit) cluster membership, 
and the empty circles represent outliers. The other panels show examples of elements with high (O, Ca, and Fe), 
and low (Si, Ti, and Mn) precision estimates. Abundances are shown versus \teff.}
\label{m67fig}
\end{figure}

Stellar clusters are ideal benchmarks to calibrate abundance determinations. Stars in clusters 
share essentially the same chemical content, though some globular clusters and/or chemical elements show some 
variations associated with multiple populations or with mixing processes. 
Abundance trends with \teff\ are an indication of systematic uncertainties, providing that 
mixing processes are not altering the chemical composition.

APOGEE observed several clusters in a wide metallicity range, including 
the very well studied solar-metallicity open cluster M67. The APOGEE 
results for the cluster show ASPCAP abundances of high precision. Our cluster membership 
is based on a combination of photometry, radial velocity, and metallicity information.  
We redefined the sample using a $\sim3.0\times\sigma$ (Gaussian) cut in both radial 
velocity and metallicity.  DR12 heliocentric radial velocities and direct ASPCAP chemical abundances 
(without the calibration offsets applied to DR12 as described by \citealt{Holtzman15})
are presented in Figure~\ref{m67fig}. 
Cluster members and outliers are presented in top panel. The cluster radial velocity 
and {\feh} derived from the Gaussian fits to the parameters distribution of the cluster members 
are 33.51~km~s$^{-1}$ (standard deviation of 0.66~km~s$^{-1}$) and 0.03 (standard deviation of 0.04~dex), respectively.
A similar metallicity value of 0.06 is obtained from iron lines. 
 
The lower panels of Figure~\ref{m67fig} plot DR12 values of [X/H] versus \teff, after eliminating non-cluster members, and 
stars flagged as {\it BAD} or with {\it BAD} abundances (treating each chemical element separately).
The O, Ca, and Fe abundances show small dispersion (0.03-0.04 dex) and little or no
trend with \teff.  The Si, Ti, and Mn abundances show clear trends with \teff\ that 
likely indicate systematic ASPCAP errors in this $4250-5500\,{\rm K}$ temperature
range.  The dispersion around this trend remains small, and even with the trend
the dispersion of values is only 0.08-0.09~dex.  Larger dispersions are found
for N and V (not shown in the figure), though the former may be affected by
mixing processes.  \cite{Holtzman15} present comparisons to a wider range of
open cluster data and derive temperature-dependent abundance calibration offsets
element by element, which are applied to the APOGEE DR12 release.

\section{Conclusions}

ASPCAP is the pipeline for deriving stellar parameters and chemical abundances 
from APOGEE spectra. The pipeline matches the observations to a set of synthetic spectrum 
templates using the {\ctwo} minimization in a multidimensional parameter space. 
Stellar parameters are derived first from the entire APOGEE spectral range, followed by the determination of individual 
chemical abundances from spectral windows optimized for each element. The precision and the level of sophistication 
of the high-dimensional analysis that ASPCAP performs is unprecedented in such a large volume of data.

ASPCAP has three main components: the model spectral libraries, the {\ferre} optimization code 
that searches for the best fit, and the IDL wrapper for book-keeping and data pre- and post-processing. 
In this paper we described each component and presented the pipeline configuration used in DR10 \citep{DR10} 
and DR12 \citep{DR12}. The employed algorithms have proven to work well with both simulations 
and observations. 

Random abundance uncertainties are expected to be typically $< 0.1$~dex, based 
on tests with simulations, and the DR12 results for M67. For accuracy, we expect typically $\lesssim0.20$~dex 
based on the comparison of our abundance results with the values of Smith et al. (2013) for a set of reference stars.

Some of the issues we have detected include:
\begin{itemize}

\item Our tests indicate that a detailed modeling of the LSF is important and the systematic 
effects associated with poor LSF matching may be appreciable. An empirical LSF has been 
used for DR12 versus a Gaussian LSF of constant $R$ for DR10.

\item PCA compression of the synthetic libraries may affect ASPCAP results for low 
metallicity spectra (\feh $\ < -1$), which lie outside the bulk of the APOGEE sample, but, nonetheless, requires further investigation.
 \item Due to the lack of information in metal-poor or warm spectra, 
ASPCAP's performance is poorer for these cases, and an alternative strategy, where fewer parameters are 
involved in the modeling, is needed in these regions of the parameter space.
\item There are significant uncertainties in the inferred nitrogen abundances, as a result of 
the modest sensitivity of CN lines to changes in the N abundance.

\end{itemize}

ASPCAP continues to evolve and efforts concentrate now on addressing issues such as
extending the parameter coverage, establishing abundance upper limits to the abundances from 
undetected spectral lines, and improving the LSF modeling. Spectral libraries for cooler stars are 
already available and will be incorporated soon. 

We plan to investigate whether individual elemental abundances can be fit independently of each other 
to deliver more accurate abundances. The larger APOGEE-2 project of SDSS-IV brings additional 
motivation for continuing the development of ASPCAP.

We acknowledge funding from NSF grants AST11-09718 and AST-907873. 
Funding for SDSS-III has been provided by the Alfred P. Sloan
Foundation, the Participating Institutions, the National Science
Foundation, and the U.S. Department of Energy Office of Science. The
SDSS-III website is http://www.sdss3.org/.
SDSS-III is managed by the Astrophysical Research Consortium for the
Participating Institutions of the SDSS-III Collaboration including the
University of Arizona, the Brazilian Participation Group, Brookhaven
National Laboratory, University of Cambridge, Carnegie Mellon
University, University of Florida, the French Participation Group, the
German Participation Group, Harvard University, the Instituto de
Astrofisica de Canarias, the Michigan State/Notre Dame/JINA
Participation Group, Johns Hopkins University, Lawrence Berkeley
National Laboratory, Max Planck Institute for Astrophysics, Max Planck
Institute for Extraterrestrial Physics, New Mexico State University,
New York University, Ohio State University, Pennsylvania State
University, University of Portsmouth, Princeton University, the
Spanish Participation Group, University of Tokyo, University of Utah,
Vanderbilt University, University of Virginia, University of
Washington, and Yale University. 

Support for A.E.G.P. was provided by SDSS-III/APOGEE. C.A.P. is thankful for support from the Spanish Ministry of Economy
and Competitiveness (MINECO) through grant AYA2014-56359-P.
Sz.M. has been supported by the J{\'a}nos Bolyai Research Scholarship of the 
Hungarian Academy of Sciences. D.A.G.H., and O.Z. acknowledge support provided by the Spanish
Ministry of Economy and Competitiveness under grant AYA-2011-27754 and AYA2014-58082-P.

\acknowledgments

\end{document}